\documentclass[twocolumn,times]{aastex62}

\usepackage{hyperref}
\usepackage{amsmath}

\newcommand{\lta}{\lesssim}

\newcommand{\kms}{km\,s$^{-1}$}
\newcommand{\masyr}{\,\mathrm{mas}\,\mathrm{yr}^{-1}}
\newcommand{\muw}{\mu_\mathrm{W}}
\newcommand{\mun}{\mu_\mathrm{N}}
\newcommand{\Msun}{\mathrm{M}_{\odot}}
\newcommand{\hst}{{\it HST}}
\newcommand{\vlos}{v_\mathrm{LOS}}
\newcommand{\vrad}{v_\mathrm{rad}}
\newcommand{\vtan}{v_\mathrm{tan}}
\newcommand{\vtot}{v_\mathrm{tot}}

\newcommand{\vx}{v_X}
\newcommand{\vy}{v_Y}
\newcommand{\vz}{v_Z}
\newcommand{\rmax}{r_\mathrm{max}}

\newcommand{\Mvir}{M_\mathrm{vir}}

\newcommand{\vcirc}{v_\mathrm{circ}}

\newcommand{\Mrmax}{M \left( < \rmax \right)}

\newcommand{\rgc}{R_\mathrm{GC}}
\newcommand{\rperi}{R_\mathrm{peri}}
\newcommand{\rapo}{R_\mathrm{apo}}

\submitjournal{ApJ}

\shorttitle{HST Proper Motions of Distant Milky Way Globular Clusters}
\shortauthors{Sohn et al.}

\begin{document}

\title{\textbf{\large Absolute \textit{HST} Proper Motion (HSTPROMO) of Distant Milky Way Globular Clusters: \\ 
       Galactocentric Space Velocities and the Milky Way Mass}} 

\author[0000-0001-8368-0221]{Sangmo Tony Sohn} 
\affiliation{Space Telescope Science Institute, 
             3700 San Martin Drive, 
             Baltimore, MD 21218, USA}

\author[0000-0002-1343-134X]{Laura L. Watkins}
\affiliation{Space Telescope Science Institute, 
             3700 San Martin Drive, 
             Baltimore, MD 21218, USA}

\author[0000-0003-4207-3788]{Mark A. Fardal}
\affiliation{Space Telescope Science Institute, 
             3700 San Martin Drive, 
             Baltimore, MD 21218, USA}

\author[0000-0001-7827-7825]{Roeland P. van der Marel}
\affiliation{Space Telescope Science Institute, 
             3700 San Martin Drive, 
             Baltimore, MD 21218, USA}
\affiliation{Center for Astrophysical Sciences, Department of Physics \& Astronomy, 
             Johns Hopkins University, 
             Baltimore, MD 21218, USA}

\author[0000-0001-6146-2645]{Alis J. Deason}
\affiliation{Institute for Computational Cosmology,
             Department of Physics, University of Durham,
             South Road, Durham DH1 3LE, UK}

\author{Gurtina Besla}
\affiliation{Department of Astronomy, University of Arizona,
             933 North Cherry Avenue, 
             Tucson, AZ 85721, USA}

\author{Andrea Bellini}
\affiliation{Space Telescope Science Institute, 
             3700 San Martin Drive, 
             Baltimore, MD 21218, USA}

\begin{abstract} 
We present {\it Hubble Space Telescope (HST)} absolute proper motion 
(PM) measurements for 20 globular clusters (GCs) in the Milky Way (MW)
halo at Galactocentric distances $\rgc \approx 10$--100 kpc, with
median per-coordinate PM uncertainty $0.06\masyr$. Young and old halo
GCs do not show systematic differences in their 3D Galactocentric
velocities, derived from combination with existing line-of-sight
velocities. We confirm the association of Arp\,2, Pal\,12, Terzan\,7,
and Terzan\,8 with the Sagittarius (Sgr) stream. These clusters and NGC\,6101 have tangential
velocity $\vtan > 290$~\kms, whereas all other clusters have $\vtan <
200$~\kms. NGC\,2419, the most distant GC in our sample, is also
likely associated with the Sgr stream, whereas NGC\,4147, NGC\,5024,
and NGC\,5053 definitely are not. We use the distribution of orbital
parameters derived using the 3D velocities to separate halo GCs that
either formed within the MW or were accreted. We also assess the
specific formation history of e.g. Pyxis and Terzan\,8. We constrain
the MW mass {\it via} an estimator that considers the full 6D
phase-space information for 16 of the GCs from $\rgc = 10$ to 40 kpc.
The velocity dispersion anisotropy parameter $\beta = 
0.609\substack{+0.130 \\ -0.229}$. The enclosed mass $M
(<39.5\,\mathrm{kpc}) = 0.61\substack{+0.18 \\ -0.12} \times
10^{12}\,\Msun$, and the virial mass $M_\mathrm{vir} =
2.05\substack{+0.97 \\ -0.79} \times 10^{12}\,\Msun$, are
consistent with, but on the high side among recent mass estimates in
the literature.
\end{abstract}

\keywords{astrometry ---
Galaxy: globular clusters: individual (Arp\,2, IC\,4499, NGC\,1261, NGC\,2298, 
NGC\,2419, NGC\,4147, NGC\,5024, NGC\,5053, NGC\,5466, NGC\,6101, NGC\,6426, 
NGC\,6934, NGC\,7006, Pal\,12, Pal\,13, Pal\,15, Pyxis, Rup\,106, 
Terzan\,7, Terzan\,8) ---
Galaxy: halo --- 
Galaxy: kinematics and dynamics ---
proper motions
}

\section{Introduction}
\label{s:intro}

Our Milky Way (MW) consists of several baryonic components: a central
bulge; thin and thick disks; and an extended metal-poor halo. These
components are embedded in a dark halo that contains most of the mass.
The structure and kinematics of the baryonic components contain
crucial information on the formation, evolution, and mass of our MW.
There are some 150 globular clusters (GCs) distributed throughout
the MW's baryonic components \citep[][2010 edition, hereafter H10]{har96}. 
These GCs are bright and they probe a large range of Galactocentric 
radii ($\rgc$). Many of their physical properties (e.g., 
distances, chemical abundances, line-of-sight velocities, and ages) 
are relatively easy to measure. These properties make GCs useful  
objects for studies of the MW's present structure and past evolution. 

GCs follow similar correlations between metallicity, spatial distribution, 
and kinematics as the other baryonic components. The metal-rich 
``bulge/disk GCs'' are found almost exclusively at $\rgc < 8$ kpc, 
and lie in a flattened distribution with significant rotation about the MW. 
The metal-poor ``halo GCs'' are found in a more spherical distribution 
extending to $\rgc \sim 120$ kpc with more pressure support from 
random motions. Most of the Galactic GCs ($\sim75$\%) belong to the 
halo component, and these are of particular interest for understanding 
the structure and evolution of the MW. Their orbital timescales are very 
long compared to the age of the Galaxy, and hence, the phase-space structure 
of the halo is intimately linked to its accretion history. Furthermore, 
their extreme radial extent makes their kinematics an excellent tracer 
for the gravitational potential of the halo.

Galaxy formation theories predict that galactic stellar halos are 
built up by cannibalizing dwarf galaxies. GCs residing in the MW halo 
provide a wealth of observational evidence supporting this hierarchical
paradigm. The pioneering work by \citet{sea78} found no metallicity 
gradient for the GCs beyond $\rgc = 8$ kpc, and later, \citet{zin93}
found that GCs can be divided into two classes based on their 
horizontal branch (HB) morphology: the old halo (OH) GCs with bluer mean HB 
color, and the young halo (YH) GCs with redder mean HB color (at given 
metallicity). The HB-based classification implies the observed differences 
are believed to be due to age. Establishing the age difference from 
main-sequence turnoff measurements has long proved difficult until high-quality 
\hst\ data allowed \citet{mar09} and \citet{dot10,dot11} to show convincingly 
that the OH GCs are indeed almost as old as the universe ($\sim 13$ Gyr), 
while the YH GCs are significantly (i.e., at least $\sim 1$ Gyr) younger 
in general. \citet{mac04} showed that the OH GCs are confined to $\rgc < 30$ 
kpc, exhibit prograde rotation about the MW, and are compact in size, while 
the YH GCs extend to $\rgc \approx 120$ kpc, show no signs of rotation about 
the MW, and are extended in structure. These results can be interpreted by 
assuming that the OH GCs formed during an early dissipative collapse 
\citep{egg62}, while the YH GCs were subsequently accreted and are of external 
origin. \citet{mar09} indeed found that GCs proposed to be accreted from either 
the Sagittarius (Sgr) dSph, the Canis Major (CMa) overdensity, or the Monoceros 
Ring are mostly identified as YH GCs. While the general paradigm of hierarchical 
galaxy formation appears validated, open questions on the details of such process 
remain, including the origin of GCs. For example, identifying which GCs are 
accreted, and establishing to which parent galaxies they are associated are 
crucial to understanding how the MW halo formed and evolved. The goal of this 
study is to provide insights into these questions using accurate proper motion 
(PM) measurements based on multi-epoch \hst\ data.

The MW mass is a fundamental quantity for understanding the MW in a 
cosmological context. Many methods have been used to estimate the MW 
mass based on the ensemble kinematics of tracers such as GCs or 
satellite galaxies \citep[e.g.,][]{wil99,wat10}, halo stars 
\citep{xue08,dea12,kin15}, using hypervelocity stars 
\citep{gne10,ros17,fra17}, the orbits of individual galaxies such as 
the Magellanic Clouds \citep{bes07,pen16,pat17}, Leo~I \citep{soh13,boy13}, 
the ensemble of satellite galaxies \citep{pat18}, or the Local Group 
timing argument \citep{vdm12}. Despite these works, the total MW mass 
remains poorly known. This is due in part to the small number of tracers at 
large radii, uncertainties on the bound/unbound status of some satellites, 
and cosmic variance. But most significantly, there is a profound lack of 3D 
kinematical information. Studies based entirely on line-of-sight velocities 
suffer from a well-known mass-anisotropy degeneracy. While attempts have 
been made using the predictions of numerical simulations to estimate the 
unknown velocity anisotropy of kinematical tracers 
\citep[e.g.,][]{xue08,dea11}, direct measurements of tangential motions 
of halo tracers are highly desirable. GCs are perhaps the best tracers 
for this purpose, since their distances and line-of-sight velocities ($\vlos$)
are well known. Furthermore, it is likely that GCs are a more relaxed population 
than satellite galaxies or individual stars. In this study, we calculate 
the anisotropy parameter of the MW halo using our PM results, and 
provide reliable estimates of the MW mass.

Almost all GCs of the MW have their $\vlos$ measured (H10), and these have
been used to provide significant insights into the properties and origin 
of the MW GC system. Full 3D velocities, which can only be accessed through 
PM measurements, are much more powerful. However, the required PM measurements 
are extremely challenging. The velocity dispersion of the MW halo is 
$\sigma \approx 120$~\kms. At distances of $\approx 10-100$ kpc, this 
corresponds to PMs of $\sim 2-0.2 \masyr$. To constrain orbits, these PMs 
must be measured to accuracies $\lta 10-20$\%. Existing PM measurements of 
GCs are mostly from ground-based photographic and/or CCD observations. 
While many PMs are available \citep[e.g.,][]{din99,din00}, their quality 
generally only reaches the required precision when distance from the Sun 
$D_{\odot} \lta 10$ kpc. Thus, there are few halo GCs at significant distance 
with accurately known PMs.

Multi-epoch \hst\ data are exquisitely well suited for astrometric and
PM science thanks to \hst's stability, high spatial resolution, and
well-determined PSFs and geometric distortions \citep[e.g.,][]{and06b}. 
In \citet{soh12,soh13,soh15,soh16,soh17}, we developed techniques 
to measure absolute PMs of resolved stellar systems by comparing the 
average shift of stars with respect to distant background galaxies. 
We applied these techniques to measure accurate PMs of M31, Leo~I, 
stars along the stellar streams, and Draco and Sculptor dwarf spheroidal 
galaxies at a wide range of distances (30--770 kpc) as part of the 
HSTPROMO collaboration. We utilize these established techniques 
to obtain PMs with unprecedented accuracies in this paper.

This paper is organized as follows. Section~\ref{s:obs_and_data} 
describes the observations and data analysis steps. 
Section~\ref{s:results} presents the PM results. In 
Section~\ref{s:spacemotions}, we combine our PM measurements with 
existing observations to explore the space motions of our 
target GCs, and in Section~\ref{s:mwmass}, we estimate the MW mass 
using the GCs as dynamical tracers.
Section~\ref{s:conclusions} presents concluding remarks.

\section{Observations and Data Analysis}
\label{s:obs_and_data}

%
\begin{deluxetable*}{lcccccccc}
\renewcommand{\arraystretch}{0.9}
\tablecaption{Basic parameters and observation summary of target 
              halo globular clusters
              \label{t:targets}
}
\tablehead{
\colhead{}        & \colhead{}                & \colhead{}       & \colhead{}                       & \multicolumn{2}{c}{Epoch~1}                              &\colhead{} & \multicolumn{2}{c}{Epoch~2} \\
\cline{5-6} \cline{8-9} 
\colhead{}        & \colhead{$\rgc$} & \colhead{}       & \colhead{}                       & \colhead{Date}    & \colhead{Exp. Time\tablenotemark{b}} &\colhead{} & \colhead{Date} & \colhead{Exp. Time\tablenotemark{b}} \\
\colhead{Cluster} & \colhead{(kpc)}  & \colhead{[Fe/H]} & \colhead{Group\tablenotemark{a}} & \colhead{(Y-M-D)} & \colhead{(s$\,\times\,N$)}           &\colhead{} & \colhead{(Y-M-D)} & \colhead{(s$\,\times\,N$)}
          }
\startdata
Arp\,2    &  22.3 & $-1.75$ & Young & 2006-04-22 & \phn345s$\,\times$\phn5 & & 2016-05-03 & \phn555s$\,\times\,$8 \\
IC\,4499  &  16.8 & $-1.53$ & Young & 2010-07-01 & \phn603s$\,\times$\phn4 & & 2016-06-24 & \phn909s$\,\times\,$6 \\
NGC\,1261 &  18.4 & $-1.27$ & Young & 2006-03-10 & \phn350s$\,\times$\phn5 & & 2017-03-12 & \phn607s$\,\times\,$8 \\
NGC\,2298 &  15.5 & $-1.92$ & Old   & 2006-06-12 & \phn350s$\,\times$\phn5 & & 2016-06-08 & \phn563s$\,\times\,$8 \\
NGC\,2419 &  95.1 & $-2.15$ & Old   & 2006-12-06 &    1107s$\,\times$\phn4 & & 2016-11-25 &    1015s$\,\times\,$8 \\
NGC\,4147 &  21.7 & $-1.80$ & Young & 2006-04-11 & \phn340s$\,\times$\phn5 & & 2017-04-14 & \phn548s$\,\times\,$8 \\
NGC\,5024 &  19.5 & $-2.10$ & Old   & 2006-03-02 & \phn340s$\,\times$\phn5 & & 2016-02-21 & \phn549s$\,\times\,$8 \\
NGC\,5053 &  18.5 & $-2.27$ & Old   & 2006-03-06 & \phn340s$\,\times$\phn5 & & 2017-03-10 & \phn549s$\,\times\,$8 \\
NGC\,5466 &  16.8 & $-1.98$ & Old   & 2006-04-12 & \phn340s$\,\times$\phn5 & & 2017-03-29 & \phn536s$\,\times\,$8 \\
NGC\,6101 &  10.6 & $-1.98$ & Old   & 2006-05-31 & \phn370s$\,\times$\phn5 & & 2016-05-13 & \phn520s$\,\times\,$9 \\
NGC\,6426 &  14.8 & $-2.15$ & Old   & 2009-08-04 & \phn500s$\,\times$\phn4 & & 2016-08-09 & \phn781s$\,\times\,$6 \\
NGC\,6934 &  13.3 & $-1.47$ & Young & 2006-03-31 & \phn340s$\,\times$\phn5 & & 2016-03-31 & \phn546s$\,\times\,$8 \\
NGC\,7006 &  37.8 & $-1.52$ & Young & 2009-10-05 & \phn505s$\,\times$\phn4 & & 2016-09-27 & \phn784s$\,\times\,$6 \\
Pal\,12   &  15.3 & $-0.85$ & Young & 2006-05-21 & \phn340s$\,\times$\phn5 & & 2016-06-11 & \phn547s$\,\times\,$8 \\
Pal\,13   &  23.5 & $-1.88$ & Old   & 2010-07-10 & \phn610s$\,\times$\phn4 & & 2016-07-15 & \phn887s$\,\times\,$6 \\
Pal\,15   &  39.2 & $-2.07$ & Old   & 2009-10-16 & \phn500s$\,\times$12    & & 2015-10-09 & \phn550s$\,\times\,$7 \\
Pyxis     &  39.5 & $-1.20$ & Young & 2009-10-11 & \phn517s$\,\times$\phn4 & & 2015-10-10 & \phn800s$\,\times\,$6 \\
Rup\,106  &  19.0 & $-1.68$ & Young & 2010-07-04 & \phn550s$\,\times$\phn4 & & 2016-07-12 & \phn843s$\,\times\,$6 \\
Terzan\,7 &  17.9 & $-0.32$ & Young & 2006-06-03 & \phn345s$\,\times$\phn5 & & 2016-05-04 & \phn555s$\,\times\,$8 \\
Terzan\,8 &  21.4 & $-2.16$ & Old   & 2006-06-03 & \phn345s$\,\times$\phn5 & & 2016-04-28 & \phn555s$\,\times\,$8 \\
\enddata
\tablenotetext{a}{Cluster groups based on relative ages from the following references: \citet{mar09} for Arp\,2, NGC\,1261, NGC\,2298, NGC\,4147, NGC\,5024, NGC\,5053, NGC\,5466, NGC\,6101, NGC\,6934, Pal\,12, Terzan\,7, and Terzan\,8; \citet{dot10} for NGC\,2419, ; and \citet{dot11} for IC\,4499, NGC\,6426, NGC\,7006, Pal\,15, Pyxis, and Rup\,106; \citet{ham13} for Pal\,13.}
\tablenotetext{b}{Integration time per individual exposure $\times$ number of exposures.}
\end{deluxetable*}
%

\subsection{Multi-epoch \hst\ Data}
\label{ss:multiepdata}

Selection of our target GCs was dictated by both observational constraints 
and scientific needs. During the initial stage of this project, we considered 
halo GCs with existing first-epoch ACS/WFC or WFC3/UVIS data in the Mikulski 
Archive for Space Telescope (MAST). We analyzed these data and selected GCs 
based on the number of compact galaxies with high signal-to-noise ($S/N$) in 
the background. For some cases, the individual exposures of the first-epoch 
images were too short, or the fields were too crowded to find a good number 
of background galaxies in the fields. Among the GCs with good first-epoch 
data, we selected a sample that covers a wide range of distance, metallicity, 
and age. We have also deliberately included GCs claimed to be associated 
with the Sgr dSph, and one of the most distant and luminous GCs NGC\,2419. 

First-epoch data for most of our target GCs, with the exception of 
Pal\,13 and NGC\,2419, were obtained through the two \hst\ survey 
programs GO-10775 \citep{sar07} and GO-11586 \citep{dot11}. These were 
originally taken for constructing high-quality color-magnitude diagrams 
(CMDs). For Pal\,13 and NGC\,2419, we used the images obtained through 
\hst\ programs GO-11680 (PI:G. Smith) and GO-10815 (PI: T. Brown), 
respectively. The Pal\,13 data were obtained to study its main sequence 
luminosity function, while the NGC\,2419 data were obtained as 
``ACS auto-parallel'' of the primary ACS/HRC observations that targeted 
the center of this cluster. All first-epoch observations were obtained 
using the ACS/WFC except for Pal\,13 which was observed with the WFC3/UVIS. 
Also, all targets were observed using two filters (F606W and F814W) 
except for NGC\,2419, which was only observed using F814W in 
the first epoch.

We obtained second-epoch data for all target clusters through 
our \hst\ program GO-14235 (PI: S. T. Sohn). We used the same detectors, 
telescope pointings, and orientations as in the first-epoch observations. 
In some cases, unavailability of guide stars due to some changes  
made in the Guide Star Catalog used by \hst\ forced our second-epoch 
orientations to be slightly offset with respect to the first-epoch ones, 
but the discrepancies were at most $\sim 1\degr$. For all target GCs but 
NGC\,2419, the second-epoch data were obtained only with F606W (F814W for 
Pal\,15). For NGC\,2419, we obtained both F606W and F814W exposures for 
constructing a CMD that was used for selecting members of NGC\,2419. 
Individual exposure times for each image of our second-epoch observations 
were mostly about $\sim 60$\% longer than those of the first-epoch. 
This was to ensure that our second-epoch data has high signal-to-noise 
ratio for constructing reliable templates for stars and galaxies 
that are later used for determining their accurate positions.
\footnote{These templates are empirically created for each star 
and galaxy, and takes into account the morphology of the source, the 
point-spread function (PSF), and the pixel binning. Details can be found 
in \citet{soh12}.} Table~\ref{t:targets} lists the summary of observations 
for the target GCs used in this paper, along with their Galactocentric 
distances, metallicities (as listed in H10), and which age group they 
belong to (see Table captions for details). The median time baseline of 
our multi-epoch observations is 10.0 years.

\subsection{Measurement of Absolute Proper Motions}
\label{ss:measurement}

Our methodology for measuring PMs largely follows that described in 
\citet{soh12,soh13,soh17}, so we refer readers interested in the details of 
the PM measurement process to these papers. Here we summarize the 
important steps and discuss specifics that are important for this study. 

We downloaded all first- and second-epoch {\tt *\_flc.fits} images from 
the MAST. These {\tt *\_flc.fits} images have been processed for the 
imperfect charge transfer efficiency (CTE) using the pixel-based 
correction algorithms of \citet{and10}. We ran the {\tt img2xym\_WFC.09x10} 
program \citep{and06a} and an equivalent version for WFC3/UVIS on each 
{\tt *\_flc.fits} image to determine a position and a flux for each star. 
The measured positions were converted to the distortion-corrected frames 
using the known geometric distortion solutions for ACS/WFC \citep{and06b} 
and WFC3/UVIS \citep{bel11}. For each target GC, we then stacked 
all the second-epoch data to create a high-resolution image. 
For Pal\,15, we used the deeper first-epoch data to create a stacked image. 
For subsequent discussion in this section regarding data from which epoch 
was used, we did all the measurements in the opposite sense for Pal\,15, 
i.e., we used the first- instead of the second-epoch data and 
\textit{vice versa}.

Stars associated with each GC were identified via CMDs constructed from 
photometry of both F606W and F814W images. Background galaxies were 
identified through a two-step process: first, an initial objective 
selection based on running SExtractor \citep{ber96} on stacked images; 
second, visually inspecting each source identified in the initial stage. 
A template was constructed from the stacked images for each star and 
background galaxy identified this way. This template was used for 
positional measurements of each star/galaxy in each exposure in each 
epoch. For images of the second epoch, templates were fitted directly, 
while for the first epoch, we included $7\times7$ pixel convolution 
kernels when fitting templates to allow for differences in PSF between 
the two epochs.

For each GC, we defined a reference frame by averaging the 
template-based positions of GC stars from repeated exposures of the 
second epoch. The positions of GC stars, in each of the first-epoch 
exposures and the reference frame, were used to transform the 
template-measured position of the galaxies in each corresponding 
first-epoch exposure into the (second-epoch) reference frame. 
Then, we measured the difference between the first- and 
second-epoch position of each galaxy with respect to the GC stars.
To control systematic PM residuals related to the detector position 
and brightness of sources, we applied a `local correction,' where each 
correction is computed by using stars of similar brightness 
($\pm 1$~mag) and within a 200 pixel region centered on the given 
background galaxy. For each individual first-epoch exposure of each GC, 
we took the error-weighted average over all displacements 
of background galaxies with respect to the GC stars to obtain an 
independent PM estimate. Therefore, for each GC we have as 
many such estimates as first-epoch exposures (compare the $N$ value 
in column 6 of Table~\ref{t:targets} and Figure~\ref{f:pmd}).
The associated uncertainty was then computed 
using the bootstrap method on the displacements of background galaxies.
Since there are far more stars (typically a few thousands to tens 
of thousands) detected in our GC fields than background galaxies 
(column 4 in Table~\ref{t:pmresults}) used for the PM measurements, 
and since positions of stars are generally better determined than those of 
galaxies, the uncertainties $\Delta\mu_\mathrm{W,i}$ and 
$\Delta\mu_\mathrm{N,i}$ in individual PM estimates are dominated 
by the galaxy measurements. The average PM $\overline{\mu_\mathrm{W}}$
and $\overline{\mu_\mathrm{N}}$, and associated errors (see 
Table~\ref{t:pmresults}) of each GC were obtained by taking the 
error-weighted mean of the individual PM estimates. These were 
converted to final PM results in units of $\masyr$ via multiplying by 
the pixel scale of our reference images (50 mas\,pix$^{-1}$ for 
ACS/WFC and 40 mas\,pix$^{-1}$ for WFC3/UVIS), and dividing by the 
time baseline of our observations. 

\section{Results}
\label{s:results}

\subsection{Proper Motion Results}
\label{ss:pmresults}

%
%
\begin{deluxetable}{lccccc}
\renewcommand{\arraystretch}{0.9}
\setlength\tabcolsep{3.2pt}
\tablecaption{Proper motion results
              \label{t:pmresults}
             }

\tablehead{
   \colhead{}        & \colhead{$\overline{\muw}$\tablenotemark{a}}   & \colhead{$\overline{\mun}$\tablenotemark{a}} & \colhead{}  & \colhead{}         & \colhead{}            \\
   \colhead{Cluster} & \colhead{(mas yr$^{-1}$)} & \colhead{(mas yr$^{-1}$)} & \colhead{$N_\mathrm{g}$\tablenotemark{b}} & \colhead{$\chi^2$\tablenotemark{c}} & \colhead{$N_\mathrm{DF}$\tablenotemark{d}}
}
\startdata
Arp\,2    & \phs$2.40 \pm 0.04$ &    $-1.54 \pm 0.03$ & 43 &   10.2 & \phn$8\pm4.0$ \\
IC\,4499  &    $-0.35 \pm 0.07$ &    $-0.35 \pm 0.07$ & 56 &\phn1.1 & \phn$6\pm3.5$ \\
NGC\,1261 &    $-1.69 \pm 0.04$ &    $-2.11 \pm 0.04$ & 36 &\phn6.0 & \phn$8\pm4.0$ \\
NGC\,2298 &    $-3.42 \pm 0.05$ &    $-2.30 \pm 0.05$ & 36 &\phn5.5 & \phn$8\pm4.0$ \\
NGC\,2419 &    $-0.05 \pm 0.03$ &    $-0.50 \pm 0.03$ & 85 &\phn0.8 & \phn$6\pm3.5$ \\
NGC\,4147 & \phs$1.78 \pm 0.04$ &    $-2.10 \pm 0.04$ & 40 &   12.4 & \phn$8\pm4.0$ \\
NGC\,5024 & \phs$0.17 \pm 0.06$ &    $-1.17 \pm 0.08$ & 15 &\phn7.7 & \phn$8\pm4.0$ \\
NGC\,5053 & \phs$0.34 \pm 0.07$ &    $-1.16 \pm 0.07$ & 33 &\phn4.5 & \phn$8\pm4.0$ \\
NGC\,5466 & \phs$4.90 \pm 0.08$ &    $-1.17 \pm 0.05$ & 61 &\phn7.8 & \phn$8\pm4.0$ \\
NGC\,6101 &    $-1.73 \pm 0.05$ &    $-0.47 \pm 0.05$ & 20 &\phn5.5 & \phn$8\pm4.0$ \\
NGC\,6426 & \phs$1.65 \pm 0.06$ &    $-3.08 \pm 0.06$ & 27 &\phn5.2 & \phn$6\pm3.5$ \\
NGC\,6934 & \phs$2.67 \pm 0.04$ &    $-4.52 \pm 0.05$ & 30 &   16.3 & \phn$8\pm4.0$ \\
NGC\,7006 & \phs$0.08 \pm 0.07$ &    $-0.73 \pm 0.08$ & 42 &\phn3.2 & \phn$6\pm3.5$ \\
Pal\,12   & \phs$3.06 \pm 0.05$ &    $-3.32 \pm 0.05$ & 66 &   18.6 & \phn$8\pm4.0$ \\
Pal\,13   &    $-1.70 \pm 0.09$ & \phs$0.08 \pm 0.06$ & 38 &\phn3.4 & \phn$6\pm3.5$ \\
Pal\,15   & \phs$0.33 \pm 0.09$ &    $-0.78 \pm 0.11$ & 32 &   15.0 &    $12\pm4.9$ \\
Pyxis     &    $-0.94 \pm 0.09$ &    $-0.13 \pm 0.09$ & 41 &\phn9.8 & \phn$6\pm3.5$ \\
Rup\,106  & \phs$1.09 \pm 0.08$ & \phs$0.48 \pm 0.06$ & 43 &\phn3.1 & \phn$6\pm3.5$ \\
Terzan\,7 & \phs$3.04 \pm 0.06$ &    $-1.71 \pm 0.05$ & 43 &\phn5.3 & \phn$8\pm4.0$ \\
Terzan\,8 & \phs$2.91 \pm 0.08$ &    $-1.63 \pm 0.06$ & 31 &\phn2.1 & \phn$8\pm4.0$ \\
\enddata
\tablenotetext{a}{$\muw$ and $\mun$ are defined as the PMs in west 
                  ($\muw = - \mu_{\alpha}\cos\delta$) and north ($\mun = \mu_{\delta}$) 
                  directions, respectively.}
\tablenotetext{b}{Number of background galaxies used for deriving the PMs.}
\tablenotetext{c}{Chi-square values as defined in Section~\ref{ss:pmresults}.}
\tablenotetext{d}{Number of degrees of freedom.}
\end{deluxetable}
%

%
%
\begin{figure*}[t]
\gridline{ \fig{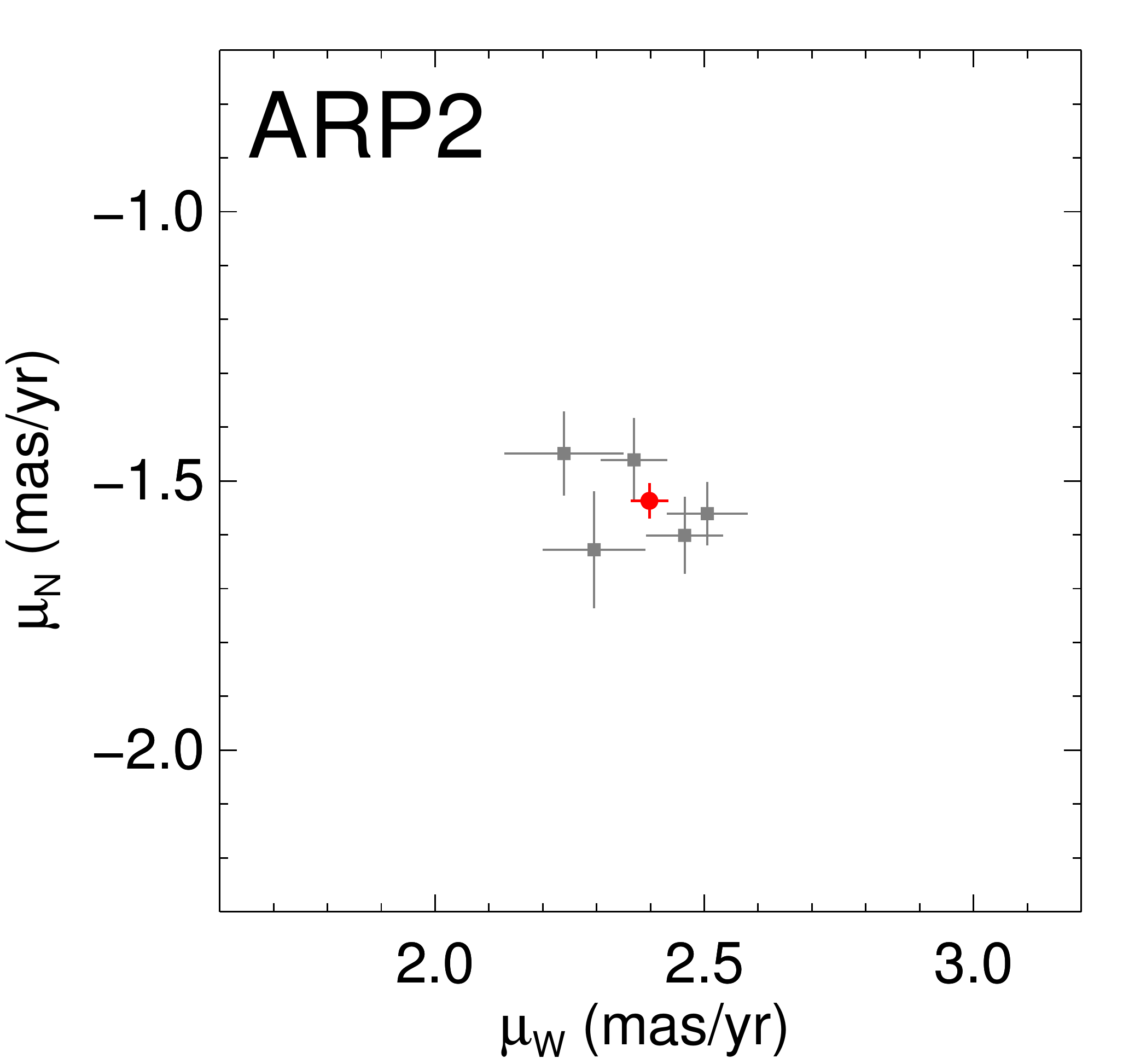}   {0.25\textwidth}{}
           \fig{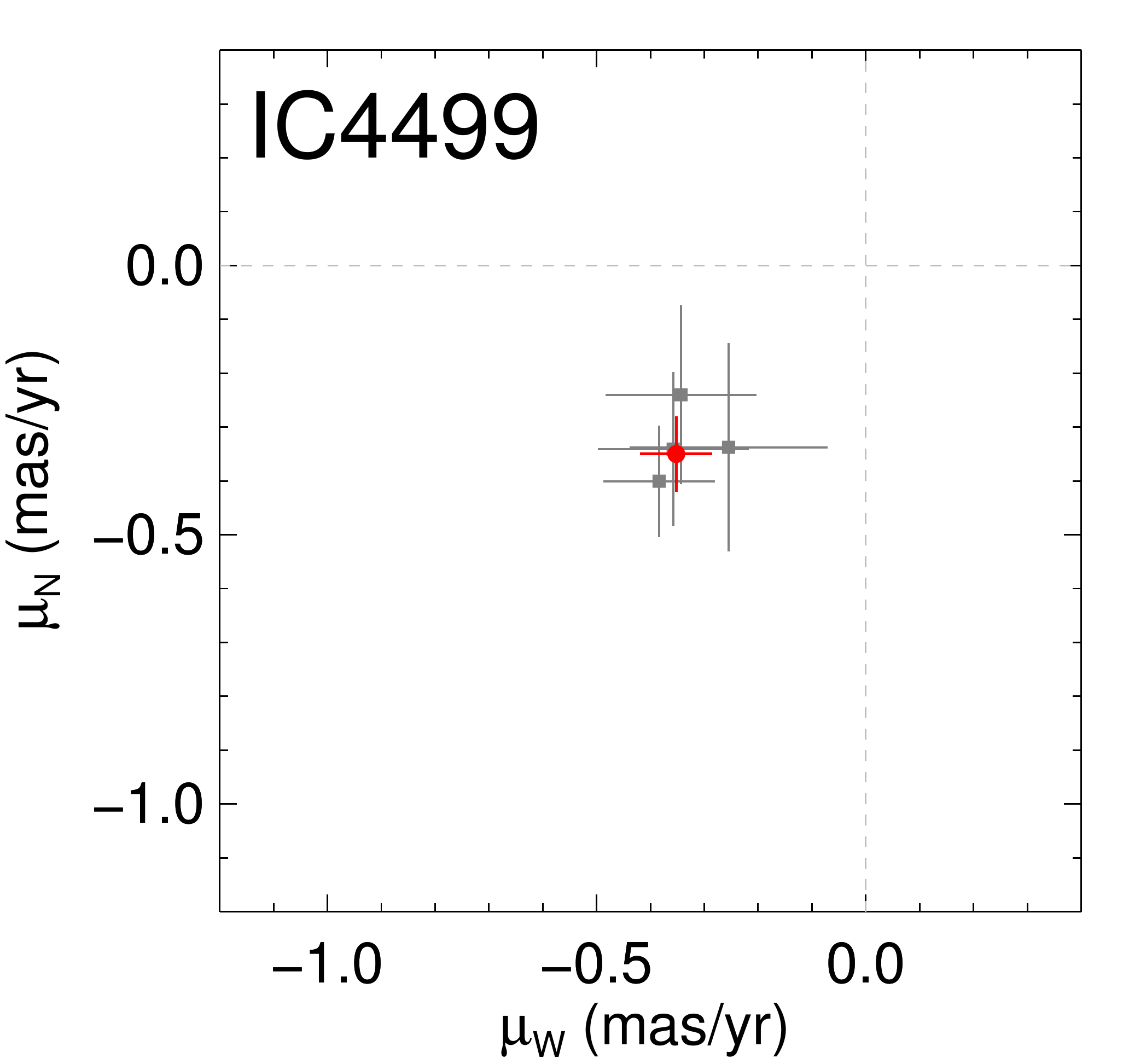} {0.25\textwidth}{}
           \fig{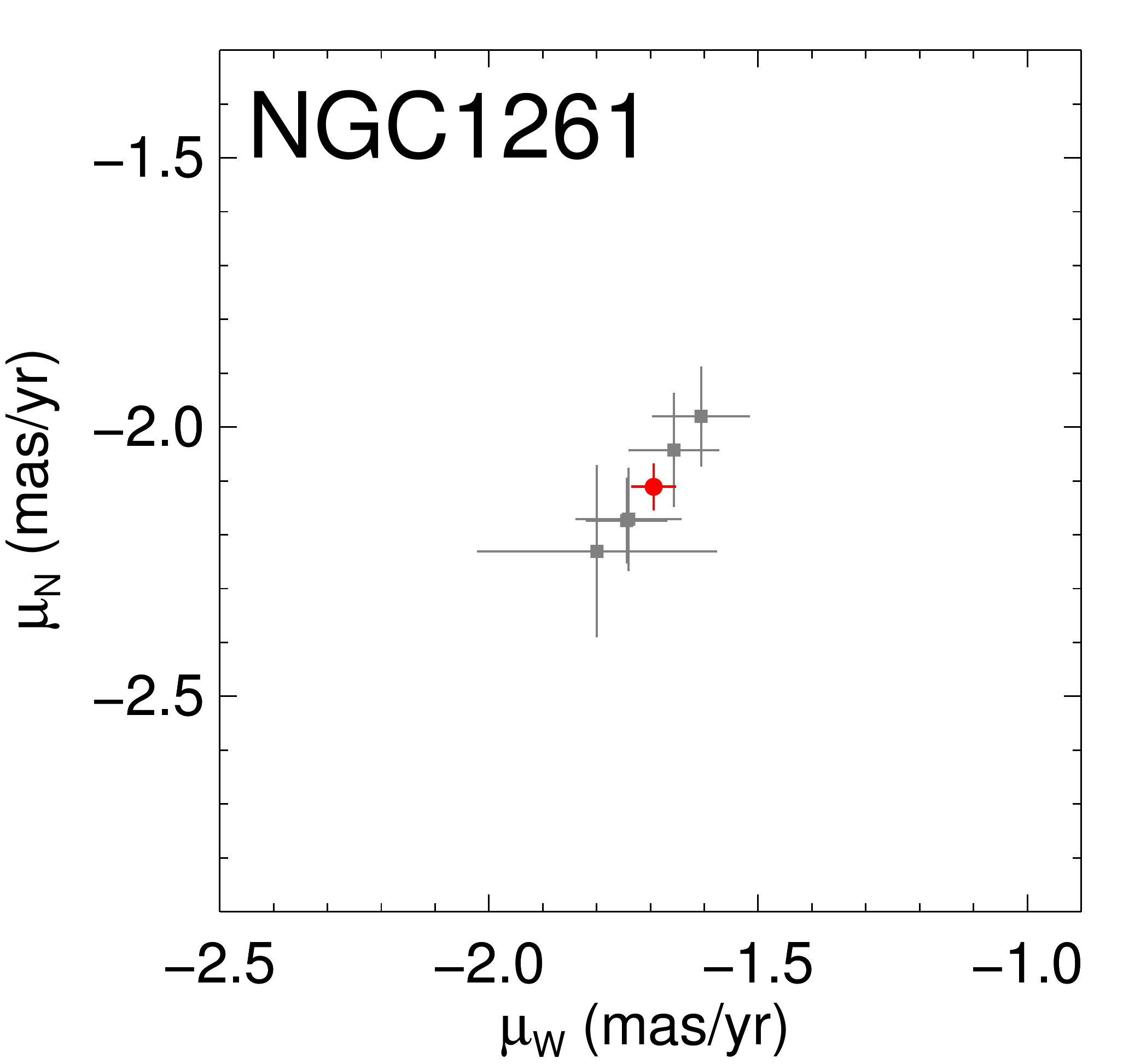}{0.25\textwidth}{} 
           \fig{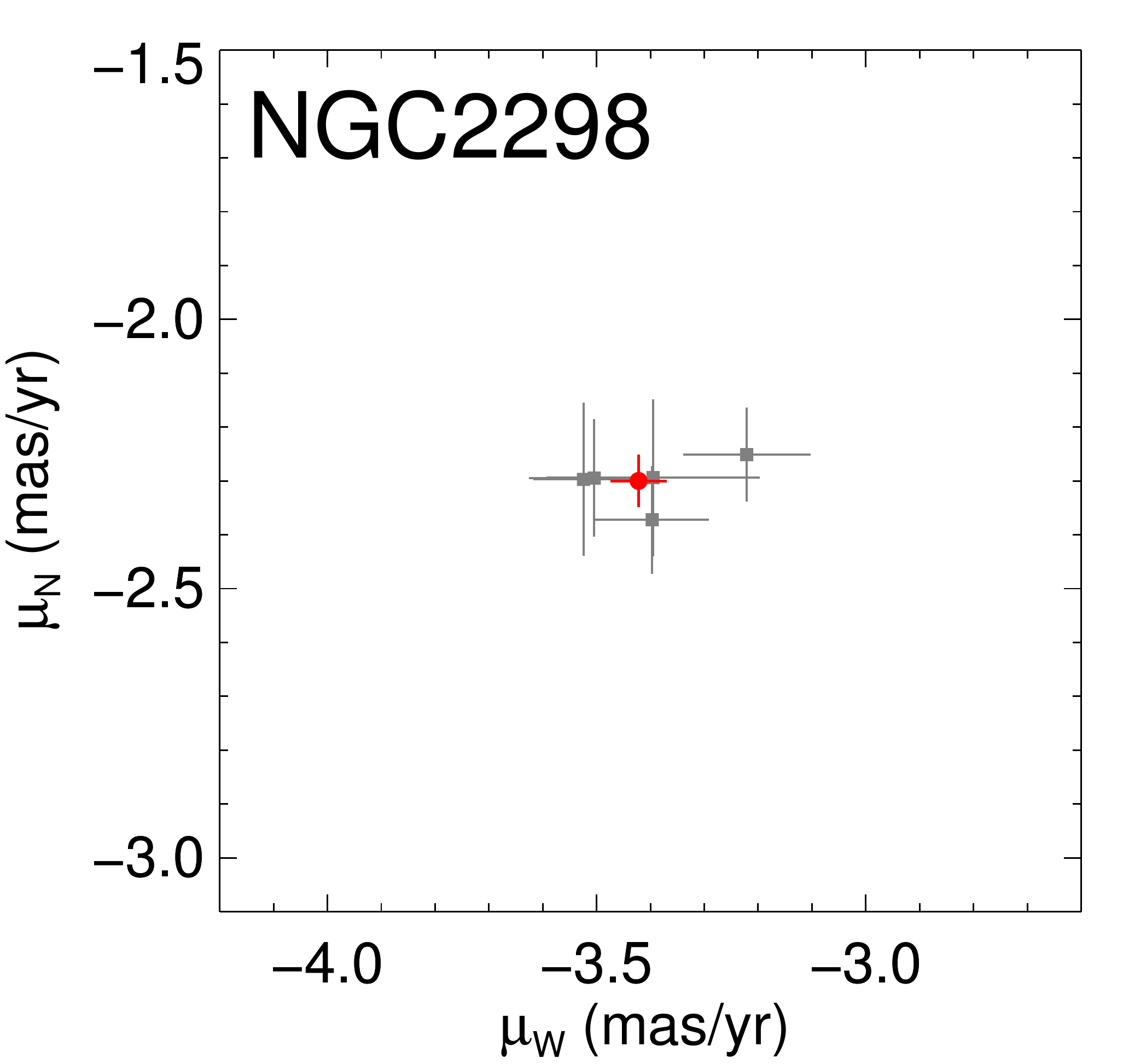}{0.25\textwidth}{} }
\vspace{-1cm}
\gridline{ \fig{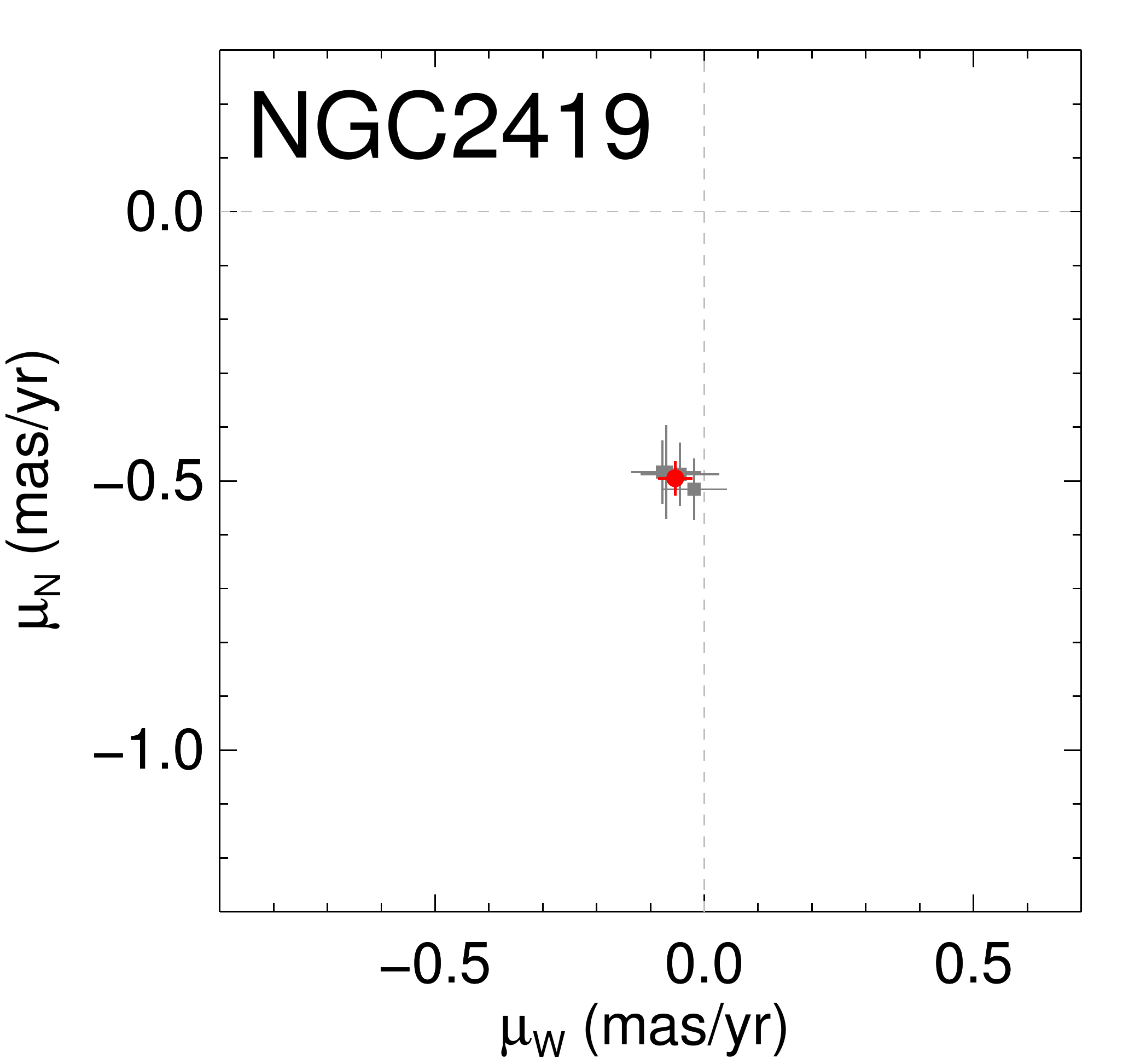}{0.25\textwidth}{}
           \fig{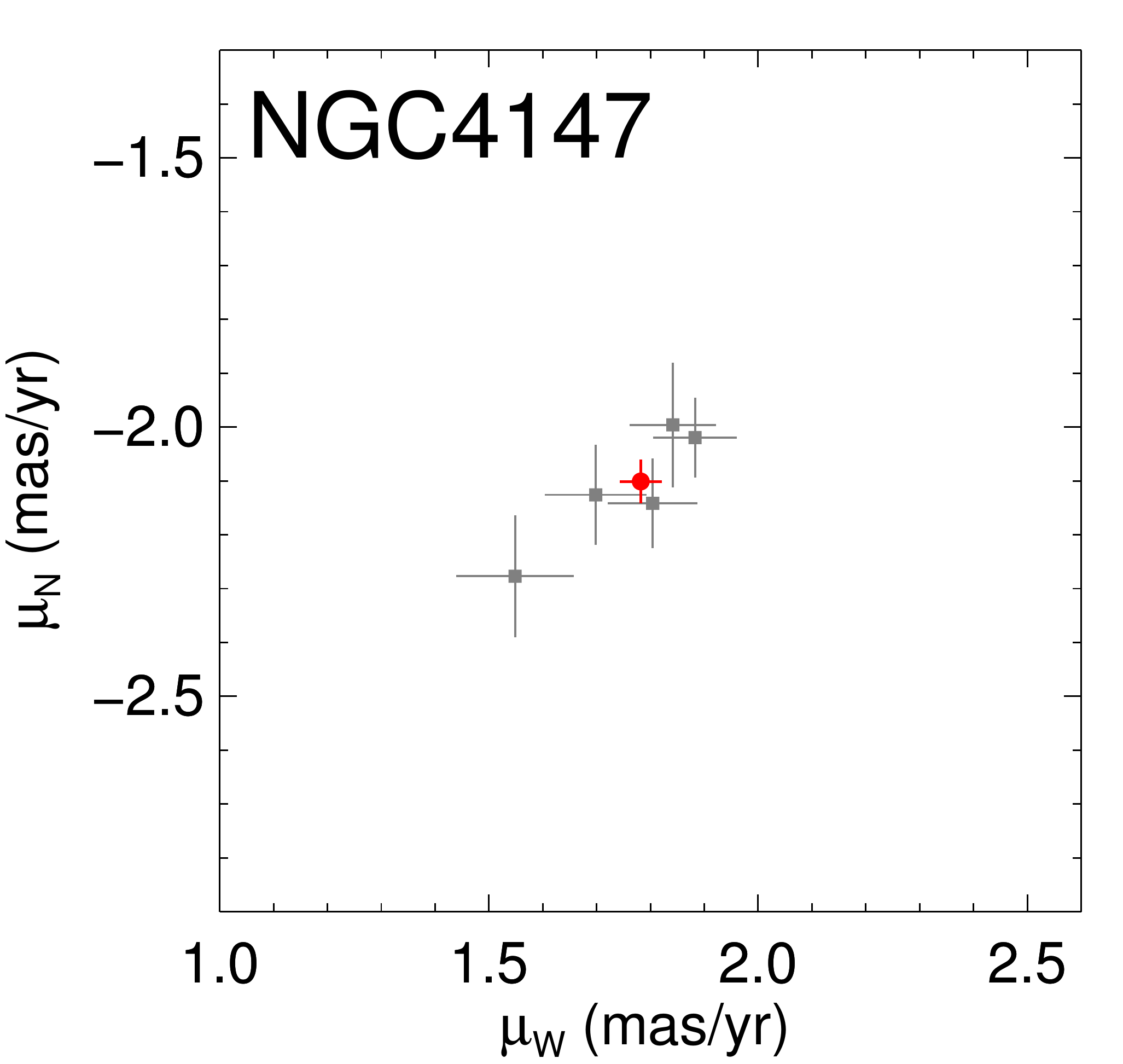}{0.25\textwidth}{} 
           \fig{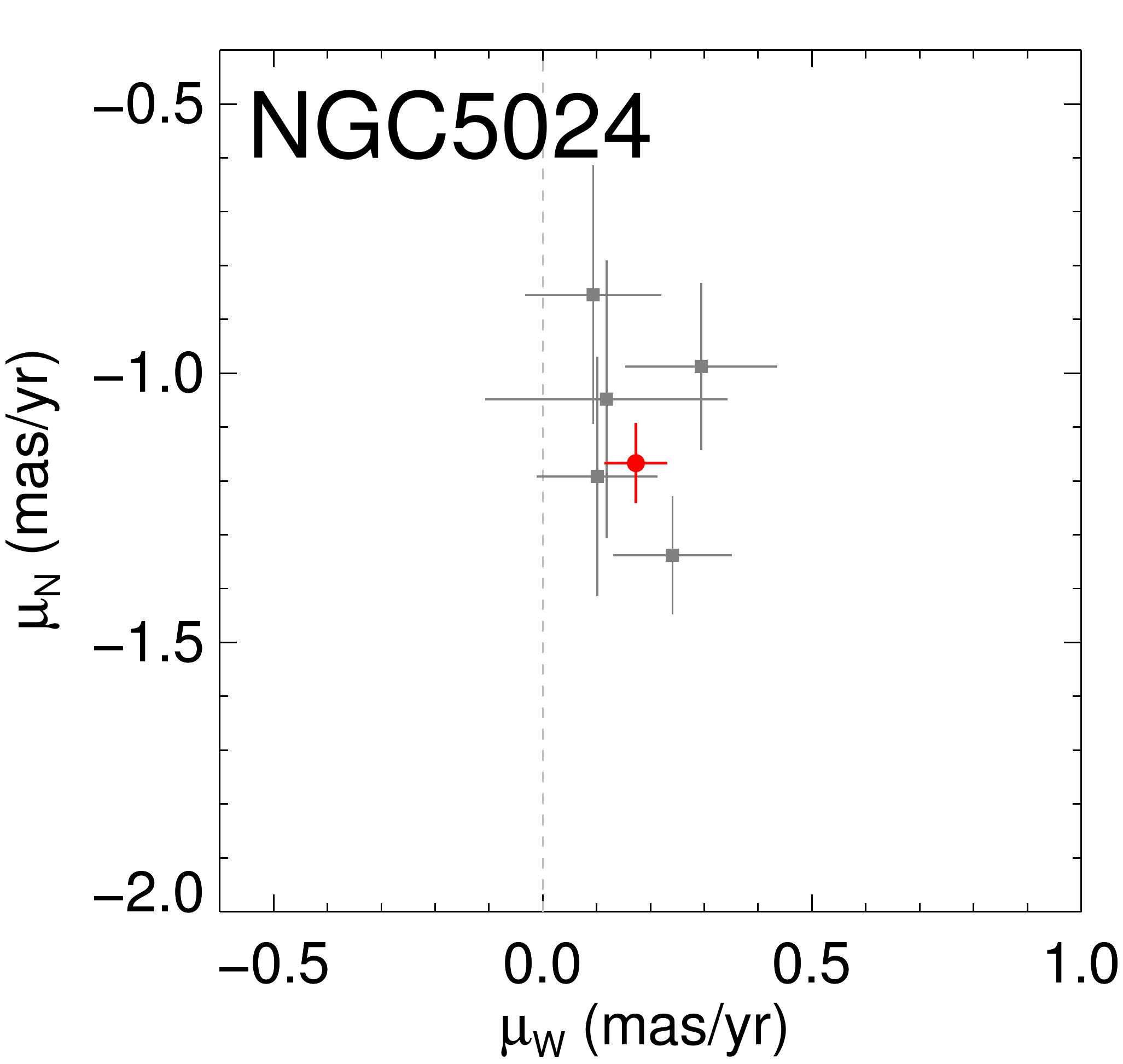}{0.25\textwidth}{}
           \fig{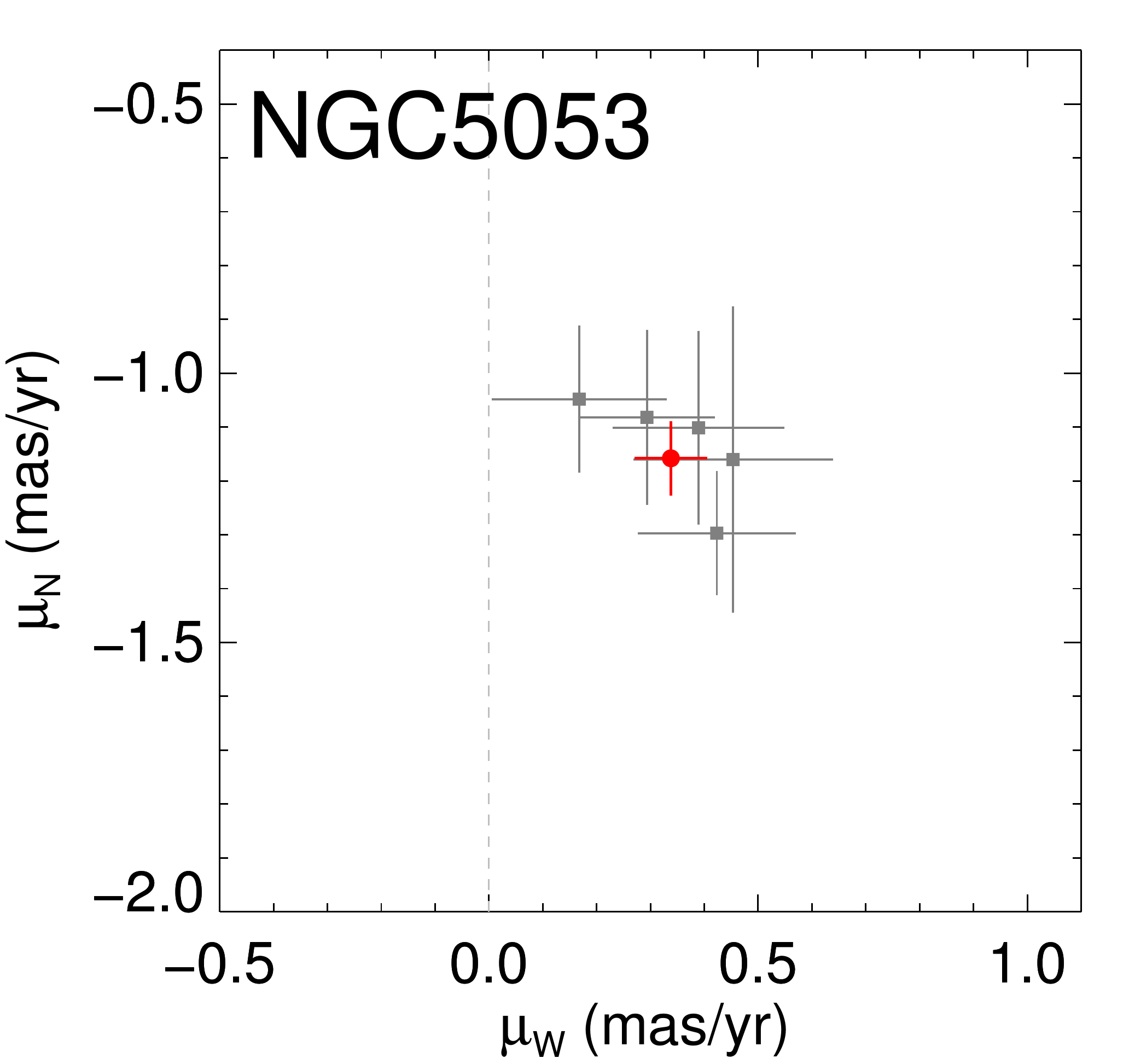}{0.25\textwidth}{} }
\vspace{-1cm}
\gridline{ \fig{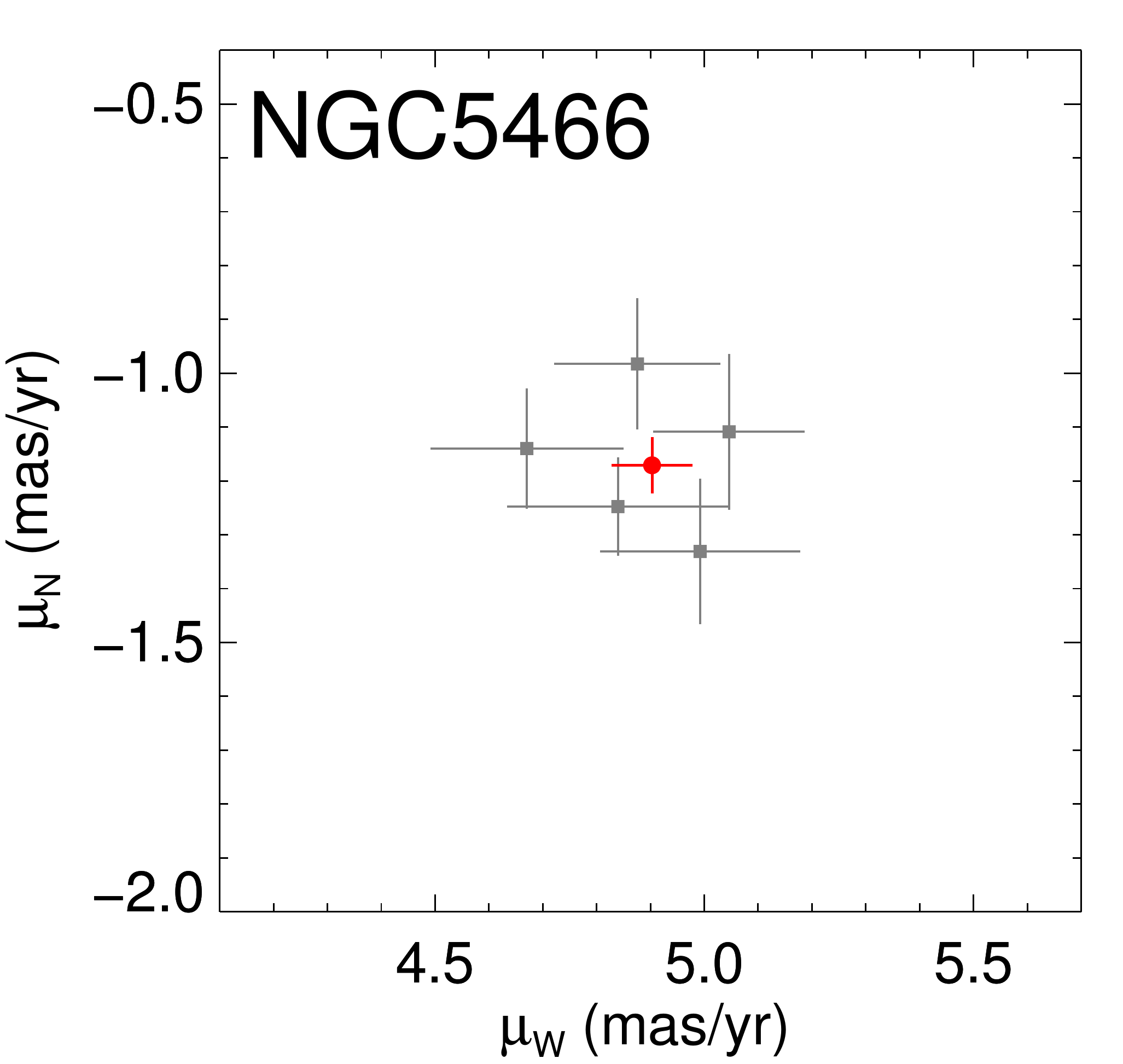}{0.25\textwidth}{} 
           \fig{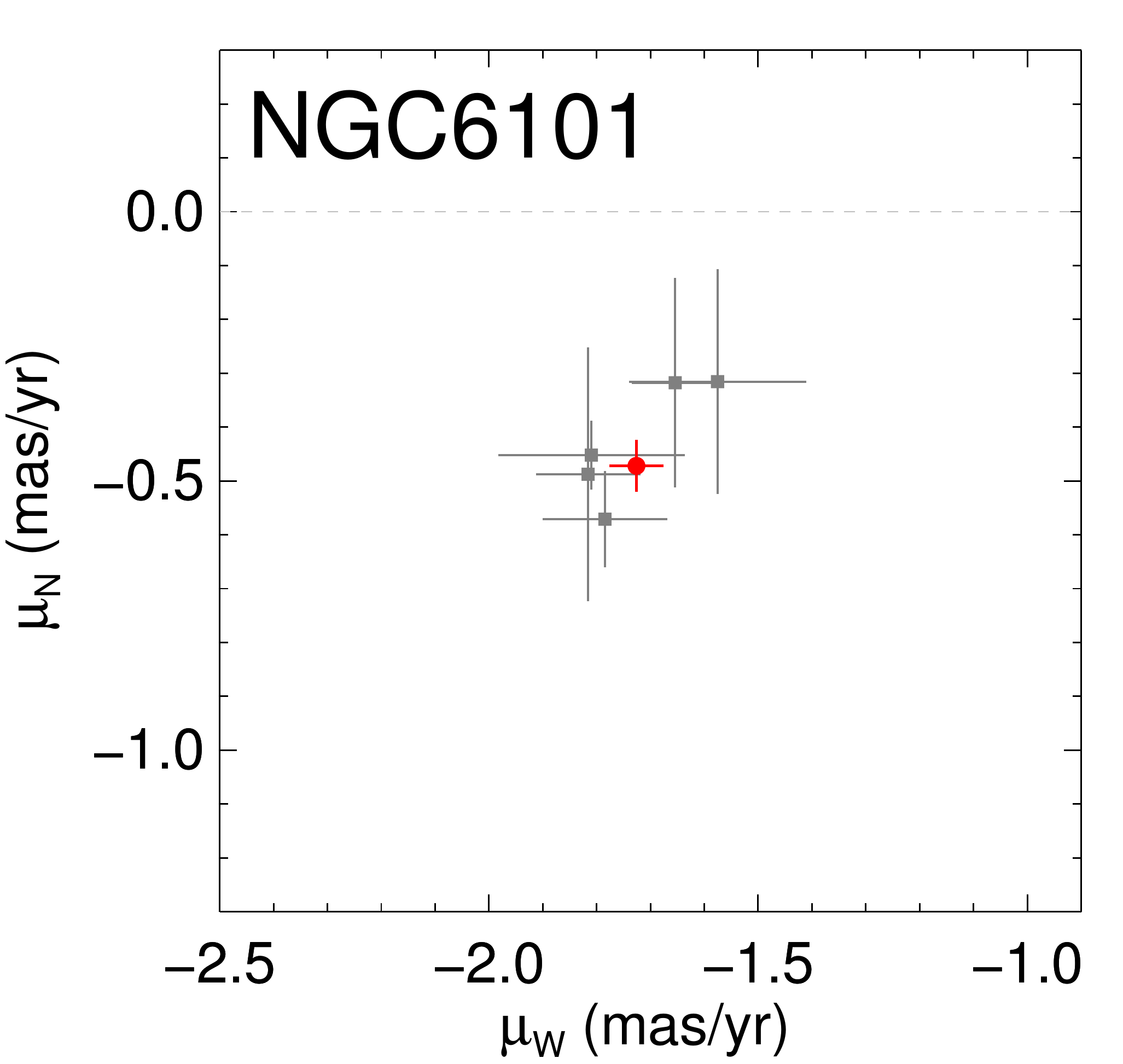}{0.25\textwidth}{}
           \fig{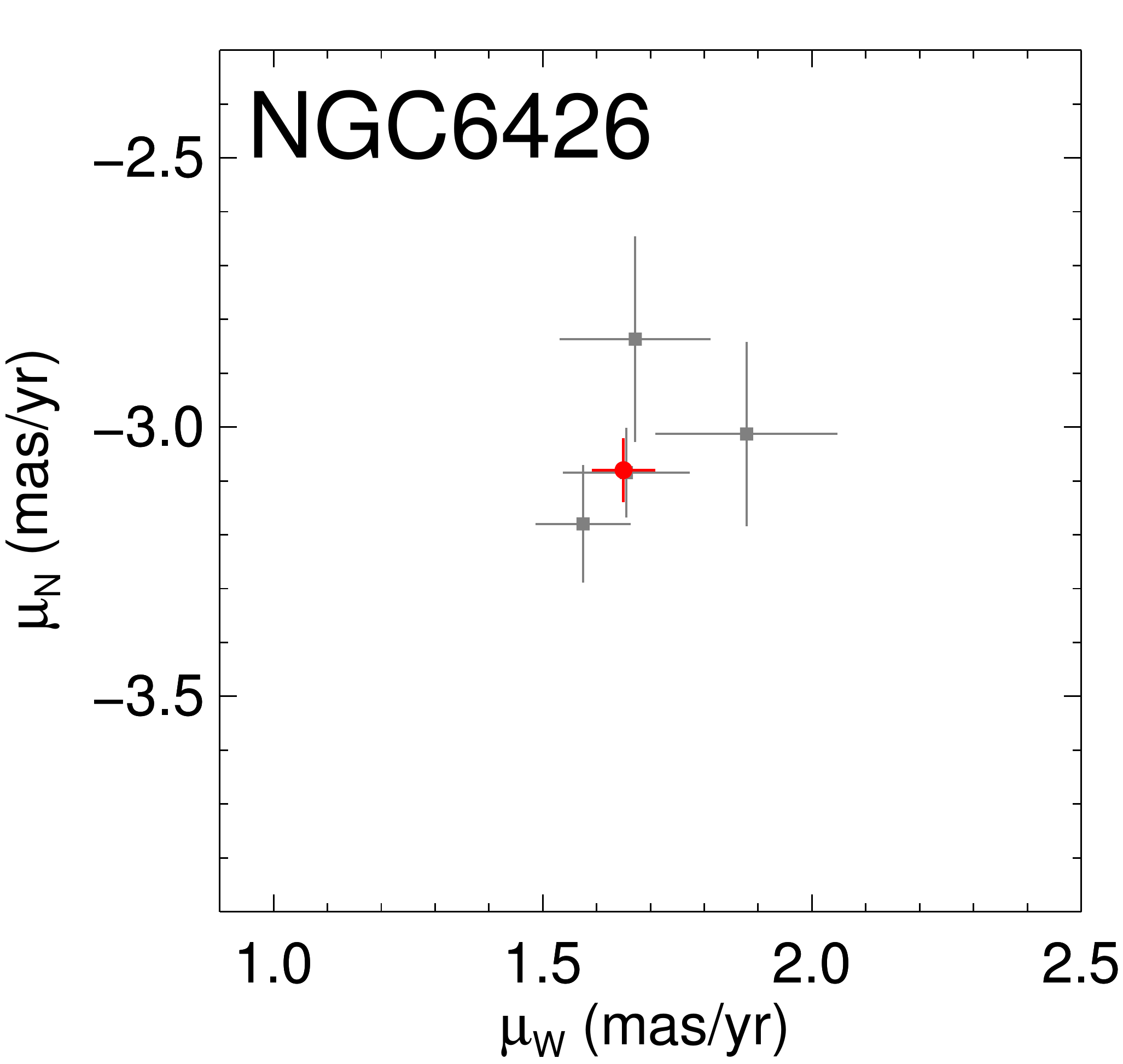}{0.25\textwidth}{}
           \fig{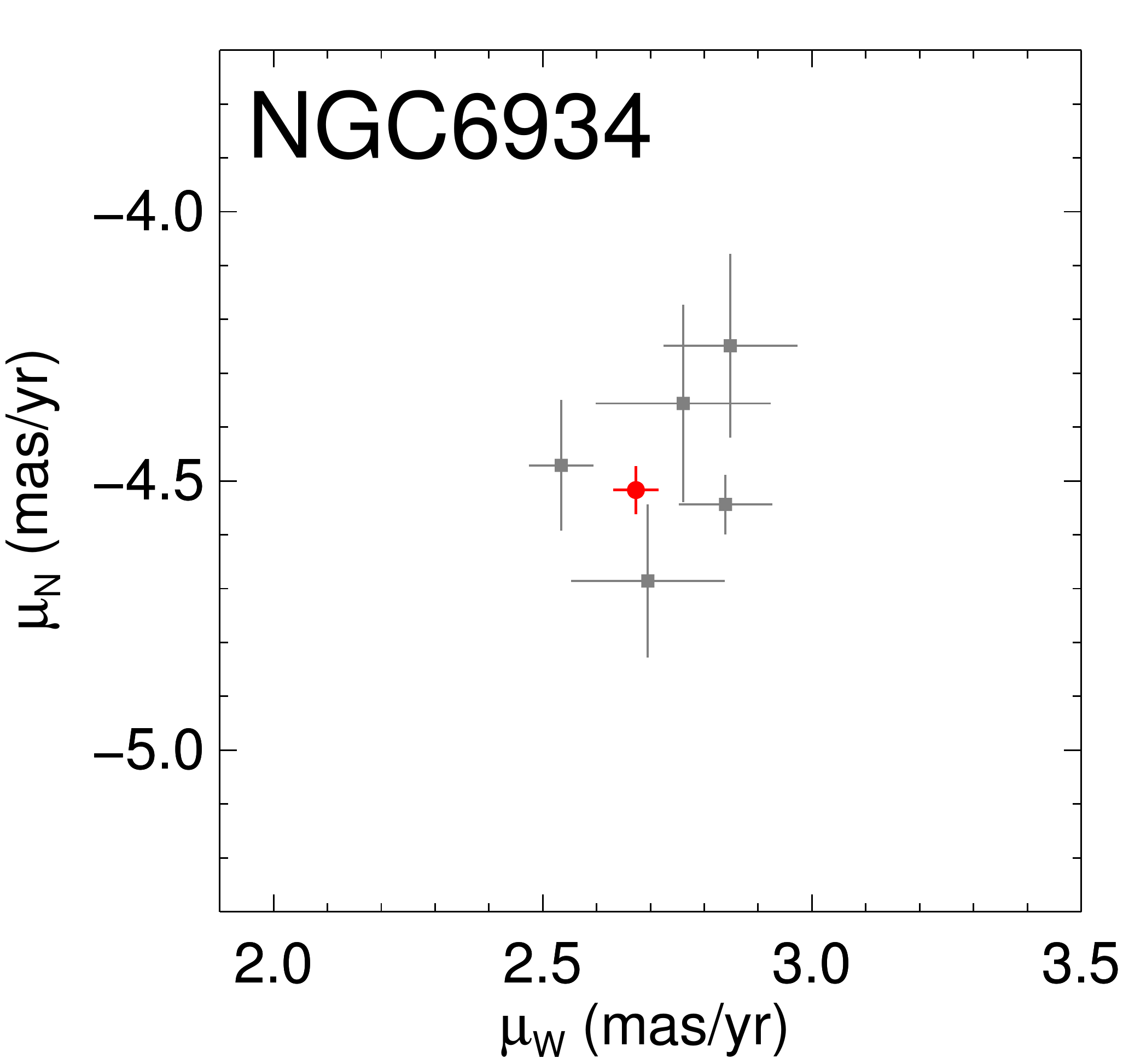}{0.25\textwidth}{} }

\vspace{-1cm}
\gridline{ \fig{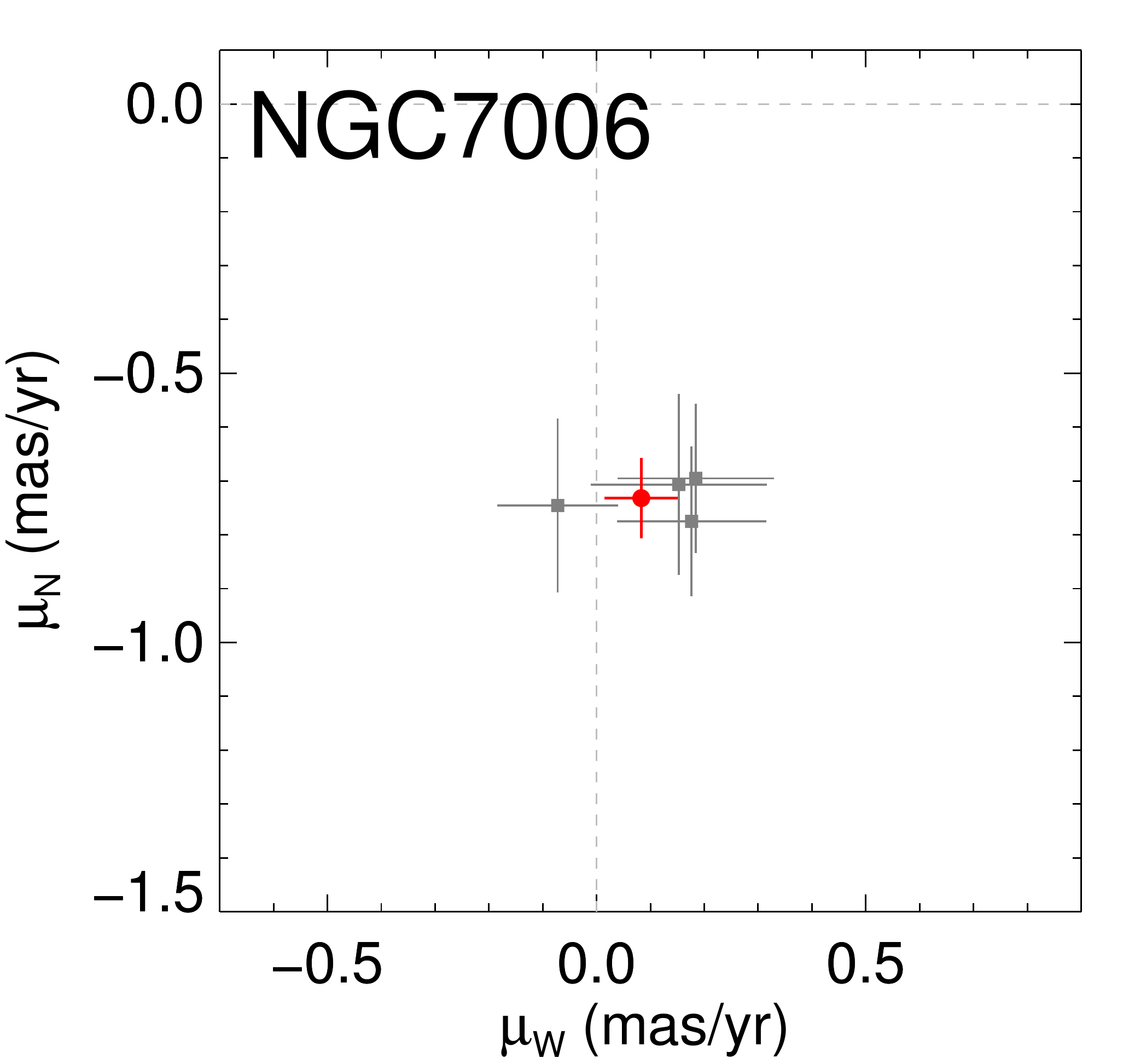}{0.25\textwidth}{}
           \fig{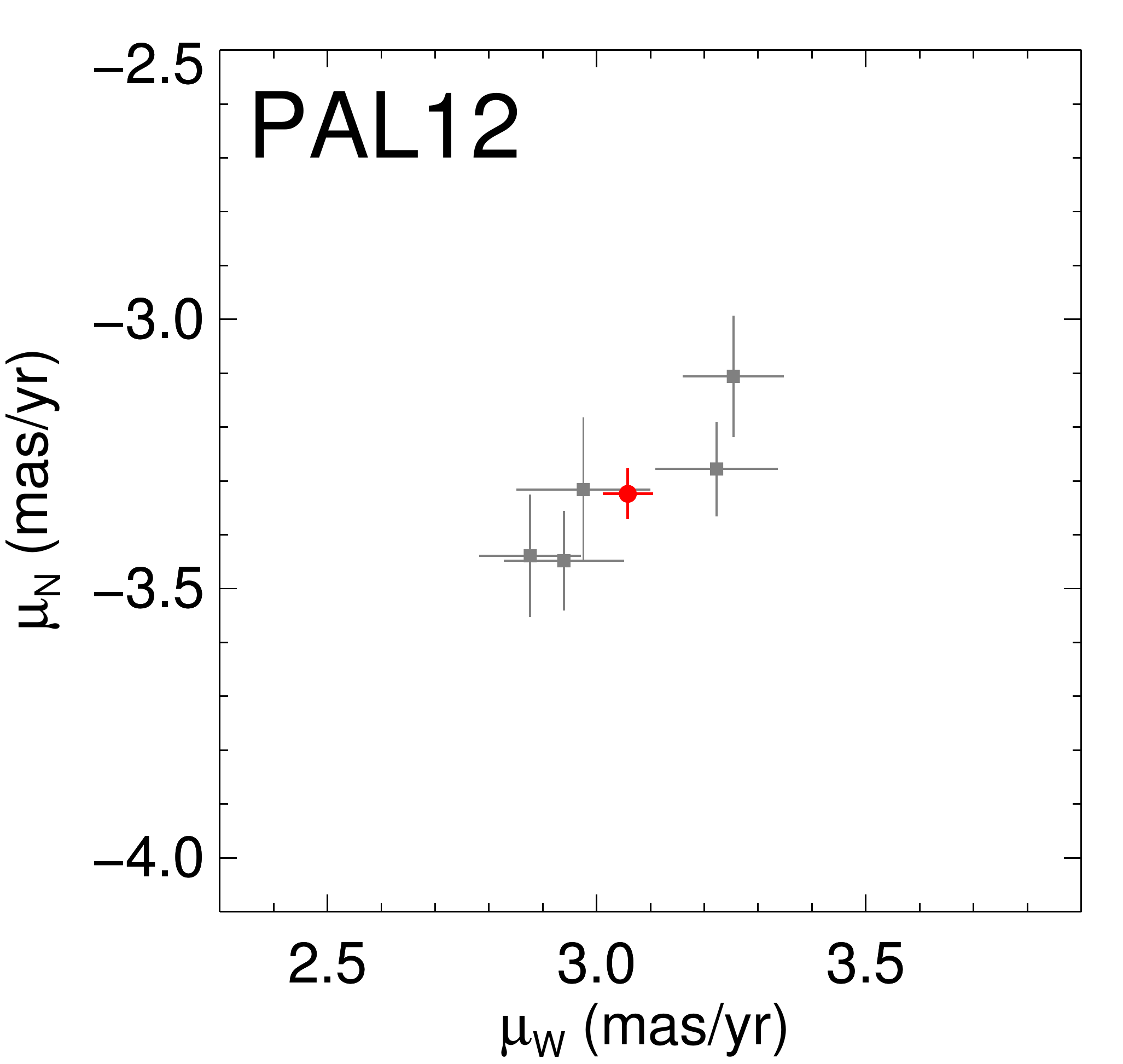}  {0.25\textwidth}{}
           \fig{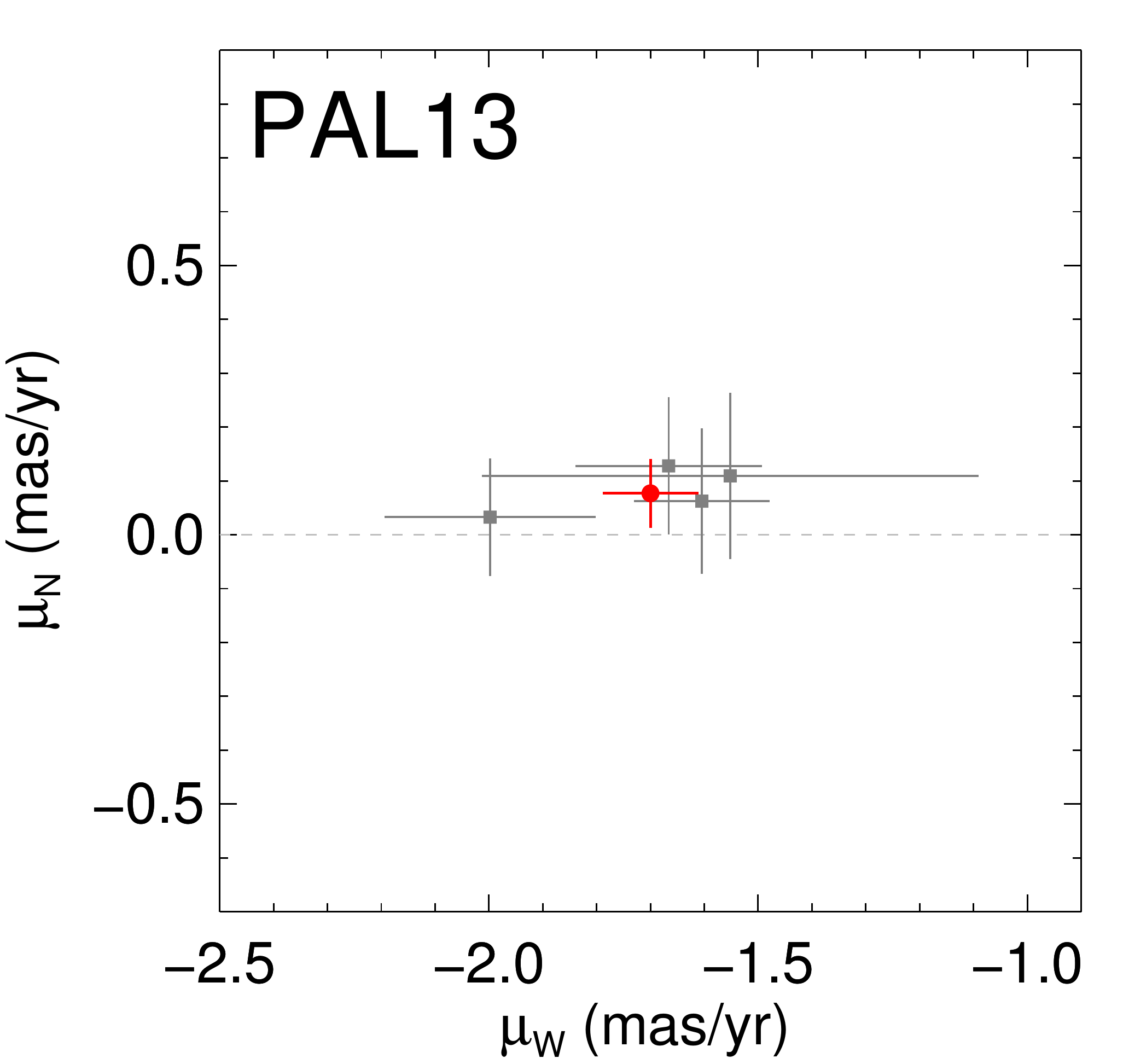}  {0.25\textwidth}{}
           \fig{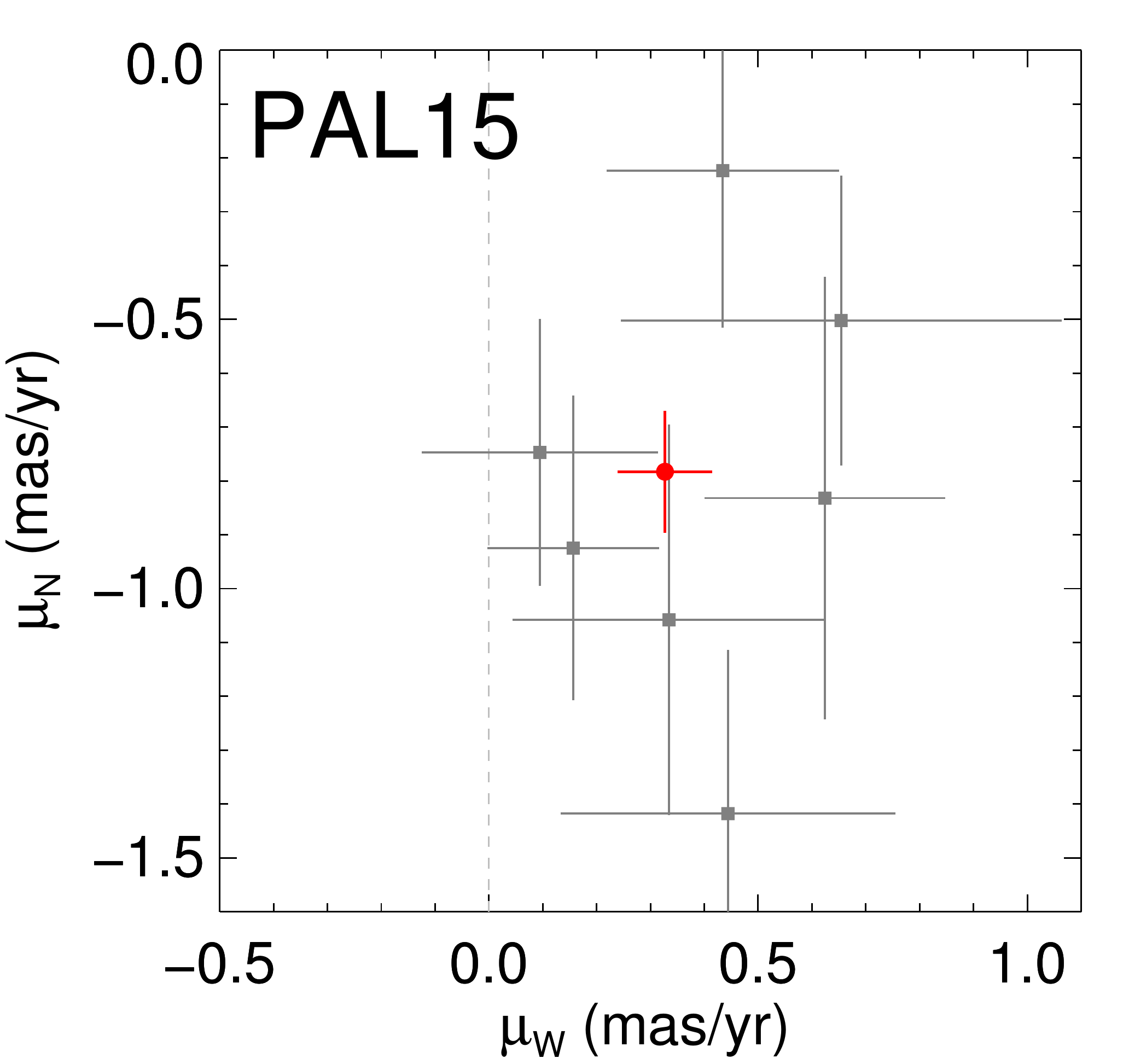}  {0.25\textwidth}{} }
\vspace{-1cm}
\gridline{ \fig{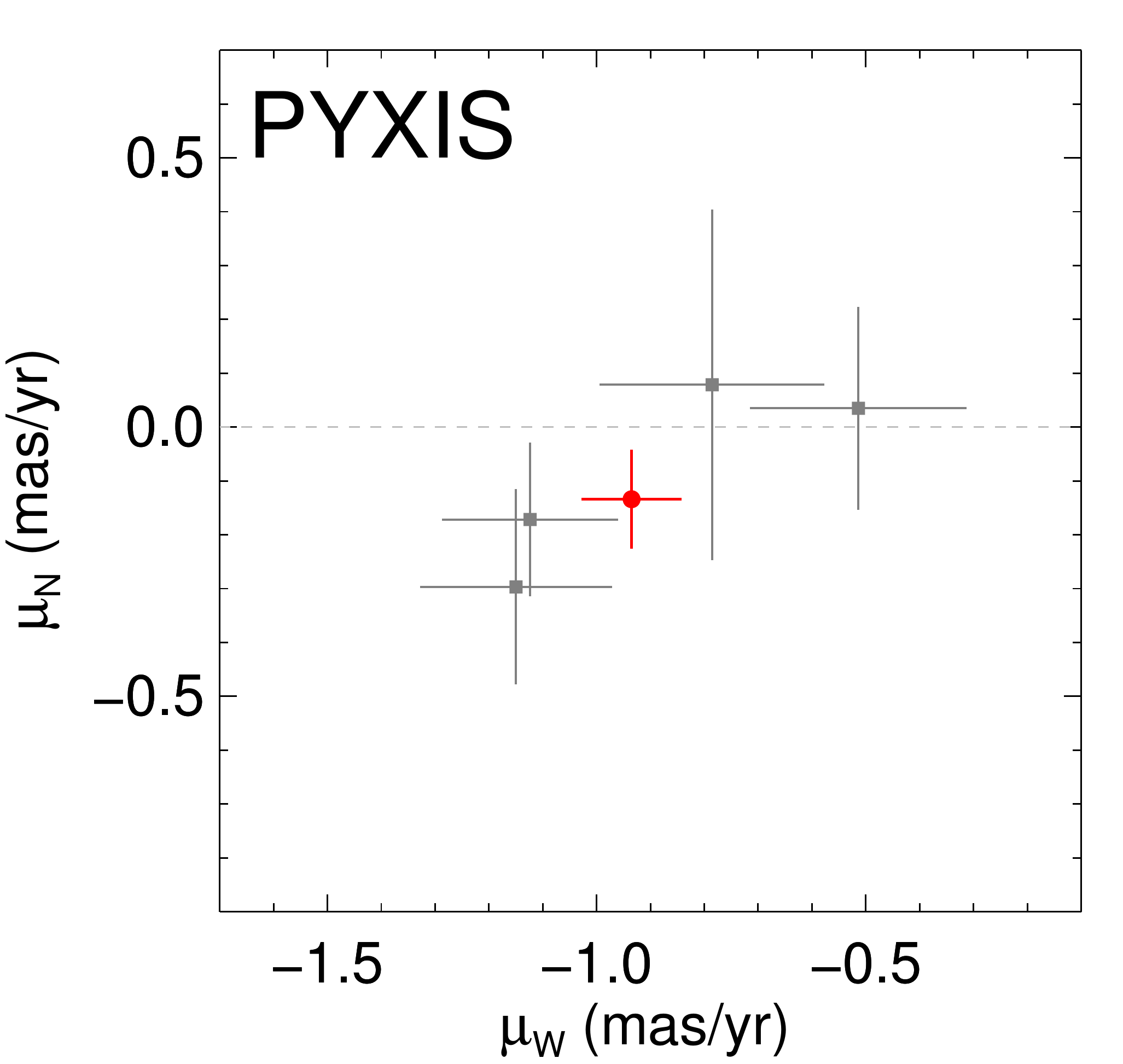}  {0.25\textwidth}{}
           \fig{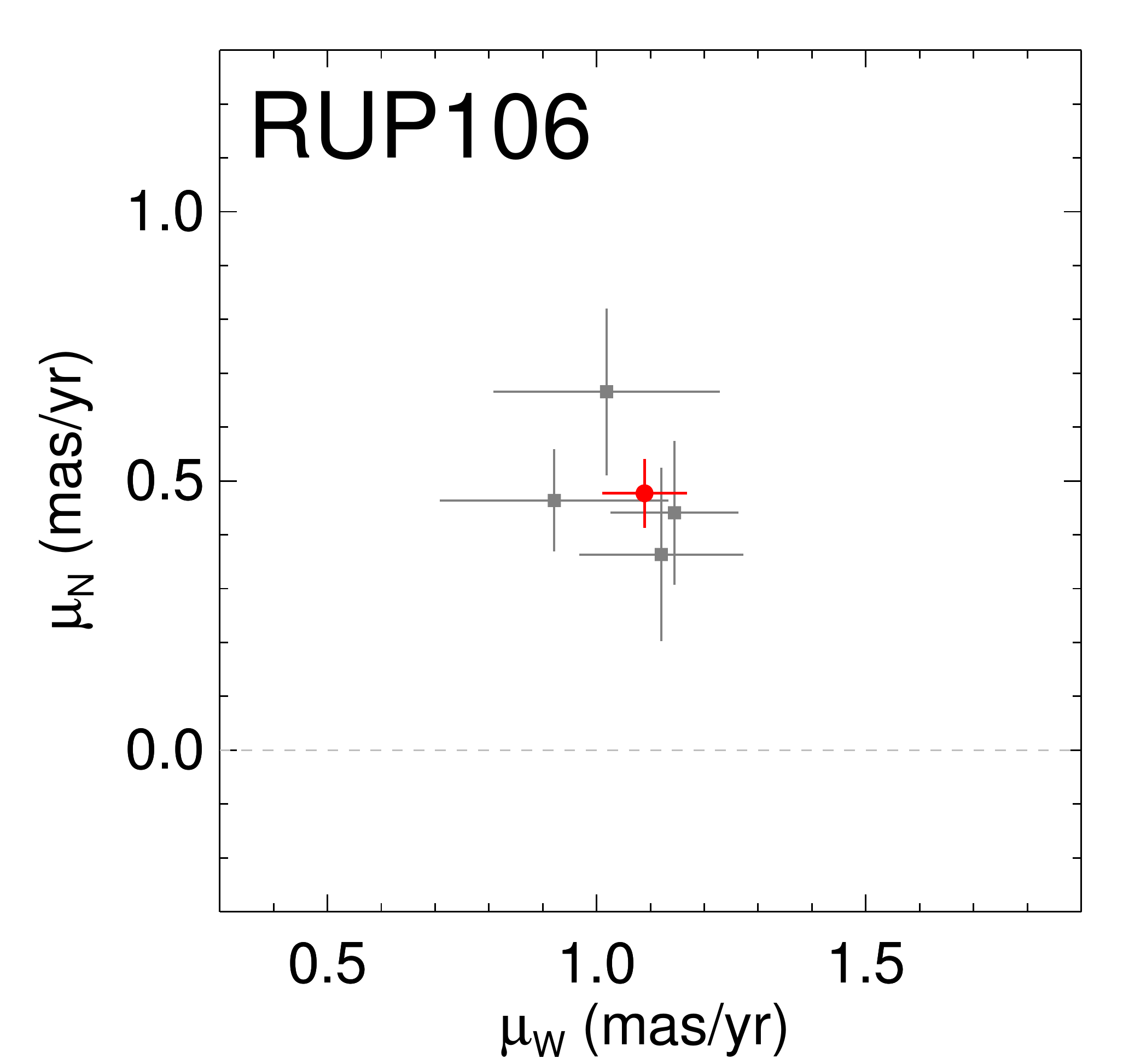} {0.25\textwidth}{} 
           \fig{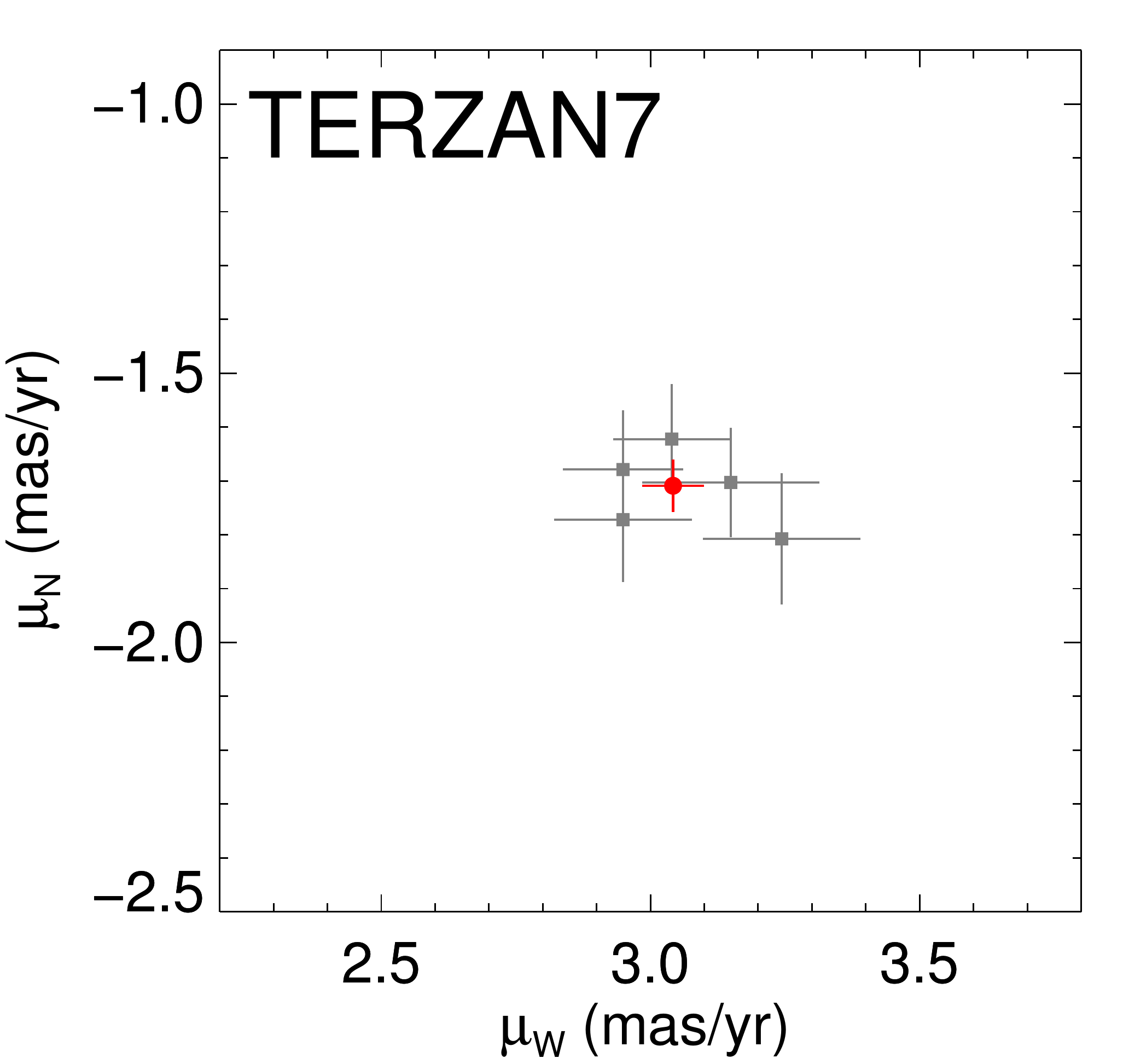}{0.25\textwidth}{}
           \fig{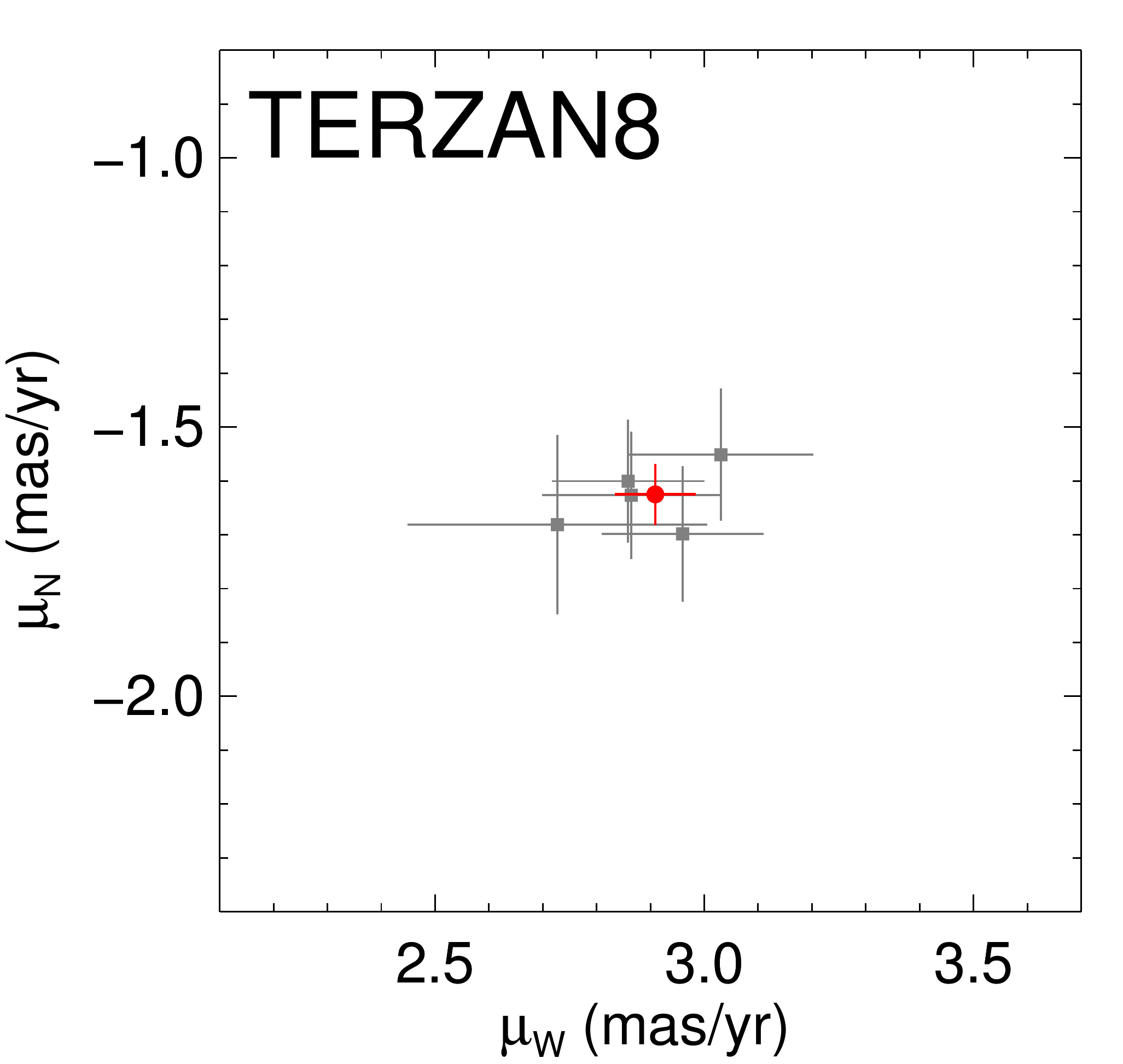}{0.25\textwidth}{} }
\caption{Proper motion diagrams for all of our target GCs with 
         cluster names indicated on the top left of each plot.
         Gray dots indicate the measurements for individual exposures, 
         and the red dot is the weighted average of the gray dots for 
         each GC. The dashed lines indicate $\muw = \mun = 0$.
         \label{f:pmd}
         }
\end{figure*}

Our PM results for the target GCs are listed in Table~\ref{t:pmresults}, 
and the corresponding PM diagrams are presented in Figure~\ref{f:pmd}. 
For each GC, we show the multiple measurements for individual exposures 
(gray squares with error bars) as described in Section~\ref{ss:measurement} 
along with the final PMs (red dots with error bars) derived by averaging 
the individual measurements. 

In the last two columns of Table~\ref{t:pmresults}, we provide our 
calculations of $\chi^2$ and $N_\mathrm{DF}$. The former is defined as 
\begin{equation}
\label{e:chisq}
  \chi^2 = \sum_{i} 
     \left[
     \left ( \frac{ \mu_{\mathrm{W},i} - {\overline{\muw}} }
               { \Delta \mu_{\mathrm{W},i} } \right )^2 
     +
     \left ( \frac{ \mu_{\mathrm{N},i} - {\overline{\mun}} }
               { \Delta \mu_{\mathrm{N},i} } \right )^2
     \right],
\end{equation}
and measures statistical agreement among the individual measurements 
for each GC. Provided that systematic errors are not present 
in our measurements, our $\chi^2$ calculated above should be 
consistent with a chi-squared probability distribution having a 
mean equal to $N_\mathrm{DF}$, the number of degrees of freedom
\footnote{The number of degrees of freedom $N_\mathrm{DF}$ is equal 
to twice the number of PM estimates per GC minus two.}, with a 
dispersion of $\sqrt{2N_\mathrm{DF}}$. In general, we find that 
the $\chi^2$ values for most clusters are within the range 
of $N_\mathrm{DF}$ with a few exceptions. For IC\,4499, NGC\,2419, 
and Terzan\,8, the $\chi^2$ values are significantly lower than 
the $N_\mathrm{DF}$ indicating that the random uncertainties for 
these clusters may be somewhat overestimated. On the other hand, 
for NGC\,6934 and Pal\,12, our final PM uncertainties may be 
underestimated based on the comparison between $\chi^2$ and 
$N_\mathrm{DF}$. Figure~\ref{f:pmd} indeed shows that for 
these two clusters, the scatter among the individual data points 
are larger with respect to the final uncertainty compared to 
other clusters.

\subsection{Effects of Internal Motions on Center-of-mass Motions}
\label{ss:internalmotions}

Because our PM measurement method relies on measuring the reflex motion 
of background galaxies found in the same fields of our GCs, any 
internal tangential motions of GC stars will affect the final PM results. 
While random internal motions will simply increase the final PM 
uncertainty, rotational motions may cause systematic offsets in 
measurements. In fact, rotation in the plane of the sky have been 
detected for a few nearby GCs. The most extreme case known to date is 
47\,Tuc which has a peak (clockwise) rotation of $\sim 6$~\kms 
\citep{bel17}. However, rotation in the plane of the sky has been 
reported to be negligible for other GCs: NGC\,6681 \citep{mas13} 
and NGC\,362 \citep{lib18a}. 
All of our target GCs, except for NGC\,2419, were imaged in the 
center of the clusters, and the background galaxies we used as 
reference objects are distributed by and large evenly on the fields. 
Therefore, even if the rotation in the plane of the sky is as 
large as the case for 47\,Tuc, these motions will cancel out when 
averaging the reflex motions of background galaxies in the field to 
obtain the final PM results\footnote{Note that these motions 
may increase the random uncertainty in PM results.}.

\citet{bau09} measured the line-of-sight rotation of NGC\,2419 
to be $3.26 \pm 0.85$~\kms. To consider an extreme case, we assume 
that similar to 47\,Tuc, the rotation in the plane of the sky 
for NGC\,2419 is twice as fast as that in the line-of-sight.
Since NGC\,2419 is at a distance of $87.5$ kpc from us, it is possible 
that our PM result will be systematically offset by $6.5$~\kms 
$= 0.016 \masyr$ in one direction. This is about half of our PM uncertainty 
for NGC\,2419, which may seem large. However, our target 
field for NGC\,2419 is located at $\sim 4.7$ arcmin away from the 
center of this cluster, which is about 5 times the half-light radius 
($r_\mathrm{h}$) of NGC\,2419. For GCs, the rotation speed generally 
increases from the center until it reaches maximum at 1--2$r_\mathrm{h}$, 
and then falls off exponentially beyond that \citep{bia13}. 
For 47\,Tuc, the best-fit model implies that the rotation speed at 
$5r_\mathrm{h}$ drops to about 40\% of the peak speed 
\citep[see Figure~6 of][]{bel17}, so a more reasonable estimate for 
the systematic offset  $0.006 \masyr$. This is significantly smaller 
than our PM uncertainty, and hence demonstrates that even in an extreme 
case, our measurement for NGC\,2419 is most likely unaffected by 
rotation in the plane of the sky.

Altogether, we conclude that the PMs listed in Table~\ref{t:pmresults} 
represent the center-of-mass (COM) motions of the target GCs, 
and therefore do not require any corrections.

\subsection{Comparison to Previous Proper Motion Measurements}
\label{ss:compare}

%
%
\begin{figure*}
\gridline{ \fig{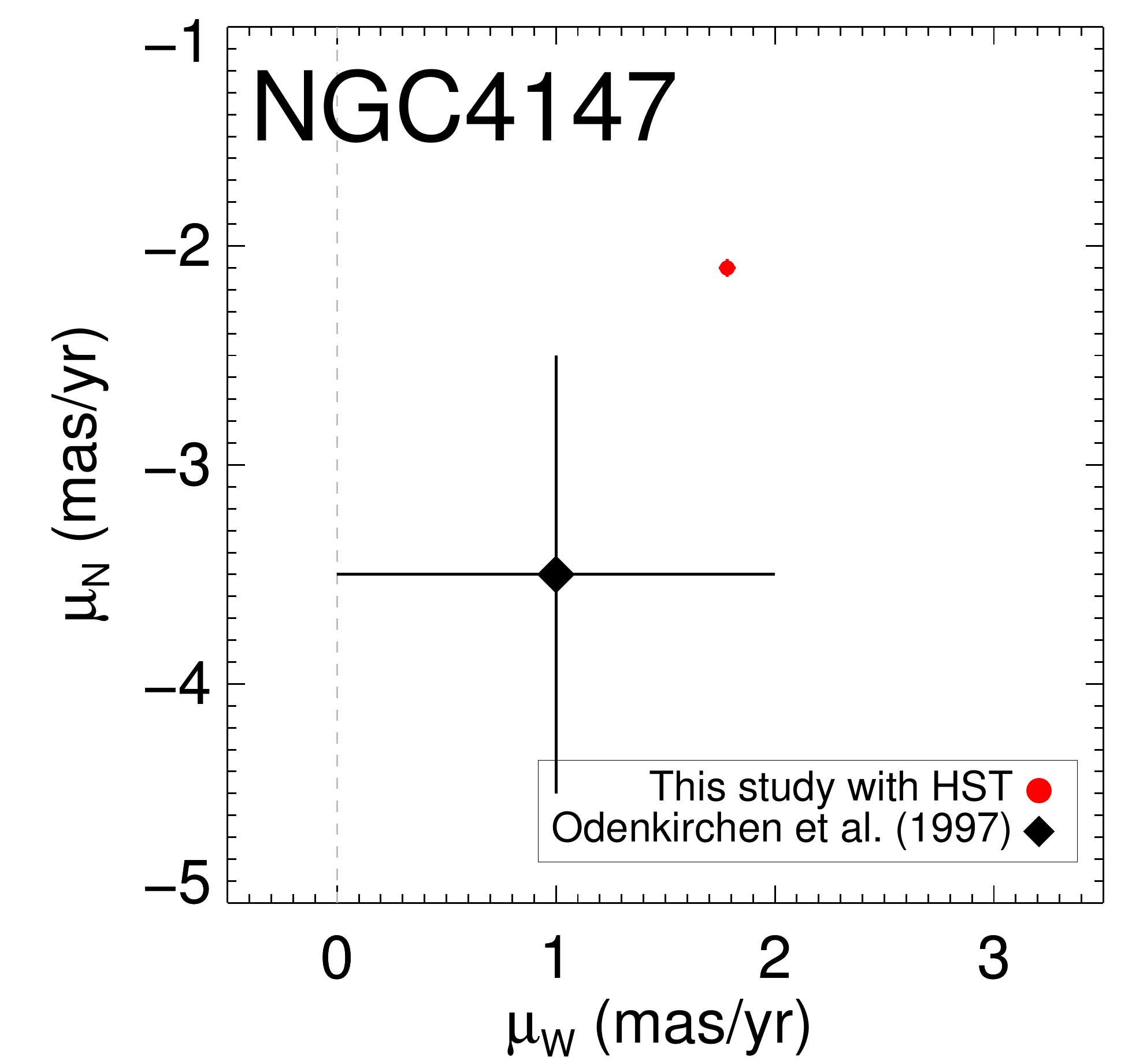}{0.25\textwidth}{(a) NGC\,4147} 
           \fig{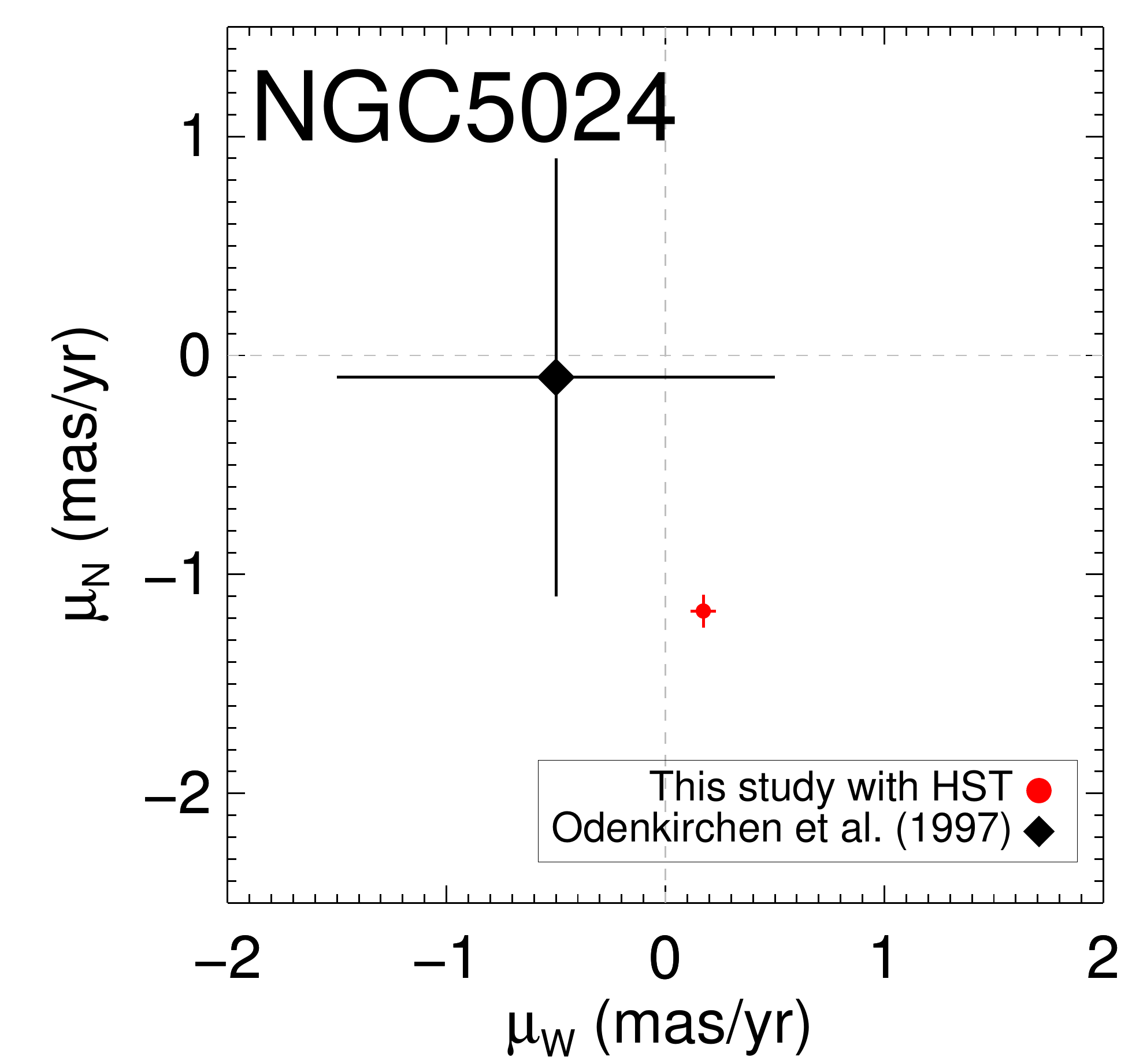}{0.25\textwidth}{(b) NGC\,5024} 
           \fig{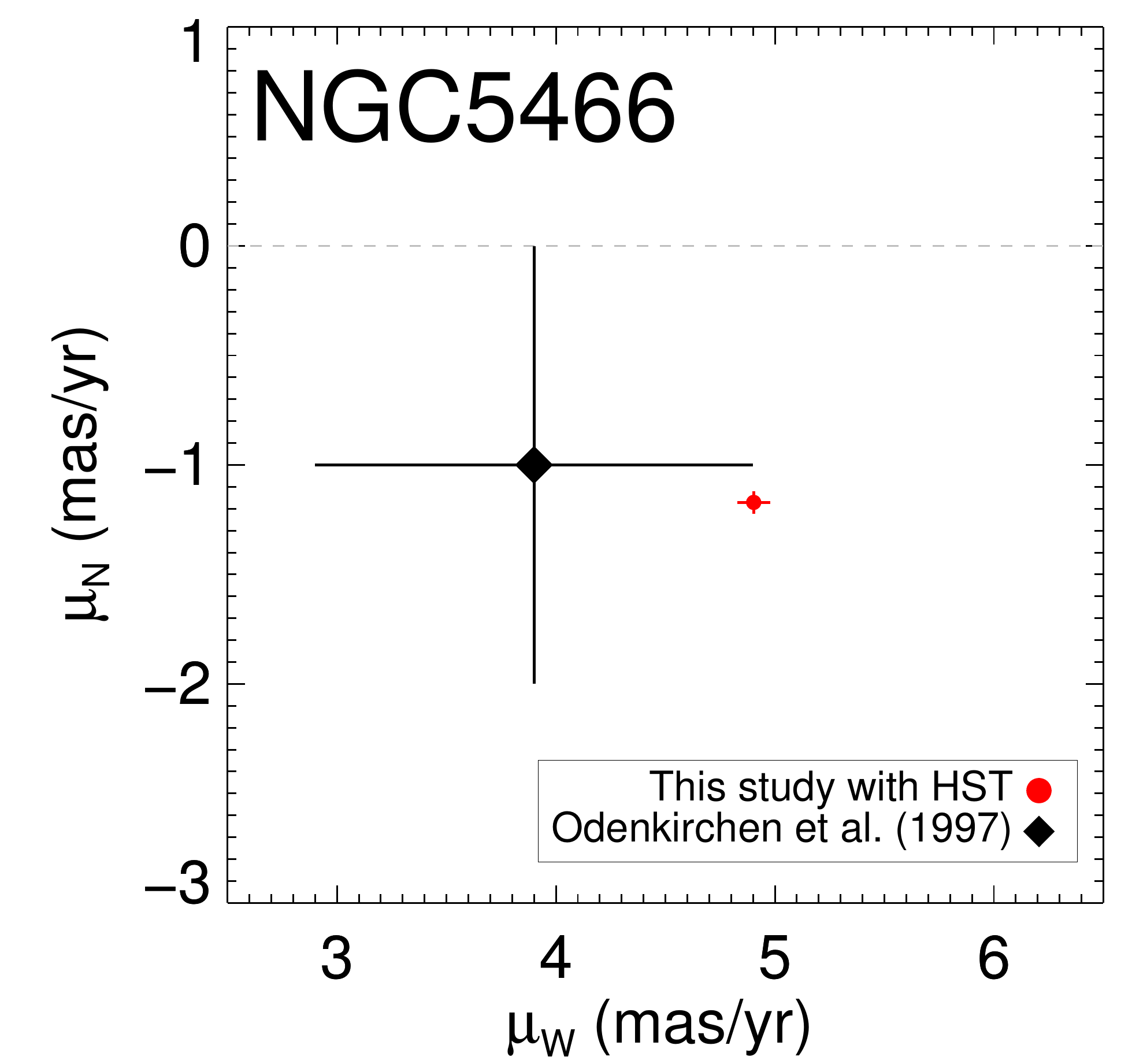}{0.25\textwidth}{(c) NGC\,5466} 
           \fig{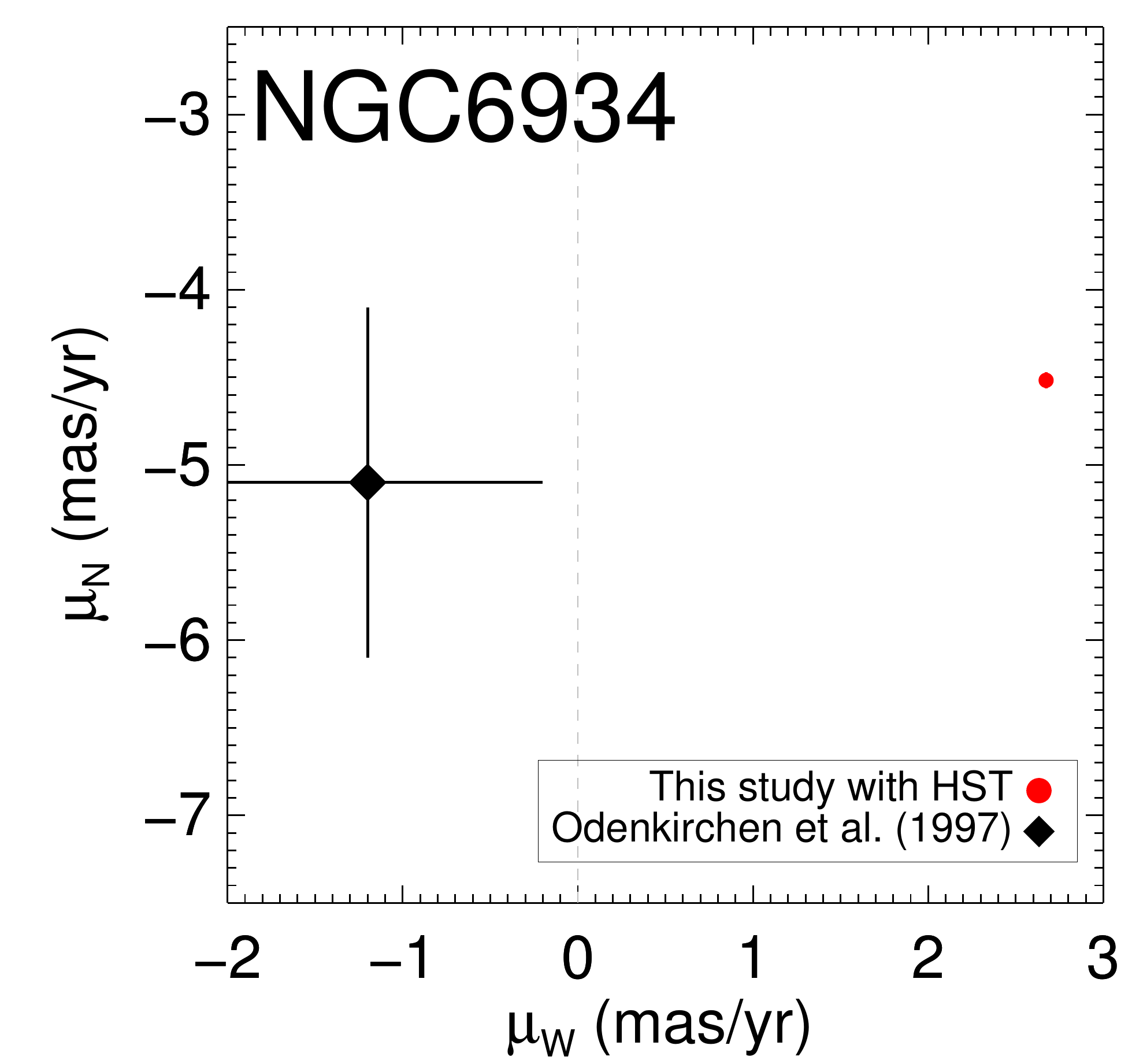}{0.25\textwidth}{(d) NGC\,6934} }
\gridline{ \fig{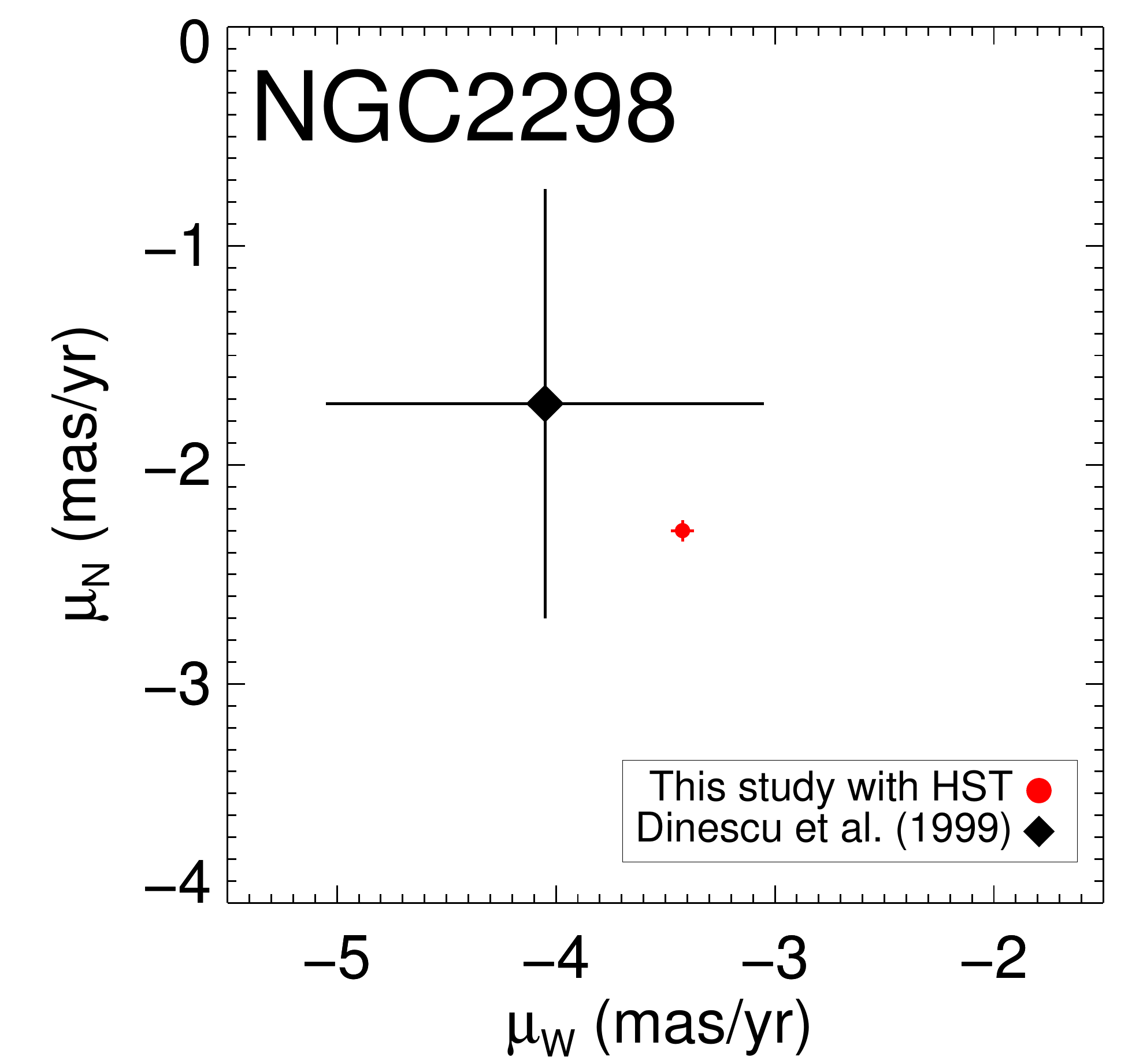}{0.25\textwidth}{(e) NGC\,2298}
           \fig{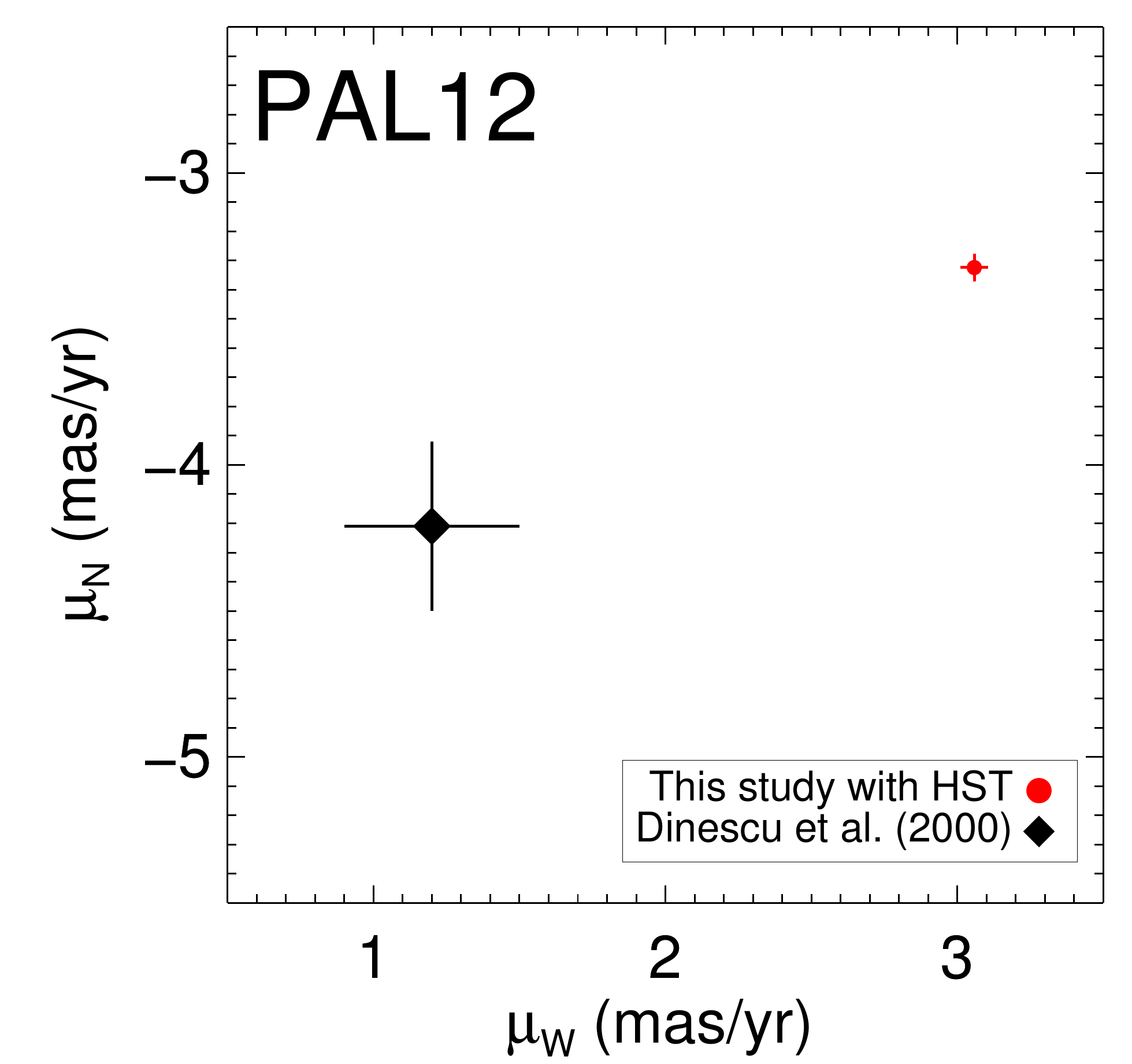}  {0.25\textwidth}{(f) Pal\,12} 
           \fig{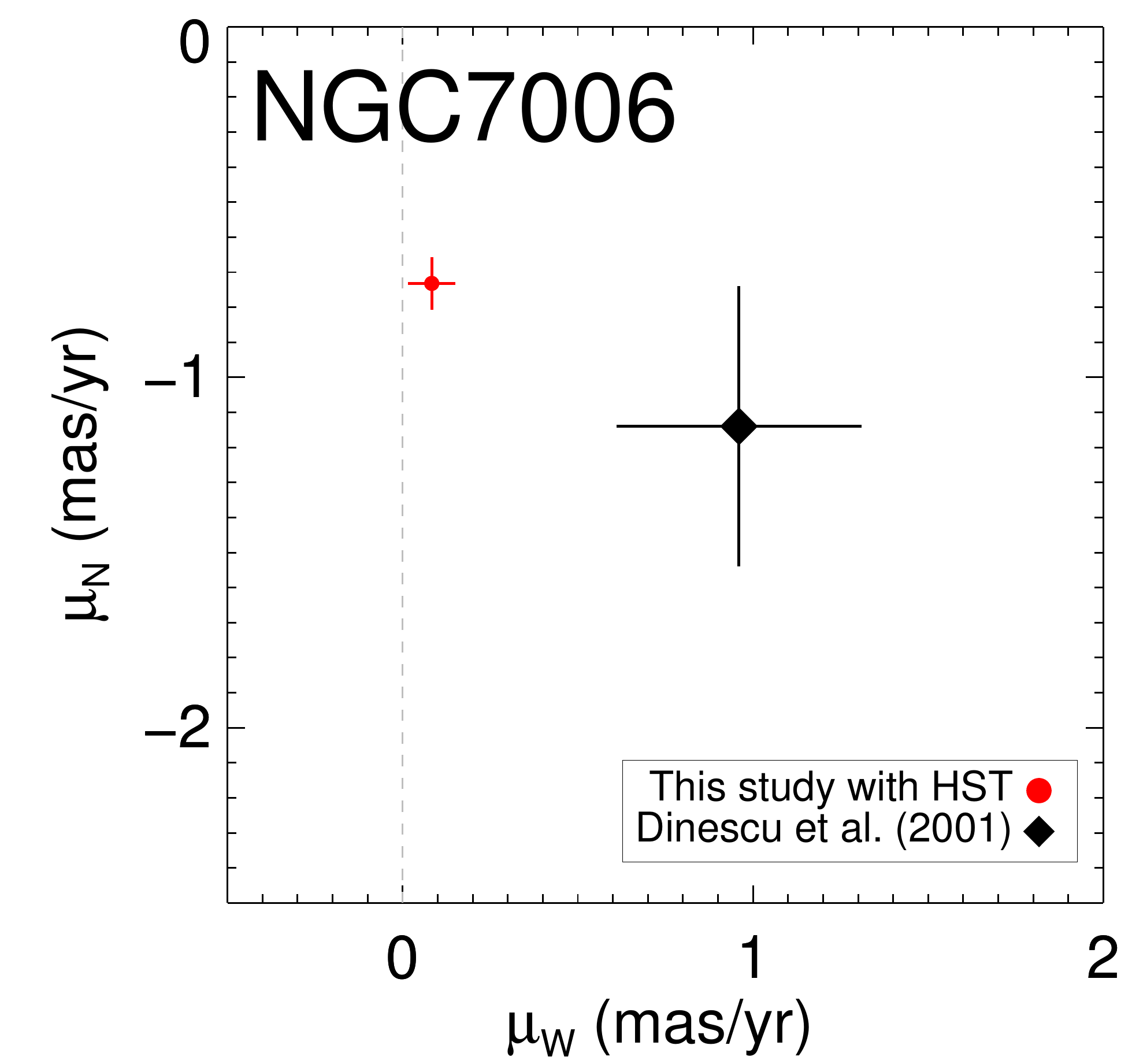}{0.25\textwidth}{(g) NGC\,7006} 
           \fig{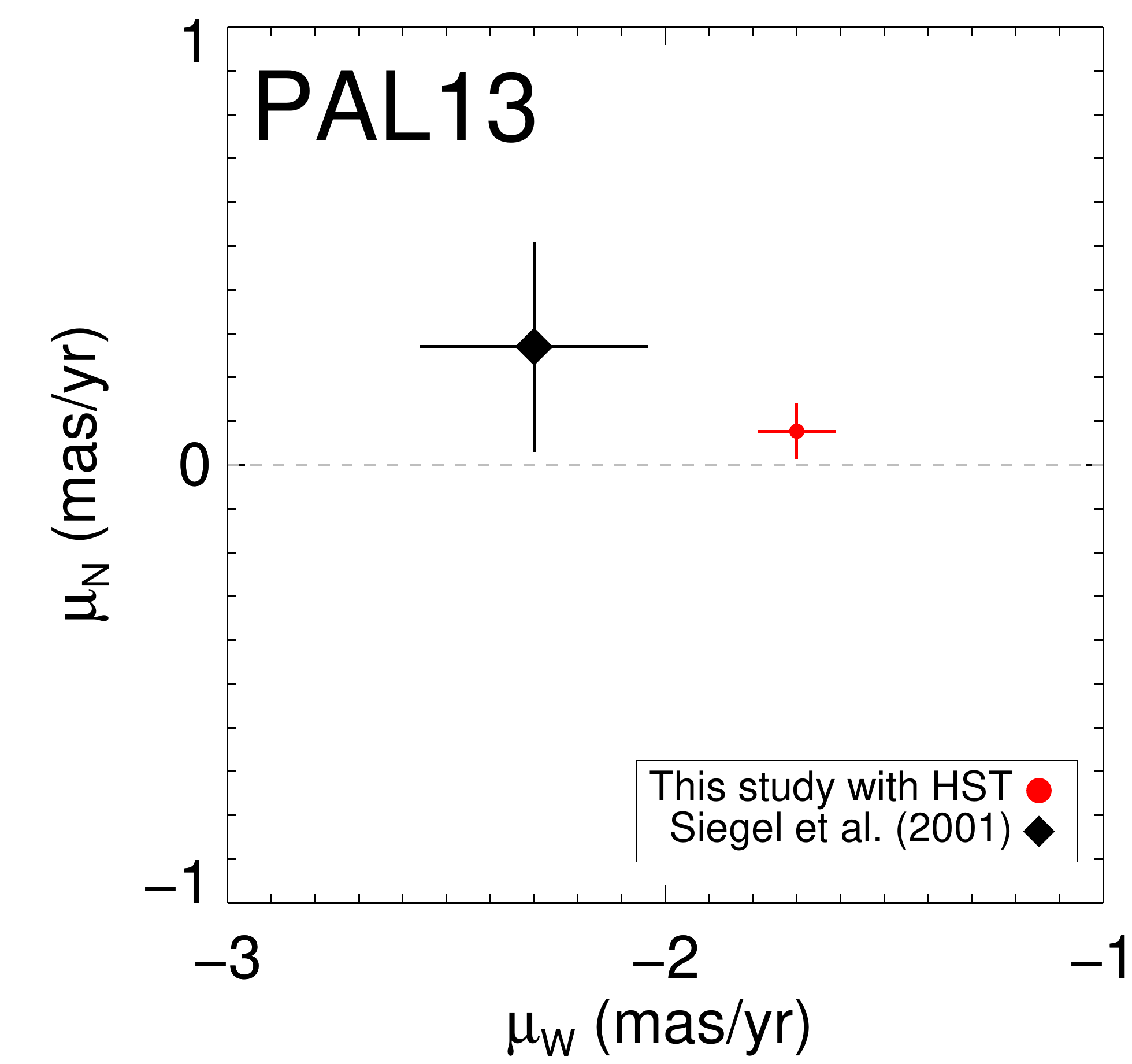}  {0.25\textwidth}{(h) Pal\,13} }
\gridline{ \fig{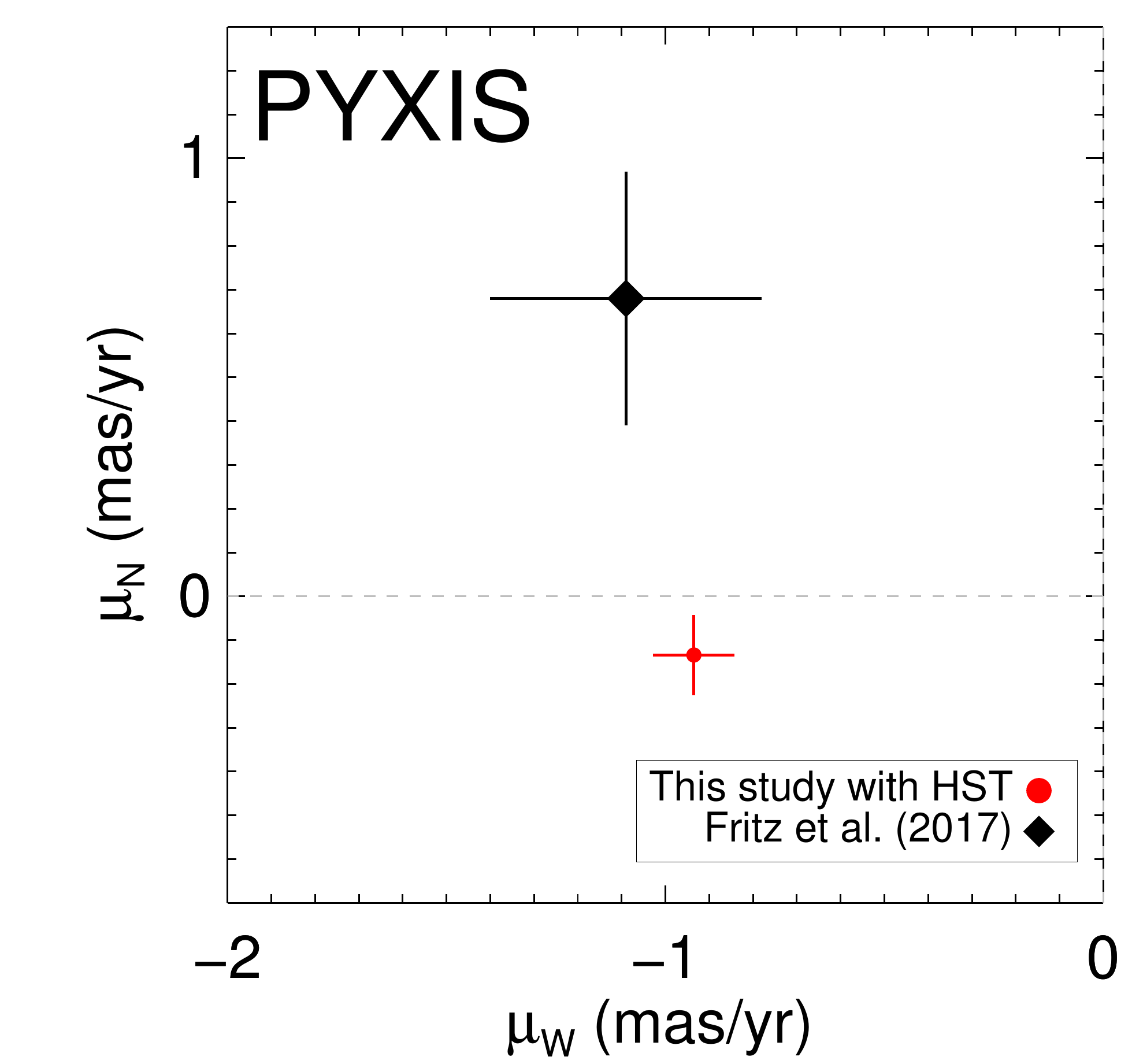}  {0.25\textwidth}{(i) Pyxis} 
           \fig{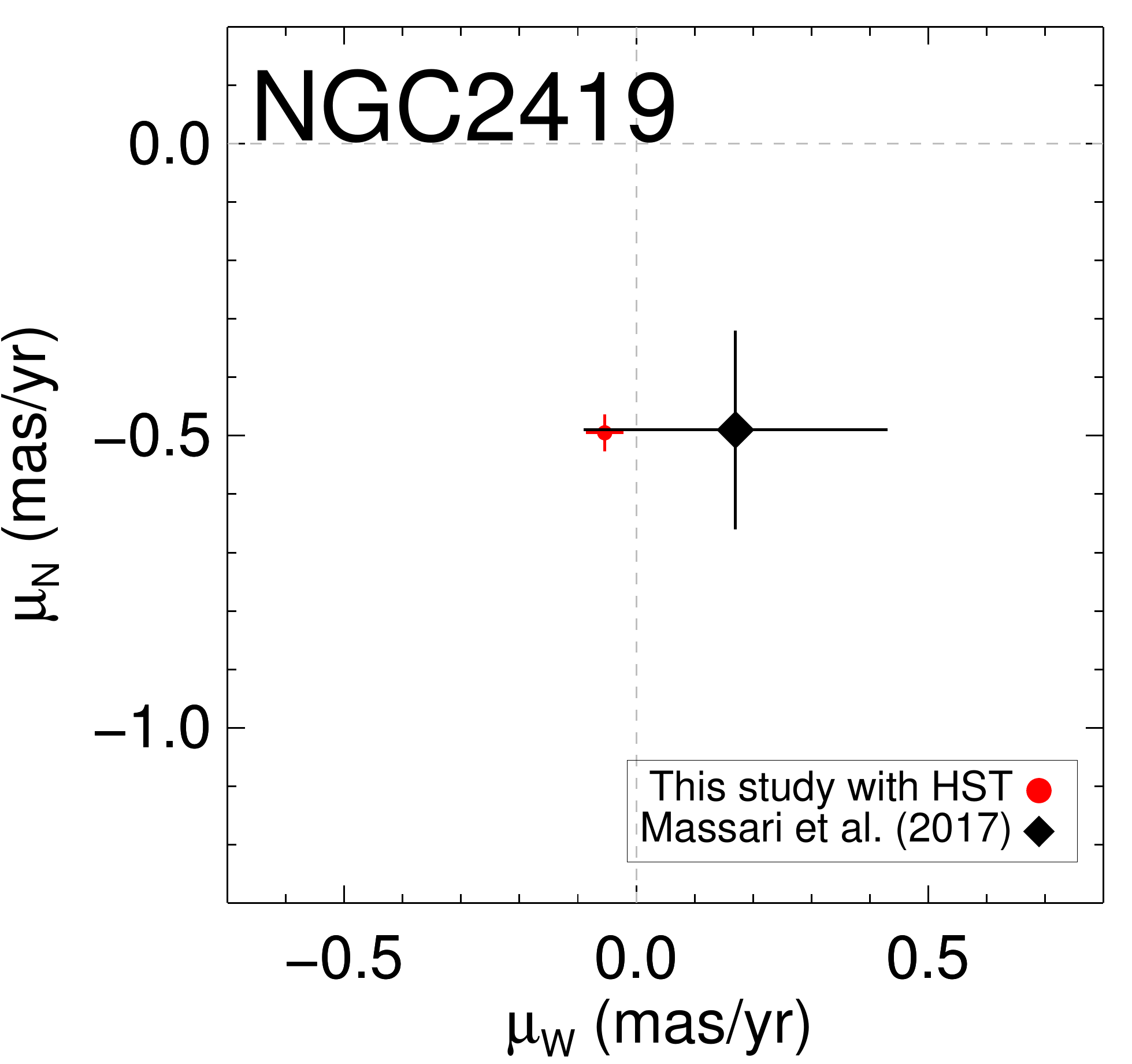}{0.25\textwidth}{(j) NGC\,2419} }
           
\caption{Comparison of our proper motions (red circles) with previous 
         measurements (black diamonds) for NGC\,4147, NGC\,5024, NGC\,5466, 
         NGC\,6934, NGC\,2298, Pal\,12, NGC\,7006, Pal\,13, Pyxis, 
         and NGC\,2419. Sources of the previous measurements are 
         indicated in the legend of each figure. The previous 
         PM measurements in the first two rows were done with 
         ground-based data only, the Pyxis one used mixed ground- and
         space-based data, and the NGC\,2419 used space-based data only.
         \label{f:pmdcompare}
         }
\end{figure*}
%

%
%
\begin{deluxetable}{lccl}
\renewcommand{\arraystretch}{0.9}
\tablecaption{Older proper motion results
              \label{t:oldpm}
             } 
\tablehead{
   \colhead{}        & \colhead{$\mu_\mathrm{W}$} & \colhead{$\mu_\mathrm{N}$} & \colhead{}      \\
   \colhead{Cluster} & \colhead{(mas yr$^{-1}$)}  & \colhead{(mas yr$^{-1}$)}  & \colhead{Ref.\tablenotemark{a}}
}
\startdata
NGC\,2298 &    $-4.05 \pm 1.00$ &    $-1.72 \pm 0.98$ & (1) \\
NGC\,2419 & \phs$0.17 \pm 0.26$ &    $-0.49 \pm 0.17$ & (2) \\
NGC\,4147 & \phs$1.00 \pm 1.00$ &    $-3.50 \pm 1.00$ & (3) \\
NGC\,5024 &    $-0.50 \pm 1.00$ &    $-0.10 \pm 1.00$ & (3) \\
NGC\,5466 & \phs$3.90 \pm 1.00$ &    $-1.00 \pm 1.00$ & (3) \\
NGC\,6934 &    $-1.20 \pm 1.00$ &    $-5.10 \pm 1.00$ & (3) \\
NGC\,7006 & \phs$0.96 \pm 0.35$ &    $-1.14 \pm 0.40$ & (4) \\
Pal\,12   & \phs$1.20 \pm 0.30$ &    $-4.21 \pm 0.29$ & (5) \\
Pal\,13   &    $-2.30 \pm 0.26$ & \phs$0.27 \pm 0.24$ & (6) \\
Pyxis     &    $-1.09 \pm 0.31$ & \phs$0.68 \pm 0.29$ & (7) \\
\enddata
\tablenotetext{a}{References from which the older PMs were adopted:  
                  (1) \citet{din99}; (2) \citet{mas17}; (3) \citet{ode97};
                  (4) \citet{din01}; (5) \citet{din00}; (6) \citet{sie01}; 
                  (7) \citet{fri17}}
\end{deluxetable}
%

Searching through the literature, we found that all of our target GCs, 
except for Pal\,13 and Pyxis, have their PMs listed in the ``Global 
survey of stars clusters in the Milky Way'' catalog by \citet{kha13}.  
These PMs are from the PPMXL catalog \citep{ros10}, which are derived 
by comparing catalog positions from the USNO-B1.0 and the Two Micron 
All Sky Survey (2MASS). Upon comparing our results with those from the 
\citet{kha13} compilation, we found that their PMs are highly unreliable 
with results significantly inconsistent (beyond their claimed 1$\sigma$ 
uncertainties) with ours and others (see below) in most cases. This is 
likely due to the PPMXL catalog positions being very uncertain in 
crowded regions. We therefore decided to ignore the PM results from 
this compilation. Ten out of our 20 target GCs have previous 
PM measurements in the literature, some in fact have multiple 
measurements, but here we considered only the most precise ones, listed 
in Table~\ref{t:oldpm}.

\citet{ode97} determined absolute PMs of 15 Galactic GCs using  
photographic observations tied to the Hipparcos reference system \citep{per97}.
Our target list includes four clusters in common with \citet{ode97}: 
NGC\,4147, NGC\,5024, NGC\,5466, and NGC\,6934. Comparison between the 
\citet{ode97} and our HST PM measurements are shown in 
Figures~\ref{f:pmdcompare}(a)--(d). Our measurement uncertainties are 
tiny compared to the older measurements as expected from the superb 
astrometric capabilities of \hst. For NGC\,5024 and NGC\,5466, 
the old and new measurements are fully consistent within their 1$\sigma$ 
uncertainties, and for NGC\,4147, they are marginally consistent. 
However, for NGC\,6934, the PM by \citet{ode97} is inconsistent 
with our result in the $\muw$ direction at $\sim 4\sigma$ level.

Using photographic plates taken 20--40 years apart, \citet{din99,din00,din01} 
measured absolute PMs of NGC\,2298, Pal\,12, and NGC\,7006. Comparisons 
of these old measurements with our new ones are presented in 
Figures~\ref{f:pmdcompare}(e)--(g). For NGC\,2298, the agreement between 
the two is excellent, but for Pal\,12 and NGC\,7006, there are 
significant inconsistencies. 

\citet{sie01} derived the PM of Pal\,13 from photographic plates separated 
by a 40-year baseline to reach a PM accuracy of $0.26 \masyr$ per coordinate. 
Figure~\ref{f:pmdcompare}(h) shows the comparison between 
the measurements made by \citet{sie01} and by us. Our \hst\ PM measurements 
are a factor of $\sim 3$ improvement over the old one in terms of 
the PM accuracy per coordinate. The two PM measurements in 
$\mun$ are consistent within the uncertainties while those in $\muw$ are 
discrepant at the $\sim 2\sigma$ level.

\citet{fri17} recently measured the PM of Pyxis for the first time 
using \hst\ ACS/WFC data as the first epoch and GeMS/GSAOI Adaptive Optics 
data as the second epoch.\footnote{The ACS/WFC data used by \citet{fri17} 
are the same as the first-epoch data used in this study.} 
Figure~\ref{f:pmdcompare}(i) shows the comparison between the 
\citet{fri17} results and our measurements. The two measurements are 
consistent in the $\muw$ direction, but inconsistent in the $\mun$ direction.

Finally, \citet{mas17} derived the PM of NGC\,2419 for the first time 
by combining data from \hst\ and {\it Gaia} DR1. As shown in 
Figure~\ref{f:pmdcompare}(j), we find that our measurement is 
consistent with the \citet{mas17} result within $1\sigma$ uncertainty.

\citet{wat17} derived PMs of five nearby GCs (no overlap with our sample) 
using the Tycho-{\it Gaia} Astrometric Solution (TGAS) catalog from 
{\it Gaia} Data Release 1 (DR1), and upon comparing their results with 
existing \hst\ PM measurements, they found excellent agreement between 
the two measurements. However, comparison with ground-based PM 
measurements showed that these measurements may have large unidentified 
systematics beyond their quoted measurements uncertainties. Additionally, 
\citet{lib18b} found significant inconsistencies between their \hst\ and 
previous ground-based measurements for $\omega$ Cen. In general, these 
are consistent with what we find above, and provides us confidence 
for our \hst\ results.

In Section~\ref{s:spacemotions}, we discuss the implications of orbital 
velocities for GCs that show large discrepancies between the old 
and new PM measurements. 

\section{Space Motions}
\label{s:spacemotions}

We now use our new PM measurements, and combine them with existing 
observational parameters to explore the space motions of our target GCs. 
We note that the goal here is not to discuss the orbital motion 
of each cluster in detail, but to provide insights into the space
motions deduced from observations. Details of orbital properties for 
some of these GCs will be presented in separate papers in the future.

\subsection{Three-dimensional Positions and Velocities}
\label{ss:3dposvel}

We start by calculating the current Galactocentric positions and 
velocities of our target GCs based on their observed parameters 
including PM results from Section~\ref{ss:pmresults}. 
To do this, we adopt the Cartesian coordinate system $(X,\,Y,\,Z)$ we 
used in our previous studies \citep[e.g.,][]{soh12}: the origin is at 
the Galactic center, and the $X$-, $Y$-, and $Z$-axes point in the 
direction from the Sun to the Galactic center, in the direction of 
the Sun's Galactic rotation, and toward the Galactic north pole, 
respectively.

For the heliocentric $\vlos$ of our target GCs, we adopted the 
most recent measurements in the literature whenever possible, but 
for GCs with no measurements available after 2010, we adopted the 
values listed in the H10 catalog. In most cases, the recent 
measurements agree well with the H10 catalog, but we found large 
discrepancies for NGC\,6426 and Terzan\,8. The $\vlos$ we used 
for each cluster are listed in the third column of 
Table~\ref{t:spacemotions}.

For the heliocentric distances to our target GCs, we took a 
careful approach for the following reason. The tangential velocity of 
a given object at distance $D$ is proportional to $\mu\times D$, where 
$\mu$ is the measured total PM. 
This implies that the uncertainties in tangential velocities 
depend on uncertainties in both distance {\it and} PM measurements.
So far, PM uncertainties have been the limiting factor when 
trying to obtain precise tangential velocities for stellar systems 
in the MW halo. However, due to our very small PM errors, uncertainties 
in distances will have comparable impacts on the tangential velocities.
To obtain distances that are as homogeneous as possible, we adopted 
the distance moduli derived in \citet{dot10} and \citet{dot11} for all 
of our target clusters except for Pal\,13 and NGC\,2419. The Dotter 
et al. studies carefully measured the HB levels of GCs using $F606W$ 
and $F814W$ photometry obtained with \hst\ ACS/WFC. The distance moduli 
provided in Table~2 of both papers were converted into physical distances 
assuming a $R_{V} = 3.1$, and adopting extinction values based on 
Table~6 of \citet{sch11}. The Dotter et al. studies do not provide 
individual uncertainties for the measured distance moduli, but 
point out that the typical measurement uncertainties are 0.05 mag 
\citep[visibly evident in Figure~6 of][]{dot10}, which results into 
distance uncertainty of $\Delta D_{\sun} = 0.023 D$. 
We adopted this relation for the distance errors.
For Pal\,13, we adopt the distance measured by \citet{ham13}, 
who used the same \hst\ WFC3/UVIS data as our first-epoch images, 
but we adopt the more conservative distance error given by 
$\Delta D_{\sun} = 0.023 D_{\sun}$, which results in $D_{\sun} = 
22.28 \pm 0.51$~kpc.\footnote{\citet{ham13} formal error is $\Delta 
D_{\sun} = 0.23$~kpc.} This measurement also relies on the same isochrones 
used for the ACS/WFC measurements above, so the Pal\,13 distance is 
considered to be in line with the distance scale used for our other 
target clusters. For NGC\,2419, we adopt the distance measured from its 
large population of RR Lyrae by \citet{cri11}: $D_{\sun} = 87.5 \pm 3.3$~kpc. 
This distance is consistent with the distance derived from the HB stars by 
\citet{bel14}, but inconsistent with the distance of $97.7$~kpc inferred 
from the isochrone fitting of \citet{dot10}.
NGC\,2419 is by far the most distant cluster in our sample, and it 
is possible that an isochrone fit through the main sequence turnoff 
and below may be less accurate for this cluster than for the others.
The heliocentric distances are listed in the second column of 
Table~\ref{t:spacemotions}.

Once we compiled the observed positions and velocities for each 
cluster, we computed their current Galactocentric positions $(X,\,Y,\,Z)$ 
and velocities $(v_{X},\,v_{Y},\,v_{Z})$ assuming the distance of the Sun from 
the Galactic center and the circular velocity of the 
local standard of rest (LSR) to be $R_{0} = 8.29 \pm 0.16$ kpc and 
$V_{0} = 239 \pm 5$~\kms, respectively \citep{mcm11}.
The solar peculiar velocity with respect to the LSR were taken 
from the estimates of \citet{sch10}: 
$(U_\mathrm{pec},\,V_\mathrm{pec},\,W_\mathrm{pec}) = (11.10,\,12.24,\,7.25)$~\kms 
with uncertainties of $(1.23,\,2.05,\,0.62)$~\kms.
Results along with the adopted distances and $\vlos$ are presented 
in Table~\ref{t:spacemotions}. The uncertainties here and hereafter were 
obtained from a Monte Carlo scheme that propagates all observational 
uncertainties and their correlations, including those for the Sun. 
In the same table, we also list the Galactocentric radial, tangential, 
and total velocities. 

%
%
\begin{deluxetable*}{lccccccccccc}
\setlength\tabcolsep{2pt}
\renewcommand{\arraystretch}{1.0}
\tabletypesize{\scriptsize}
\tablecaption{Adopted distances, radial velocities, and calculated 
              Galactocentric positions and velocities of our target GCs.
              \label{t:spacemotions}
             } 
\tablehead{
   \colhead{}        & \colhead{$D_\mathrm{\odot}$} & \colhead{$v_\mathrm{hel}$\tablenotemark{a}} & \colhead{$X$}   & \colhead{$Y$}   & \colhead{$Z$}    & \colhead{$\vx$} & \colhead{$\vy$} & \colhead{$\vz$} & \colhead{$v_\mathrm{rad}$} & \colhead{$v_\mathrm{tan}$} & \colhead{$v_\mathrm{tot}$} \\
   \colhead{Cluster} & \colhead{(kpc)}           & \colhead{(\kms)}      & \colhead{(kpc)}  & \colhead{(kpc)} & \colhead{(kpc)} & \colhead{(\kms)}    & \colhead{(\kms)}    & \colhead{(\kms)}    & \colhead{(\kms)}      & \colhead{(\kms)}      & \colhead{(\kms)}  
}
\startdata
Arp\,2    &  $29.8\pm0.7$ &    \phs$115.0\pm   10.0$ &    \phs$19.2$ & \phs\phn$4.1$ &       $-10.6$ &    \phs$250.3\pm\phn9.9$ &    \phn$-36.0\pm   10.1$ &    \phs$194.0\pm\phn7.7$ &       \phs$117.1\pm\phn9.9$ &    $296.4\pm\phn8.4$ &    $318.7\pm\phn9.2$ \\ 
IC\,4499  &  $20.1\pm0.5$ & \phs\phn$31.5\pm\phn0.4$ & \phs\phn$3.1$ &       $-14.9$ &    \phn$-7.0$ & \phs\phn$28.9\pm\phn5.5$ &    \phs$247.8\pm\phn6.9$ &    \phn$-46.5\pm\phn6.3$ &          $-195.6\pm\phn6.2$ &    $161.6\pm\phn6.8$ &    $253.7\pm\phn7.4$ \\ 
NGC\,1261 &  $16.4\pm0.4$ & \phs\phn$68.2\pm\phn4.6$ &    \phn$-8.2$ &       $-10.1$ &       $-13.0$ & \phs\phn$78.0\pm\phn3.9$ & \phs\phn$51.6\pm\phn7.6$ & \phs\phn$76.6\pm\phn5.1$ &          $-117.2\pm\phn5.4$ & \phn$29.3\pm\phn6.0$ &    $120.9\pm\phn4.9$ \\ 
NGC\,2298 &  $10.2\pm0.2$ &    \phs$148.9\pm\phn1.2$ &       $-12.3$ &    \phn$-8.9$ &    \phn$-2.8$ & \phs\phn$90.4\pm\phn4.1$ & \phs\phn$24.5\pm\phn6.1$ & \phs\phn$73.3\pm\phn3.6$ &       \phn$-99.4\pm\phn4.1$ & \phn$65.2\pm\phn5.6$ &    $118.9\pm\phn4.4$ \\ 
NGC\,2419 &  $87.5\pm3.3$ &    \phn$-20.3\pm\phn0.7$ &       $-87.4$ &    \phn$-0.5$ &    \phs$37.3$ & \phs\phn$16.6\pm\phn5.8$ & \phs\phn$48.5\pm   16.1$ &    \phn$-31.4\pm   12.0$ &       \phn$-27.9\pm\phn1.4$ & \phn$53.2\pm   15.1$ & \phn$60.1\pm   13.1$ \\ 
NGC\,4147 &  $19.5\pm0.4$ &    \phs$179.5\pm\phn0.5$ &    \phn$-9.6$ &    \phn$-4.1$ &    \phs$19.0$ &    \phn$-49.2\pm\phn4.0$ &    \phn$-30.5\pm\phn8.6$ &    \phs$126.1\pm\phn1.7$ &       \phs$138.1\pm\phn2.1$ & \phn$13.4\pm\phn4.4$ &    $138.7\pm\phn1.9$ \\ 
NGC\,5024 &  $19.0\pm0.4$ &    \phn$-62.8\pm\phn0.3$ &    \phn$-5.3$ &    \phn$-1.5$ &    \phs$18.7$ & \phs\phn$47.5\pm\phn6.0$ &    \phs$161.9\pm\phn8.6$ &    \phn$-69.8\pm\phn1.4$ &       \phn$-92.5\pm\phn2.6$ &    $157.4\pm\phn8.2$ &    $182.7\pm\phn7.3$ \\ 
NGC\,5053 &  $18.1\pm0.4$ & \phs\phn$42.6\pm\phn0.3$ &    \phn$-5.1$ &    \phn$-1.4$ &    \phs$17.7$ & \phs\phn$51.0\pm\phn6.0$ &    \phs$150.7\pm\phn8.3$ & \phs\phn$35.5\pm\phn1.4$ & \phs\phn\phn$8.2\pm\phn2.9$ &    $162.8\pm\phn7.9$ &    $163.0\pm\phn7.9$ \\ 
NGC\,5466 &  $16.5\pm0.4$ &    \phs$106.9\pm\phn0.2$ &    \phn$-4.8$ & \phs\phn$3.1$ &    \phs$15.8$ &       $-180.3\pm\phn7.1$ &    \phn$-40.6\pm   10.2$ &    \phs$218.2\pm\phn3.0$ &       \phs$249.3\pm\phn3.6$ &    $140.0\pm   12.8$ &    $285.9\pm\phn7.7$ \\ 
NGC\,6101 &  $14.8\pm0.3$ &    \phs$361.4\pm\phn1.7$ & \phs\phn$2.2$ &    \phn$-9.6$ &    \phn$-4.0$ &    \phs$283.6\pm\phn3.0$ & \phs\phn$79.5\pm\phn6.3$ &       $-199.1\pm\phn4.2$ &    \phs\phn$63.6\pm   10.9$ &    $349.8\pm\phn4.7$ &    $355.5\pm\phn4.0$ \\ 
NGC\,6426 &  $21.1\pm0.5$ &       $-212.2\pm\phn0.5$ & \phs\phn$9.6$ & \phs\phn$9.6$ & \phs\phn$5.9$ & \phn\phn$-5.1\pm\phn5.1$ &       $-154.1\pm   10.4$ &    \phn$-47.3\pm\phn5.7$ &          $-121.8\pm\phn3.5$ &    $105.7\pm   11.6$ &    $161.2\pm\phn9.8$ \\ 
NGC\,6934 &  $16.3\pm0.4$ &       $-406.5\pm\phn0.5$ & \phs\phn$1.2$ &    \phs$12.1$ &    \phn$-5.3$ & \phs\phn$92.0\pm\phn7.8$ &       $-303.2\pm\phn8.2$ &    \phs$129.4\pm\phn3.2$ &          $-320.4\pm\phn4.8$ &    $120.4\pm   13.5$ &    $342.3\pm\phn8.8$ \\ 
NGC\,7006 &  $40.5\pm0.9$ &       $-384.1\pm\phn0.4$ & \phs\phn$8.6$ &    \phs$34.2$ &       $-13.4$ &    \phn$-50.8\pm   12.7$ &       $-148.9\pm\phn9.6$ & \phs\phn$66.6\pm   12.9$ &          $-170.3\pm\phn5.8$ & \phn$14.5\pm   11.2$ &    $170.9\pm\phn5.2$ \\ 
Pal\,12   &  $18.6\pm0.4$ & \phs\phn$27.8\pm\phn1.5$ & \phs\phn$2.5$ & \phs\phn$6.4$ &       $-13.7$ &    \phs$318.0\pm\phn7.7$ & \phs\phn$14.0\pm\phn8.8$ &    \phs$100.7\pm\phn4.0$ &       \phn$-32.8\pm\phn3.1$ &    $332.2\pm\phn8.1$ &    $333.8\pm\phn8.1$ \\ 
Pal\,13   &  $22.3\pm0.5$ & \phs\phn$25.2\pm\phn0.3$ &    \phn$-7.5$ &    \phs$16.4$ &       $-15.1$ &       $-146.9\pm\phn9.6$ &    \phs$217.3\pm\phn7.6$ &    \phn$-75.3\pm\phn5.8$ &       \phs$246.5\pm\phn5.4$ &    $117.2\pm\phn7.1$ &    $272.9\pm\phn6.8$ \\ 
Pal\,15   &  $46.2\pm1.1$ & \phs\phn$68.9\pm\phn1.1$ &    \phs$31.5$ &    \phs$13.6$ &    \phs$19.0$ &    \phs$139.6\pm   12.4$ &    \phs$100.7\pm   22.7$ & \phs\phn$13.1\pm   19.0$ &       \phs$153.3\pm\phn3.5$ & \phn$79.3\pm   19.3$ &    $172.6\pm\phn9.4$ \\ 
Pyxis     &  $37.4\pm0.9$ & \phs\phn$34.3\pm\phn1.9$ &       $-13.9$ &       $-36.7$ & \phs\phn$4.6$ &    \phs$135.2\pm   16.6$ &    \phs$211.1\pm\phn6.5$ &    \phs$117.7\pm   16.5$ &          $-230.0\pm\phn6.5$ &    $154.2\pm   16.9$ &    $276.9\pm   11.6$ \\ 
Rup\,106  &  $21.7\pm0.5$ &    \phn$-44.0\pm\phn3.0$ & \phs\phn$2.6$ &       $-18.3$ & \phs\phn$4.4$ &       $-113.8\pm\phn7.6$ &    \phs$236.9\pm\phn7.5$ & \phs\phn$40.9\pm\phn6.6$ &          $-234.5\pm\phn6.0$ &    $125.6\pm\phn7.2$ &    $266.0\pm\phn6.1$ \\ 
Terzan\,7 &  $25.5\pm0.6$ &    \phs$166.0\pm\phn4.0$ &    \phs$15.6$ & \phs\phn$1.4$ &    \phn$-8.7$ &    \phs$280.2\pm\phn5.3$ &    \phn$-51.9\pm   10.8$ &    \phs$209.8\pm\phn8.8$ &       \phs$137.4\pm\phn4.4$ &    $326.1\pm\phn9.8$ &    $353.9\pm\phn9.2$ \\ 
Terzan\,8 &  $28.6\pm0.7$ &    \phs$145.3\pm\phn0.2$ &    \phs$17.6$ & \phs\phn$2.6$ &       $-11.9$ &    \phs$304.4\pm\phn5.8$ &    \phn$-48.7\pm   11.9$ &    \phs$230.5\pm   11.1$ &       \phs$116.5\pm\phn2.4$ &    $366.9\pm   12.8$ &    $384.9\pm   11.9$ \\ 
\enddata
\tablenotetext{a}{References for adopted $v_\mathrm{hel}$: IC\,4499 -- \citet{han11}; NGC\,2419 -- \citet{bau09}; 
                  NGC\,4147, NGC\,5024, NGC\,5053, NGC\,5466, NGC\,6934 -- \citet{kim15}; NGC\,6426 -- \citet{han17};
                  Terzan\,8 -- \citet{sol14}; all others -- H10}
\end{deluxetable*}
%

%
\begin{figure}
\includegraphics[width=\columnwidth]{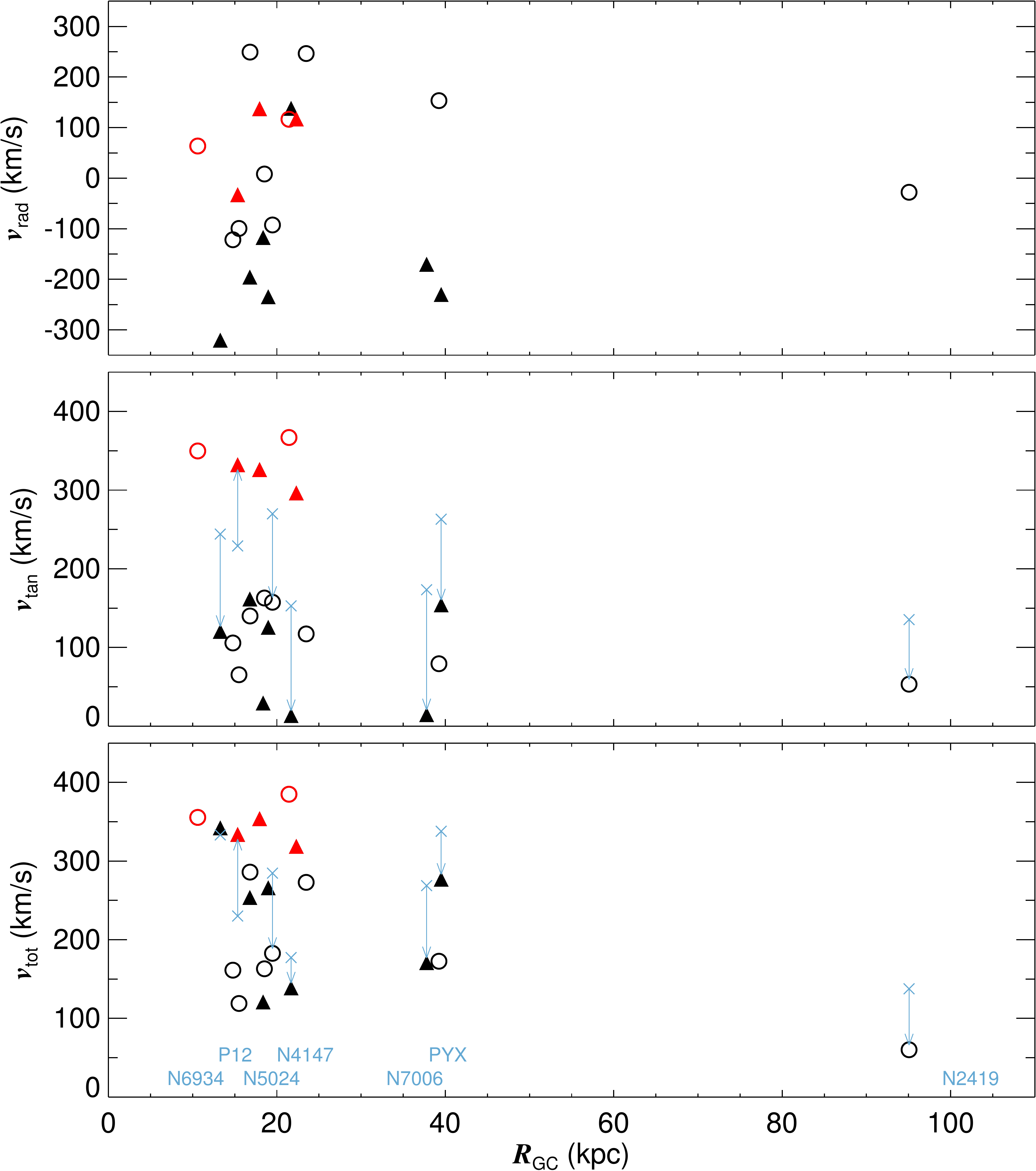}
\caption{Galactocentric radial (top), tangential (middle), and total 
         velocities (bottom) versus Galactocentric distance of our 
         target GCs. GCs that belong to the young (old) group as 
         listed in Table~\ref{t:targets} are plotted as  
         closed triangles (open circles). The five GCs with 
         $\vtan > 295$~\kms\ are plotted in red, while all other GCs 
         are plotted in black to highlight the dichotomy seen in 
         the $\vtan$ distribution. In the middle and bottom 
         panels, velocities implied from the older PM measurements are 
         plotted as light blue $\times$ symbols with arrows pointing 
         to the new measurements for NGC\,6934, Pal\,12, NGC\,5024, 
         NGC\,4147, NGC\,7006, Pyxis, and NGC\,2419 (from left to 
         right as indicated in the bottom panel).
         \label{f:vel_rgc}
         }
\end{figure}
%

Figure~\ref{f:vel_rgc} shows the Galacticentric velocities 
($\vrad$, $\vtan$, and $\vtot$ taken from Table~\ref{t:spacemotions}) 
as a function of distance from the Galactic Center ($\rgc = 
\sqrt{X^2 + Y^2 + Z^2}$) for our target GCs. GCs that belong to 
different age groups (see Table~\ref{t:targets}) are plotted in 
different symbols. Overall, we do not find any systematic difference 
in the velocity distribution between the young and old GCs. 

The $\vrad$ vs. $\rgc$ plot in the top panel shows a hint of the 
well known trend of decreasing velocity dispersion as a function of 
Galactocentric distance \citep[see e.g., Figure~2 of][]{bat05}.
In the middle panel of Figure~\ref{f:vel_rgc}, we find a clear 
dichotomy in $\vtan$ among our sample: five GCs have $\vtan \gtrsim 295$~\kms\ 
(red), whereas all other GCs have $\vtan < 170$~\kms (black). 
These high $\vtan$ GCs are also among the GCs with highest $\vtot$ (bottom panel).
Among the five GCs with high $\vtan$, Arp\,2, Terzan\,7, and Terzan\,8 
are known to be associated with the Sgr dSph based on their positional 
and kinematical properties, and Pal\,12 is also known to be a likely 
candidate \citep{law10a}. A detailed analysis testing associations with 
the Sgr dSph is presented in Section~\ref{ss:Sgr}. The remaining cluster, 
NGC\,6101 (which is not associated with the Sgr dSph), has $\vtan \simeq 
350$~\kms which is comparable to those of Sgr clusters, but very high 
compared to other halo GCs in our sample. Based purely on the high $\vtan$ 
and $\vtot$ of this cluster compared to other halo GCs, we propose that 
NGC\,6101 has an origin outside the MW, like the Sgr GCs. Indeed, 
\citep{mar04} pointed out a possibility that NGC\,6101 may be associated 
with the CMa overdensity. 

In the middle and bottom panels of Figure~\ref{f:vel_rgc}, we have 
included $\vtan$ and $\vtot$ based on the older PM measurements 
in Table~\ref{t:oldpm} to demonstrate how the Galactocentric 
velocities change when different PM measurements are adopted. 
For all but Pal\,12, the older PMs implies significantly higher 
$\vtan$ than those for our \hst\ measurements. This is likely due to 
the nature of PM measurements in the sense that when systematic 
uncertainties dominate, there is a tendency of overestimating the 
measurement therefore yielding higher-than-real energy orbits.
A larger PM generally translates to a higher $\vtan$, and as a 
result, the $\vtot$ is overestimated (see lower panel of Figure~\ref{f:vel_rgc}).
For example, based on their PM measurements for NGC\,7006, \citet{din01} 
concluded that this cluster is on a highly energetic orbit with the 
apocentric distance reaching beyond $\rgc = 100$ kpc. However, our 
new PM results imply that NGC\,7006 is on an orbit with much lower energy. 
The most distant cluster in our sample, NGC\,2419, also seems to 
be on a less energetic orbit than implied by the PM measurement by \citet{mas17}. 

\subsection{Globular Clusters Associated with the Sagittarius dSph/Stream}
\label{ss:Sgr}

We now focus on GCs that have been claimed to be associated with the 
Sgr Stream, and illustrate how the PM results can aid in the 
identification. Comprehensive analyses of GCs possibly associated 
with the Sgr dSph were first performed by \citet{pal02} and 
\citet{bel03}. Later, \citet{law10a} carried out a nearly complete 
census of Sgr GCs by comparing the $N$-body model of the Sgr stream 
\citep{law10b} to observed parameters (positions, distances, $\vlos$, 
and PMs) of GCs residing in the MW halo. They concluded that 
Arp\,2, M54, NGC\,5634, Terzan\,8, and Whiting~1 are high-probable 
members of the Sgr dSph, and Berkeley~29, NGC\,5053, Pal\,12, 
and Terzan\,7 are also likely to be associated with Sgr.
Some of the clusters discussed in these papers are included in 
our sample, so we revisit their identification using our new PM 
measurements.

\begin{figure}[t]
\includegraphics[width=1\columnwidth]{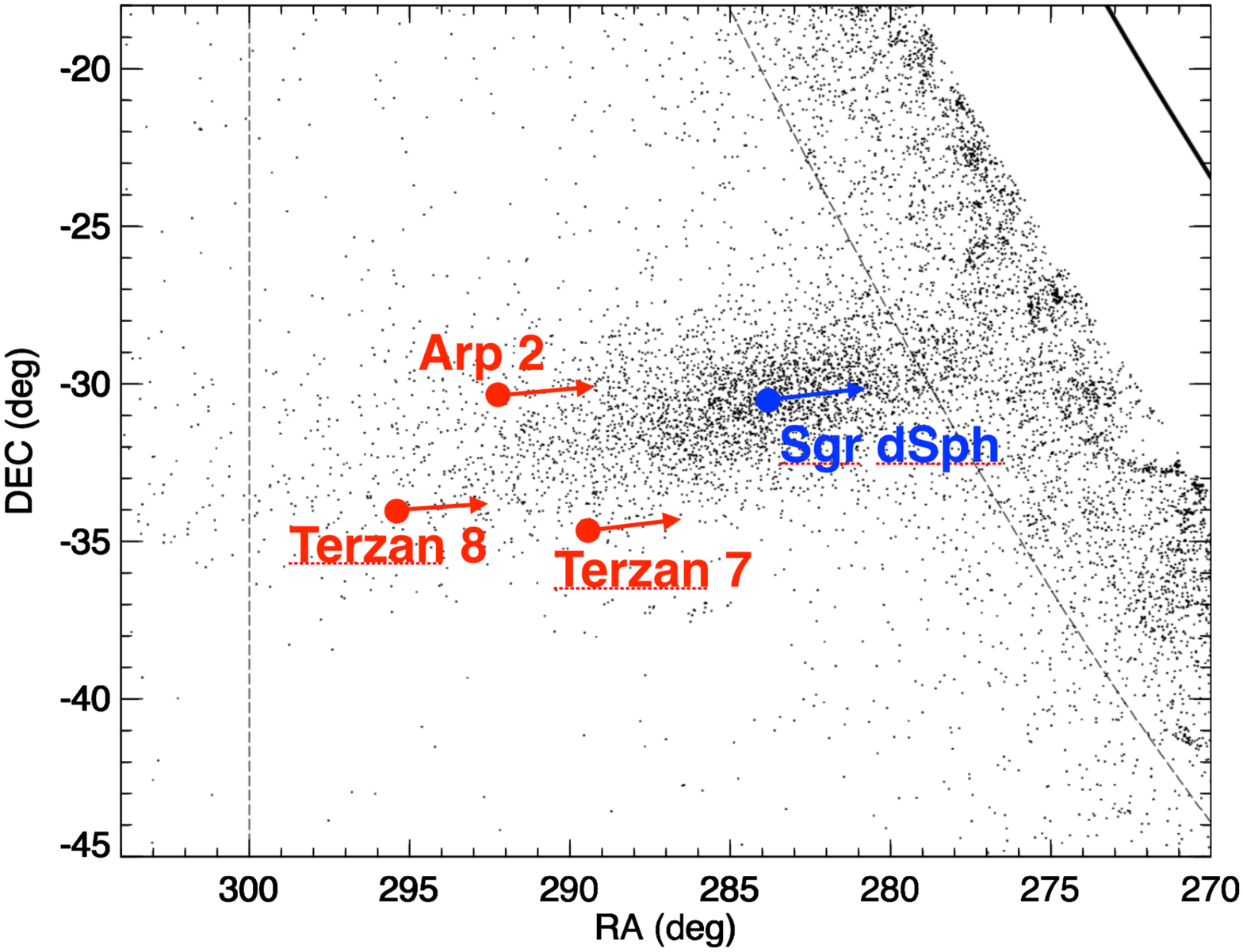}
\caption{Solar motion-corrected PMs of clusters associated with 
         the Sgr dSph (red arrows), and the Sgr dSph itself (blue arrow).
         For the blue arrow, we used the perspective-corrected center 
         of mass (COM) motion of $(\muw, \mun) = (2.82, -1.51) \masyr$ 
         by \citet{soh15}. For the solar motion correction of PMs, 
         we assumed the heliocentric distances $D_{\sun}$ listed in 
         Table~\ref{t:spacemotions}, and for the Sgr dSph 
         $D_{\sun} = 28$~kpc. Black dots are the Sgr dSph and stream 
         M-giant stars selected using the 2MASS catalog 
         by \citet{maj03}. The relative magnitude of each vector 
         represent the actual difference in space velocity of each 
         component. The dashed lines indicate the region used by 
         \citet{maj03} to fit King profiles to the M-dwarf star counts. 
         The solid line on the upper right marks the Galactic midplane.
         \label{f:Sgr_near}
        }
\end{figure}

Among the clusters we measured PMs for, Arp\,2, Terzan\,7, and 
Terzan\,8 are the three GCs that are located closest to the Sgr 
dSph on the sky. These clusters have been associated with the 
Sgr dSph since the discovery of the dwarf galaxy itself primarily 
based on their proximities to the dSph's core \citep{iba94}.
Our PMs allow us to compare the motions between these clusters 
and the Sgr dSph on the sky as illustrated in Figure~\ref{f:Sgr_near}.
The vector directions and lengths of the three GCs in 
Figure~\ref{f:Sgr_near} agree with those of the Sgr dSph within 
3\degr\ and $0.33 \masyr$ (44~\kms\ at the Sgr dSph distance of 
28 kpc), respectively. We therefore conclude that the 
solar motion-corrected PMs of Arp\,2, Terzan\,7, and Terzan\,8 
are consistent with that of the Sgr dSph, which confirms the association.

%
\begin{figure}[t]
\includegraphics[width=\columnwidth]{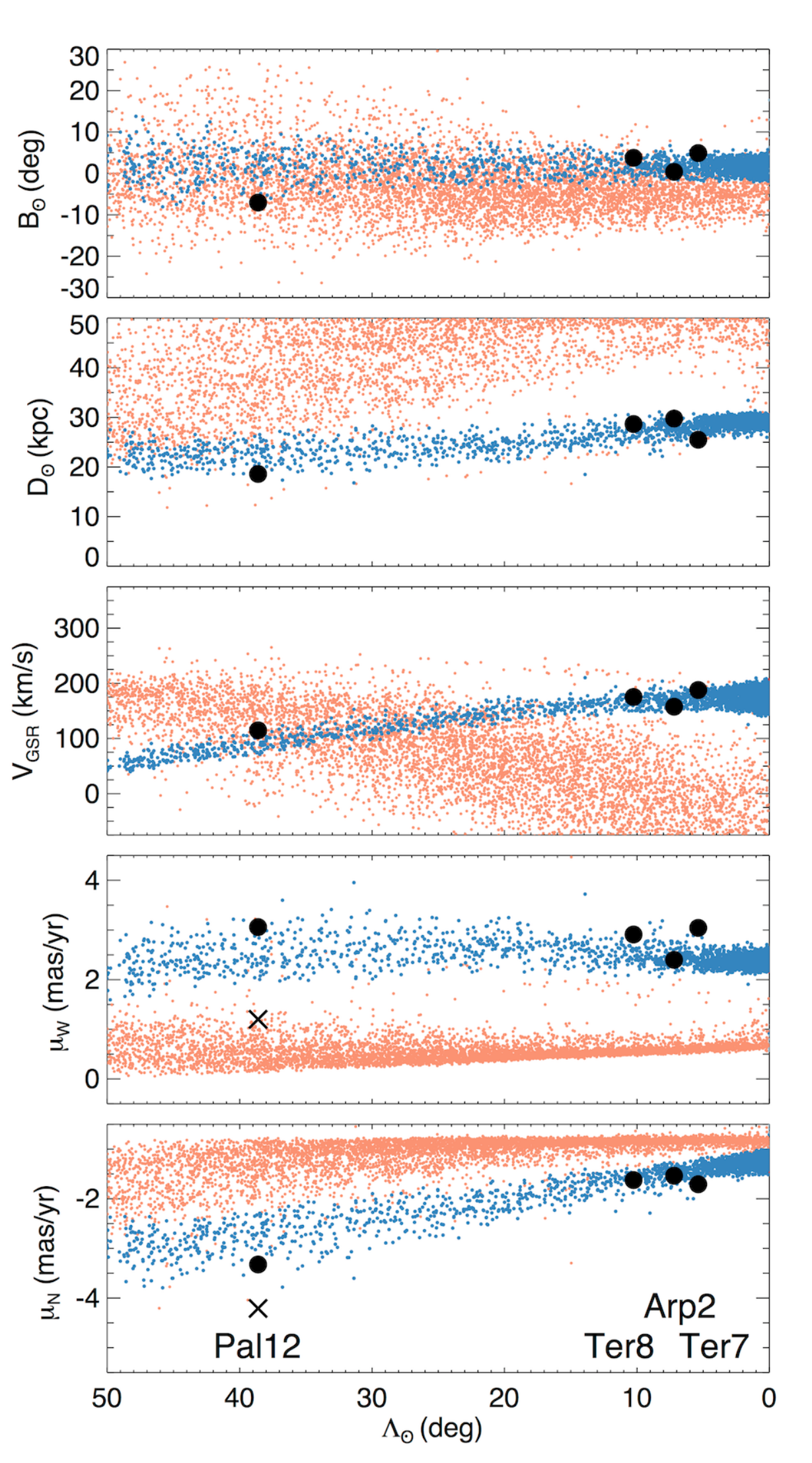}
\caption{Comparison of observed positions, distances, 
         radial velocities (as measured from the Galactic Standard 
         of Rest or GSR), and PMs (top to bottom) between 
         Pal\,12, Terzan\,8, Arp\,2, and Terzan\,7 (black circles) and 
         $N$-body particles of \citet{law10a} model (red and blue dots).
         Cluster identifications are labeled in the bottom panel.
         For Pal\,12, we also plot the old PM results of \citet{din00} 
         in \texttt{x} symbols. 
         $(\Lambda_{\odot}, B_{\odot})$ is the coordinate system as 
         defined by \citet{maj03} that runs along and across the Sgr 
         stream. Coloring scheme of the model particles in 
         this and Figure~\ref{f:id_nosgr} are the same as 
         in Figure~7 of \citet{soh15}: light red and dark blue 
         dots indicate particles on the secondary leading and primary 
         trailing arms, respectively.
         \label{f:id_sgr}
        }
\end{figure}
%

%
\begin{figure}[t]
\includegraphics[width=\columnwidth]{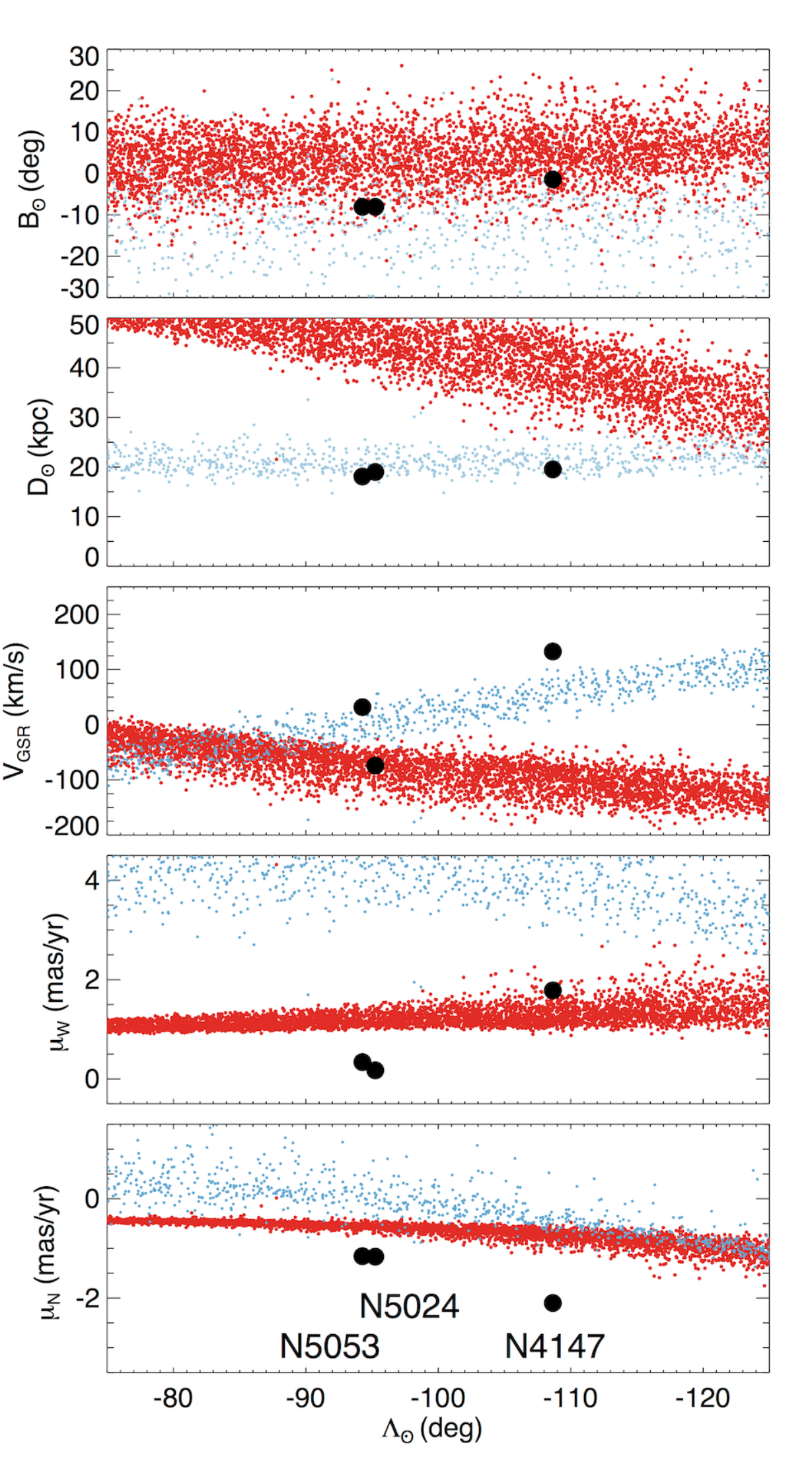}
\caption{Same as in Figure~\ref{f:id_sgr}, but for NGC\,5053, NGC\,5024, 
         and NGC\,4147. Dark red and light blue dots indicate 
         model particles on the primary leading and secondary trailing 
         arms, respectively.
         \label{f:id_nosgr}
        }
\end{figure}
%

Pal\,12 has also been suggested as a cluster associated with the 
Sgr dSph based on its observed parameters \citep[see][and references 
therein]{law10b}. In Section~\ref{ss:3dposvel}, we showed that Pal\,12 
belongs to the group with higher $\vtan$ among our sample, together 
with Arp\,2, Terzan\,7, and Terzan\,8, hinting that this cluster 
is associated with the Sgr dSph. In Figure~\ref{f:id_sgr}, we provide 
a more detailed analysis by comparing our results to the $N$-body 
models of \citet{law10b}. We first note that all observed parameters of 
Arp\,2, Terzan\,7, and Terzan\,8 indicate that these three clusters 
follow the motion on the sky of the Sgr trailing arm 
(as seen in Figure~\ref{f:Sgr_near}).
For Pal\,12, the position, distance, and radial velocity (top three panels) 
do not make it clear whether this cluster follows the trailing- (blue) or 
leading-arm (red) model particles due to the overlap of phase-space 
information for the two tidal arms in these coordinates. 
The $N$-body model separates the two tidal arms better in the
PM vs. $\Lambda_{\odot}$ phase-space (bottom two panels). Nevertheless, 
our new PMs make clear that Pal\,12 follows the primary Sgr trailing arm 
debris, at least for this model of the stream. We conclude that Pal\,12 
is associated with the Sgr dSph, and probably has been brought in 
together with other Sgr clusters. Our results put to rest contradictory 
results based on old PM measurements \citep[][\texttt{x} symbols]{din00} 
regarding this issue.

In addition to the four Sgr clusters discussed above, NGC\,4147, 
NGC\,5024, and NGC\,5053 have also been discussed as GCs possibly 
associated with the Sgr dSph \citep{law10b}. We repeat the analysis 
above for these three clusters and show the results in Figure~\ref{f:id_nosgr}.
The position, distance, and radial velocities (top three panels) all 
indicate a possible connection to the Sgr trailing arm. However, our 
PM results {\it completely rule out} associations with the Sgr dSph.
We note that the recent study by \citet{tan18} also rules out 
NGC\,5053 Sgr association based on its chemical composition and 
orbital characteristics.

Finally, NGC\,2419 has also been suggested as a cluster associated with 
the Sgr stream \citep{new03,bel14}. However, this cluster is located in a 
region where the \citet{law10b} model is a poor fit to the observed Sgr 
stream distance, so we cannot use this $N$-body model as a reference here. 
Instead, we use the observed characteristics of the stream to check whether 
NGC\,2419 is associated. \citet{bel14} already showed that NGC\,2419 
is consistent with the sky position, distance, and $\vlos$ of the trailing 
arm. The RR Lyrae sample of \citet{ses17} and \citet{her17} show that at 
this longitude, the trailing arm splits into two components. The denser 
inner component is centered on $91$~kpc with a width of $7$~kpc, while the 
lower-density outer component spans roughly 105--120 kpc \citep{her17}. 
The distance $87.5\pm3.3$~kpc of \citet{cri11} places NGC\,2419 in the inner 
component. At a distance this large, the density of the general GC population 
is quite low and a chance coincidence in all these variables is extremely 
unlikely. Our PM results offer a chance to revisit this assessment.

To explore this in detail, we
convert the Galactocentric space motion of NGC\,2419 into that of the
Sgr coordinate system defined by \citet{maj03}. Expressed in spherical
coordinates, the radial velocity is $\vrad = -28 \pm 1.4$~\kms, the azimuthal velocity
(along the longitude direction $\Lambda$) is $v_\mathrm{azi} = -31 \pm 13$~\kms, and the polar
velocity (in the latitude direction $B$) is $v_\mathrm{polar}-43 \pm 16$~\kms.
Here the negative sign on the azimuthal
velocity indicates agreement with the orbital
direction of the Sgr galaxy and its stream.
For comparison, the angular momentum
of Sgr itself around the pole of its coordinate system is approximately $L_{z} = -5400$~kpc\,\kms,
derived based on the COM PM of the Sgr dSph of \citet{soh15}. We expect
the angular momentum of stream material to be roughly comparable (though not
quite equal) to that of Sgr, so the expected tangential velocity at the
cluster's Galactocentric radius $r \approx 95$~kpc is
roughly $v_\mathrm{tan,s} \approx -57$~\kms.
The observed velocity dispersion in the stream is about
$\sigma_\mathrm{intr} = 13$~\kms\ \citep{gib17},
which we take as typical
even though the dispersion for a globular cluster in the stream at NGC\,2419's location could be
somewhat different. 
Combining the observational error and intrinsic velocity dispersion, the longitudinal motion
appears quite consistent with membership in the stream.
The motion of NGC\,2419 in the polar
direction (transverse to the stream path on the sky) of $-43$~\kms\ is however
somewhat surprising in that it is not centered on zero as we might expect.
The vast majority of orbits allowed by our measurement put the cluster moving
toward the northern Sgr coordinate pole, while a mere 0.4\% of the orbits
have the opposite sign. Another way to quantify this is the distribution of
the orbital poles for this cluster. The angular offset of the pole from that
of the Sgr galaxy is $54 \pm 16\degr$. Given the $\sim10\degr$ offset of
NGC\,2419 from the plane, it is not surprising that there is some offset,
but it seems surprisingly large. We intend to investigate this issue further
in a future work.
 
In any case, from a broader perspective, we see that both
PM dimensions show agreement with the expected velocities of the Sgr stream
material within a few tens of \kms. The level of agreement expected from a
random coincidence is instead of the order $\sigma_h \simeq 120$~\kms, the velocity
dispersion of the diffuse halo component.
We can compute a formal Bayes factor $P_s(D) / P_h(D)$ representing the relative
probability of the observed data with models where NGC\,2419 comes from
stream and smooth halo components respectively.
We fold the observed and intrinsic dispersions together into a total dispersion
$\sigma' = (\sigma^2 + \sigma_\mathrm{intr})^{1/2}$, and assume normal distributions
to get
\begin{equation}
P_s(D) = \mathcal{N}(v_\mathrm{azi}-v_\mathrm{azi,s},\sigma_\mathrm{azi}'^2)
         \mathcal{N}(v_\mathrm{polar},\sigma_\mathrm{polar}'^2) \; ,
\end{equation}
\begin{equation}
P_h(D) = \mathcal{N}(v_\mathrm{azi},\sigma_h'^2)
         \mathcal{N}(v_\mathrm{polar},\sigma_h'^2)
\end{equation}
For the values given above we find $P_s(D) / P_h(D) \approx 2$.
The somewhat surprising value of the polar velocity
is outweighed by the fact we found values
somewhat close to expectations for the stream, when
for a halo population they did not have to be.
The prior odds based on the cluster's sky position, distance, and radial 
velocity already favored an association with the stream.
Our PM results appear to confirm this association.

\subsection{Orbital Parameters}
\label{ss:orbparams}

To better understand the origin of our target GCs, we carried out orbital
integrations based on our PM measurements. This was done using the code
\textsc{galpy} \citep{bov15}. For the MW potential, we created our 
own model that is consistent with our MW mass estimated in Section~\ref{ss:mass}. 
We rescaled the NFW halo of \textsc{mwpotential2014} (included as the default 
potential in \textsc{galpy}) such that the enclosed mass at \replaced{$\rgc = 250$~kpc}{$R = 250$~kpc} is 
$1.5\times10^{12}\,\Msun$. Since our target GCs spend most of their time in 
the halo, scaling the bulge and disk was not necessary. 
Adopting a critical overdensity of 100 and the \citet{pla16} cosmological 
parameters, the corresponding virial mass of our MW model is 
$\Mvir = 1.7\times 10^{12}\,\Msun$, which is consistent with what we derive in 
Section~\ref{ss:mass}. For the position and velocity of the Sun, we used the 
same values as in Section~\ref{ss:3dposvel}. For better sampling the orbital 
parameters, we integrated orbits of each cluster backward in time for a duration 
of 7.5 times its orbital period using the observed positions and velocities as 
initial conditions. We measured a set of orbital parameters based on this 
integration span. To calculate uncertainties of each orbital parameter, the 
process above was repeated using 10,000 Monte Carlo drawings for each GC 
that samples the observational uncertainties. We note that our uncertainties 
do not account for the systematics arising from different choice of potentials.
Table~\ref{t:orbparams} lists the orbital parameters and associated 1$\sigma$ 
uncertainties as calculated above for each GC in the following order: total 
energy, angular momentum in the $Z$-axis direction, magnitude of the angular 
momentum, pericenter, apocenter, and orbital period. 

%
%
\begin{deluxetable*}{lcccccc}
\renewcommand{\arraystretch}{0.9}
\tabletypesize{\small}
\tablecaption{Orbital parameters
              \label{t:orbparams}
             } 
\tablehead{
   \colhead{}        & \colhead{$E_\mathrm{tot}$}         & \colhead{$L_{Z}$}           & \colhead{$|L|$}             & \colhead{$\rperi$} & \colhead{$\rapo$} & \colhead{$P$} \\
   \colhead{Cluster} & \colhead{($10^4$ km$^2$ s$^{-2}$)} & \colhead{($10^3$ kpc~\kms)} & \colhead{($10^3$ kpc~\kms)} & \colhead{(kpc)}    & \colhead{(kpc)}   & \colhead{(Gyr)}  
}
\startdata
Arp\,2    & \phn$-9.1\pm0.4$ & \phs$1.7 \pm 0.2$ & $6.6 \pm 0.3$ &    $19.2 \pm 0.9$ & $49.3 \pm$\phn$5.3$ & $0.69 \pm 0.07$ \\
IC\,4499  & $   -12.5\pm0.2$ &    $-1.2 \pm 0.1$ & $2.7 \pm 0.1$ & \phn$6.3 \pm 0.4$ & $27.6 \pm$\phn$1.2$ & $0.33 \pm 0.02$ \\
NGC\,1261 & $   -14.5\pm0.1$ &    $-0.4 \pm 0.1$ & $0.5 \pm 0.1$ & \phn$1.1 \pm 0.3$ & $20.9 \pm$\phn$0.5$ & $0.22 \pm 0.01$ \\
NGC\,2298 & $   -15.5\pm0.1$ &    $-0.5 \pm 0.1$ & $1.0 \pm 0.1$ & \phn$1.3 \pm 0.3$ & $17.1 \pm$\phn$0.4$ & $0.18 \pm 0.00$ \\
NGC\,2419 & \phn$-6.8\pm0.1$ & \phs$4.2 \pm 1.4$ & $5.1 \pm 1.2$ &    $10.5 \pm 3.9$ & $96.1 \pm$\phn$3.3$ & $1.20 \pm 0.05$ \\
NGC\,4147 & $   -13.3\pm0.1$ &    $-0.1 \pm 0.1$ & $0.3 \pm 0.1$ & \phn$0.4 \pm 0.3$ & $25.9 \pm$\phn$0.5$ & $0.28 \pm 0.01$ \\
NGC\,5024 & $   -13.2\pm0.2$ & \phs$0.8 \pm 0.1$ & $3.1 \pm 0.2$ & \phn$9.1 \pm 0.7$ & $22.3 \pm$\phn$0.5$ & $0.30 \pm 0.01$ \\
NGC\,5053 & $   -13.8\pm0.2$ & \phs$0.7 \pm 0.1$ & $3.0 \pm 0.1$ & \phn$9.9 \pm 0.8$ & $18.5 \pm$\phn$0.4$ & $0.27 \pm 0.01$ \\
NGC\,5466 & $   -11.5\pm0.3$ &    $-0.8 \pm 0.1$ & $2.4 \pm 0.2$ & \phn$5.3 \pm 0.6$ & $34.6 \pm$\phn$1.9$ & $0.40 \pm 0.03$ \\
NGC\,6101 & $   -12.0\pm0.3$ &    $-2.9 \pm 0.1$ & $3.7 \pm 0.1$ &    $10.3 \pm 0.2$ & $28.2 \pm$\phn$1.5$ & $0.37 \pm 0.02$ \\
NGC\,6426 & $   -15.2\pm0.3$ & \phs$1.4 \pm 0.2$ & $1.6 \pm 0.2$ & \phn$3.8 \pm 0.6$ & $17.3 \pm$\phn$0.8$ & $0.20 \pm 0.01$ \\
NGC\,6934 & $   -11.2\pm0.4$ & \phs$1.5 \pm 0.2$ & $1.6 \pm 0.2$ & \phn$3.1 \pm 0.4$ & $38.0 \pm$\phn$3.0$ & $0.42 \pm 0.04$ \\
NGC\,7006 & \phn$-9.9\pm0.2$ &    $-0.5 \pm 0.5$ & $0.5 \pm 0.4$ & \phn$1.0 \pm 0.8$ & $50.3 \pm$\phn$1.7$ & $0.55 \pm 0.02$ \\
Pal\,12   & $   -10.6\pm0.4$ & \phs$2.0 \pm 0.1$ & $5.1 \pm 0.2$ &    $15.2 \pm 0.4$ & $36.0 \pm$\phn$3.4$ & $0.49 \pm 0.04$ \\
Pal\,13   & $   -10.1\pm0.2$ &    $-0.8 \pm 0.2$ & $2.8 \pm 0.2$ & \phn$5.6 \pm 0.5$ & $46.1 \pm$\phn$2.2$ & $0.53 \pm 0.03$ \\
Pal\,15   & \phn$-9.6\pm0.2$ &    $-1.3 \pm 0.8$ & $3.1 \pm 0.7$ & \phn$6.3 \pm 2.1$ & $50.9 \pm$\phn$1.6$ & $0.59 \pm 0.03$ \\
Pyxis     & \phn$-7.3\pm0.4$ &    $-2.0 \pm 0.7$ & $6.1 \pm 0.6$ &    $13.8 \pm 2.0$ & $83.7 \pm$\phn$7.4$ & $1.06 \pm 0.11$ \\
Rup\,106  & $   -11.6\pm0.2$ & \phs$1.5 \pm 0.1$ & $2.4 \pm 0.1$ & \phn$4.7 \pm 0.4$ & $34.7 \pm$\phn$1.4$ & $0.39 \pm 0.02$ \\
Terzan\,7 & \phn$-9.1\pm0.5$ & \phs$1.2 \pm 0.2$ & $5.9 \pm 0.3$ &    $15.5 \pm 0.7$ & $51.8 \pm$\phn$5.9$ & $0.68 \pm 0.08$ \\
Terzan\,8 & \phn$-7.0\pm0.6$ & \phs$1.7 \pm 0.3$ & $7.9 \pm 0.4$ &    $19.9 \pm 0.7$ & $87.3 \pm$   $14.7$ & $1.17 \pm 0.21$ \\
\enddata
\end{deluxetable*} 
%
%

%
\begin{figure*}
\centering
\includegraphics[width=0.98\textwidth]{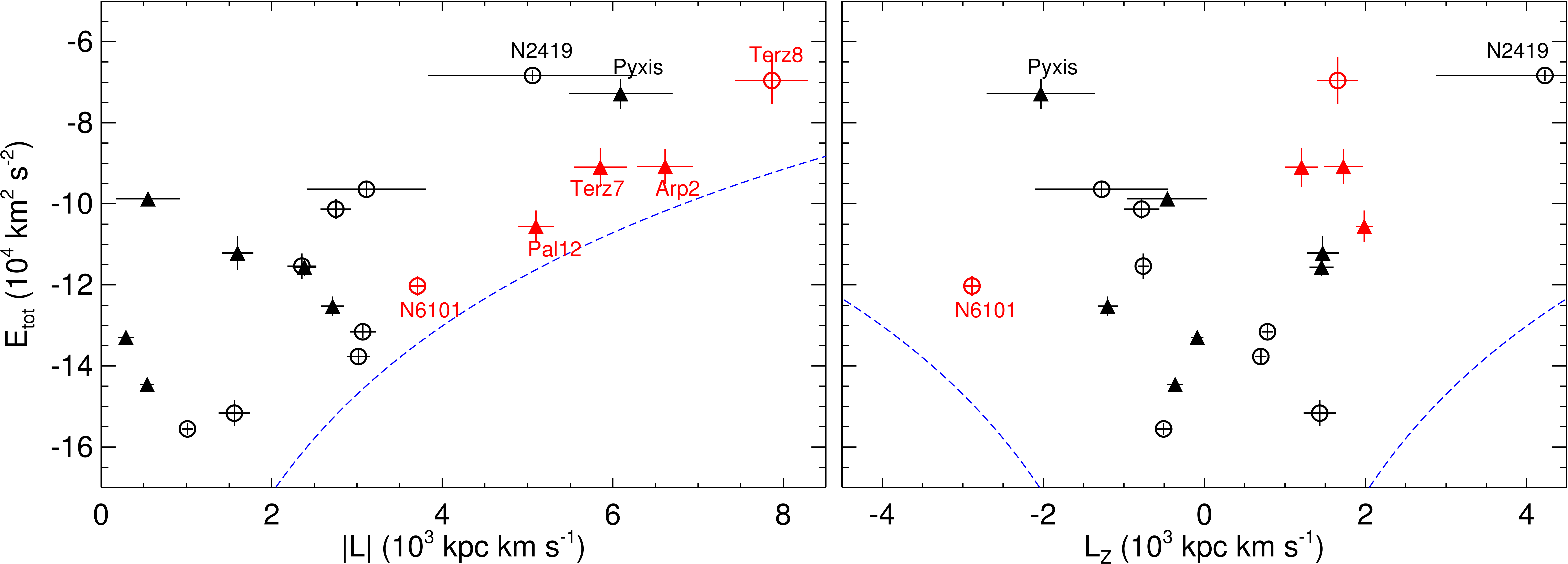}
\caption{Total orbital energy as a function of the magnitude angular 
         momentum $|L|$ (left panel), and angular momentum in the 
         $Z$-axis direction $L_{\mathrm Z}$ (right panel). As in 
         Figure~\ref{f:vel_rgc}, young and old GCs are plotted 
         in closed triangles and open circles, respectively.
         Also, GCs with $\vtan > 290$~\kms are plotted in red, 
         while all other GCs are plotted in black.
         The blue dashed lines indicate the maximum angular momentum 
         allowed by the potential for a given orbital energy.
         Uncertainties of all parameters plotted here and in 
         Figure~\ref{f:apo_peri} were propagated by sampling the 
         observed uncertainties in a Monte Carlo fashion, and 
         do not include the uncertainty in the MW mass model.
         \label{f:etot_vs_l}
        }
\end{figure*}
%

Figure~\ref{f:etot_vs_l} shows the total orbital energy as a function of 
angular momentum. Although the sample is small, the young GCs in the left 
panel appear to fall into two groups: one group lies at higher energy and 
angular momentum; and another group lies at low angular momentum, even 
lower than the older GCs at similar orbital energy range. The separation 
of these two groups likely indicates that different formation mechanisms exist 
among the young GCs, e.g., accreted (higher angular momentum) versus 
within the MW (lower angular momentum). Pyxis is part of the high energy, 
high angular momentum group, which suggests an external origin 
\citep[see a detailed discussion in][]{fri17}. From the right panel of 
Figure~\ref{f:etot_vs_l}, we can infer that NGC\,6101 has a strong 
retrograde motion with respect to the rotational direction of the Galactic 
disk. This strengthens our claim in Section~\ref{ss:3dposvel} that this 
cluster has an external origin. 

\begin{figure}
\includegraphics[width=0.98\columnwidth]{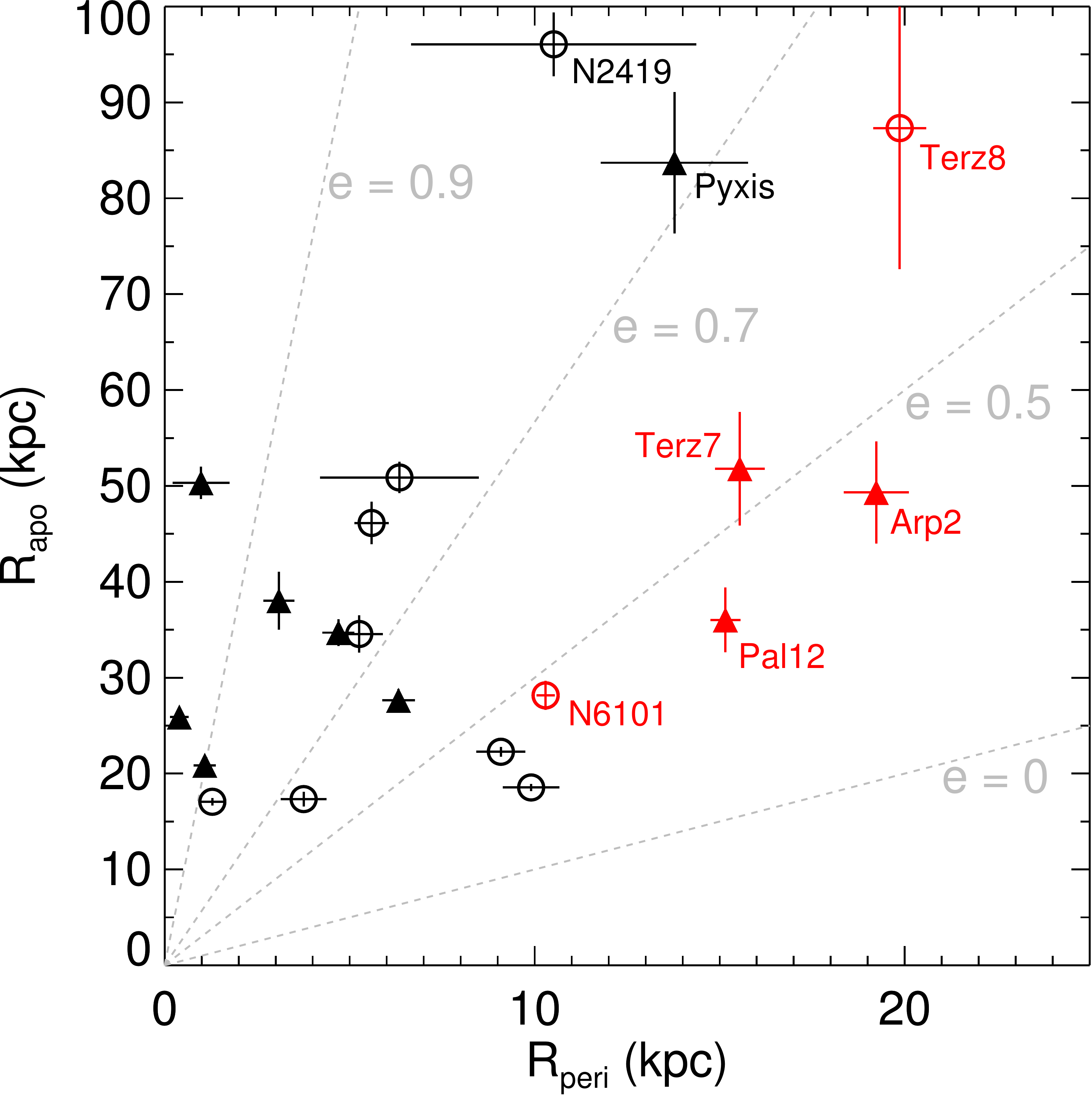}
\caption{Apocentric versus pericentric distances of our target GCs.
         Orbital eccentricities 
         of 0, 0.5, 0.7, and 0.9 are plotted as gray dashed lines.
         For each GC, we used the same symbols and colors as  
         in Figure~\ref{f:vel_rgc}.
         \label{f:apo_peri}
        }
\end{figure}

In Figure~\ref{f:apo_peri}, we plot the apocenter versus pericenter 
of our target GCs. Overall, we find that GCs with $\rperi \lesssim 10$~kpc 
have orbits only going out to $\rapo \simeq 50$~kpc, while those with 
$\rperi \gtrsim 10$~kpc have apocenter in the range $30 < \rapo < 100$~kpc. 
Related to the two groups of young GCs in the left panel of 
Figure~\ref{f:etot_vs_l}, the GCs in the low energy, low angular momentum 
have smaller pericenters than the older GCs at similar apocenter range. 
Among the Sgr GCs, Arp\,2, Terzan\,7, and Pal\,12 (three red triangles) 
have similar orbital eccentricities of $e \sim 0.5$, whereas the orbit of 
Terzan\,8 is significantly more radial at $e = 0.63\pm0.06$ with the 
apocenter extending out to $\rapo \sim 90$~kpc. Interestingly, Terzan\,8 
has been established as being older and more metal-poor than the other 
Sgr GCs \citep{mon98,mar09}, which suggests that this cluster may 
have a distinct formation history.  
The 3D geometry of the Sgr stream recently presented by \citet{bel14} using 
blue-HB stars, and later confirmed by \citet{ses17} and \citet{her17} using 
RR Lyrae stars, reveals that the trailing arm extends out to an extreme 
distance of $\rgc \sim 100$~kpc. 
The apocenter of Terzan\,8 coinciding with the distance of this extension 
suggests that this cluster will eventually reside among the extreme trailing 
debris.

We note that Pal\,12 has the lowest apocenter of the Sgr-associated
clusters at only $\rapo = 36 \pm 3$~kpc, despite our earlier conclusion 
based on comparison to the \citet{law10b} stream model that it belongs 
to the trailing stream, which is composed of objects with high energies 
and apocenters. This puzzle may be resulting from the mismatch between 
the potential we used for the orbital integrations and that employed for 
the Law \& Majewski model of the stream. As Pal\,12 is very near pericenter, 
its calculated apocenter is maximally sensitive to the potential. 
Alternatively, Pal\,12 could be a member of the leading stream if the 
actual stream deviates from the Law \& Majewski model in this region, 
which is wrapped $320\degr$ around the sky from Sgr. PMs along the Sgr 
stream derived from future {\it Gaia} data may provide a more robust way 
of assessing membership in the leading and trailing arms of the stream. 
All in all, the orbital properties of the Sgr GCs will help constrain 
the models of Sgr disruption.

%
%

\section{Mass of the Milky Way using Globular Clusters as Dynamical Tracers}
\label{s:mwmass}

\subsection{Method}
\label{ss:method}

Traditionally, the spherical Jeans equation \citep{bin08} has been applied 
to dynamical tracers to estimate the total MW mass enclosed within a certain 
radius. \citet{wat10} introduced a variety of mass estimators based on this 
Jeans equation that work with different types of distance and velocity data 
depending on which observed parameters are available. Since we have the full 
6D phase-space information of our cluster sample, we adopt the estimator 
that uses intrinsic distances and total velocities. This is written as 
\begin{equation}
\label{e:tme}
    \Mrmax = \frac{1}{G} \frac{\alpha+\gamma-2\beta}{3-2\beta} \rmax^{1-\alpha}
        \left\langle v^2 r^\alpha \right\rangle .
\end{equation}
where $\rmax$ is the Galactocentric distance of the most distant tracer object, 
$\alpha$ is the power-law index for the underlying potential ($\Phi \propto r^{-\alpha}$), 
$\gamma$ is the power-law index for the radial number density profile ($ \rho \propto r^{-\gamma}$),
and $\beta$ is the velocity anisotropy parameter of the tracer objects. 
In this section, we calculate these parameters and provide our best estimate 
of the MW mass using Equation~\ref{e:tme}.

\subsection{Anisotropy Parameter}
\label{ss:beta}

The anisotropy parameter is defined as 
\begin{equation}
   \beta \equiv 1 - \frac{\sigma_{\theta}^2 + \sigma_{\phi}^2}{2 \sigma_{r}^2},
\end{equation}
where $\sigma_{r}$, $\sigma_{\theta}$, and $\sigma_{\phi}$ denote the velocity 
dispersions in the radial, polar, and azimuthal directions in spherical 
coordinates. In general, $\beta$ has been the limiting factor for mass 
estimations due to the lack of tangential velocities, but our PMs of GCs combined 
with existing $\vlos$ allows us to {\it directly} calculate $\beta$. To do this, 
we first consider all GCs in our sample except for NGC\,2419. The majority of our 
cluster sample lie between 10 and 40~kpc, however, NGC\,2419 is considerably more 
distant at $\sim$95~kpc. To first order, $M \propto v^2 r$, so points at large 
$r$ have the most weight in the mass estimate; this makes it particularly 
important that the RMS velocity is accurately estimated at large distance, and 
this cannot be guaranteed when there is only a single tracer. Were we to include 
NGC\,2419 in the following analysis, there is a risk that our mass estimate would 
be significantly biased by the presence of this single cluster at a distance far 
greater than the others. As such, we remove this cluster from our sample in 
subsequent analyses.

%
%
\begin{deluxetable*}{ccccccccc}
\renewcommand{\arraystretch}{1.1}
\tabletypesize{\small}
\tablecaption{Velocity mean, dispersions, and anisotropy parameters
              \label{t:beta}
             } 
\tablehead{
   \colhead{} & \colhead{}                    & \colhead{$\overline{v_{r}}$} & \colhead{$\overline{v_{\theta}}$} & \colhead{$\overline{v_{\phi}}$} & \colhead{$\sigma_{r}$} & \colhead{$\sigma_{\theta}$} & \colhead{$\sigma_{\phi}$} & \colhead{} \\
   \colhead{} & \colhead{$N_\mathrm{sample}$} & \colhead{(\kms)}             & \colhead{(\kms)}                  & \colhead{(\kms)}                & \colhead{(\kms)}       & \colhead{(\kms)}            & \colhead{(\kms)}          & \colhead{$\beta$} 
}
\startdata
Sample1\tablenotemark{a} & 19 & $-20.2\pm39.3$ & $-65.9\pm36.0$ &       $-21.5\pm29.1$ & $171.1\pm27.3$ & $156.7\pm25.5$ & $126.2\pm20.2$ & $0.309\substack{+0.213 \\ -0.377}$ \\
Sample2\tablenotemark{b} & 16 & $-37.9\pm45.3$ & $-26.2\pm28.6$ & \phs\phn$3.3\pm28.2$ & $181.1\pm32.0$ & $114.0\pm20.0$ & $112.3\pm20.1$ & $0.609\substack{+0.130 \\ -0.229}$ \\
\enddata
\tablenotetext{a}{All GCs except for NGC\,2419}
\tablenotetext{b}{Sample1 minus Terzan\,7, Terzan\,8, and Pal\,12}
\end{deluxetable*}
%

%
\begin{figure*}
\centering
\includegraphics[width=0.97\textwidth]{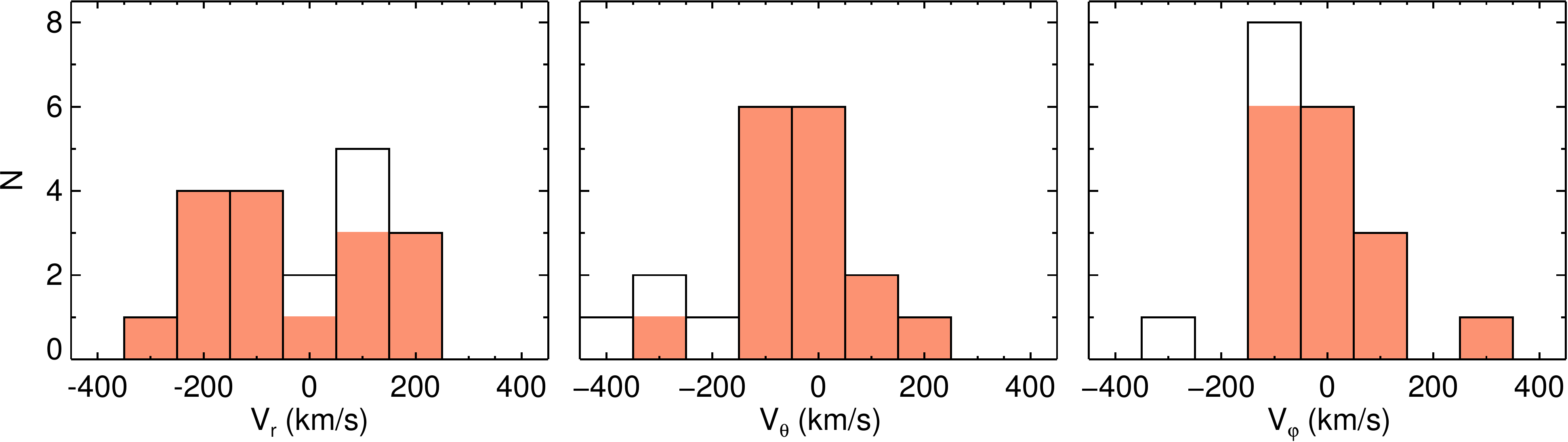}
\caption{Histogram of spherical velocity components for Sample1 
         (black outline) and Sample2 (red blocks) as defined in 
         Table~\ref{t:beta}.
         \label{f:hist_vrtp}
        }
\end{figure*}
%

We converted the space velocities in the Cartesian coordinate system 
$(\vx,\,\vy,\,\vz)$ of our GCs as listed in Table~\ref{t:spacemotions} 
to velocities in the spherical coordinate system 
$(v_{r},\,v_{\theta},\,v_{\phi})$. We then calculated the mean 
velocities $(\overline{v_{r}},\,\overline{v_{\theta}},\,\overline{v_{\phi}})$, 
velocity dispersions $(\sigma_{r},\,\sigma_{\theta},\,\sigma_{\phi})$, 
and the resulting velocity anisotropy parameter $\beta$. The uncertainties 
were calculated using the maximum likelihood estimation method assuming 
the uncertainty distribution in each coordinate is Gaussian. Results for the 
19 GCs (excluding NGC\,2419) are shown in Table~\ref{t:beta} (Sample1), 
and the distribution of each velocity component is plotted in 
Figure~\ref{f:hist_vrtp}.

For a pressure-supported system such as the halo GCs, one expects the 
mean velocity of each component to be zero. This is the case for the 
mean radial ($\overline{v_{r}}$) and azimuthal velocities 
($\overline{v_{\phi}}$), but we find that the mean polar velocity is 
non-zero at $\sim 2\sigma$ level ($\overline{v_{\theta}} = -65.9\pm 35.9$~\kms). 
Figure~\ref{f:hist_vrtp} indeed shows an excess 
in negative $v_{\theta}$. As discussed in Section~\ref{ss:Sgr}, our 
sample contains four Sgr clusters. The Sgr dSph is known to have a 
nearly polar orbit about the MW disk, and so the non-zero mean polar  
velocity (i.e., motion in vertical direction w.r.t. the disk) for 
Sample1 is likely due to the presence of Sgr GCs. This indicates that 
Sample1 does not correctly represent the halo GC population, and using 
this sample will significantly bias our mass estimates. There are at 
least 5 {\it confirmed} Sgr GCs \citep[Arp\,2, Terzan\,7, Terzan\,8, 
Pal\,12 analyzed in this study, and M54 which resides near the core of the 
Sgr dSph;][]{bel08}, so one out of every $\sim 20$ halo GCs is associated 
with the Sgr dSph. As such, to correctly represent the halo GC population, 
we decided to exclude three of the four Sgr GCs from our sample, keeping 
only Arp\,2.\footnote{We repeated all subsequent calculations keeping 
each Terzan\,7, Terzan\,8, or Pal\,12 at a time, instead of Arp\,2, 
and for all cases, results are fully consistent within uncertainties. 
This indicates that our choice of keeping a specific Sgr GC has negligible 
effect on the final results.} We denote this as Sample2 and repeat the 
calculations above. Results are shown in the second row of 
Table~\ref{t:beta}, and we now find that the mean velocities of all three 
components are consistent with zero. Our final anisotropy parameter 
adopted for the purpose of MW mass calculations below is 
$\beta = 0.609\substack{+0.130 \\ -0.229}$.

%
\begin{figure}
\includegraphics[width=\columnwidth]{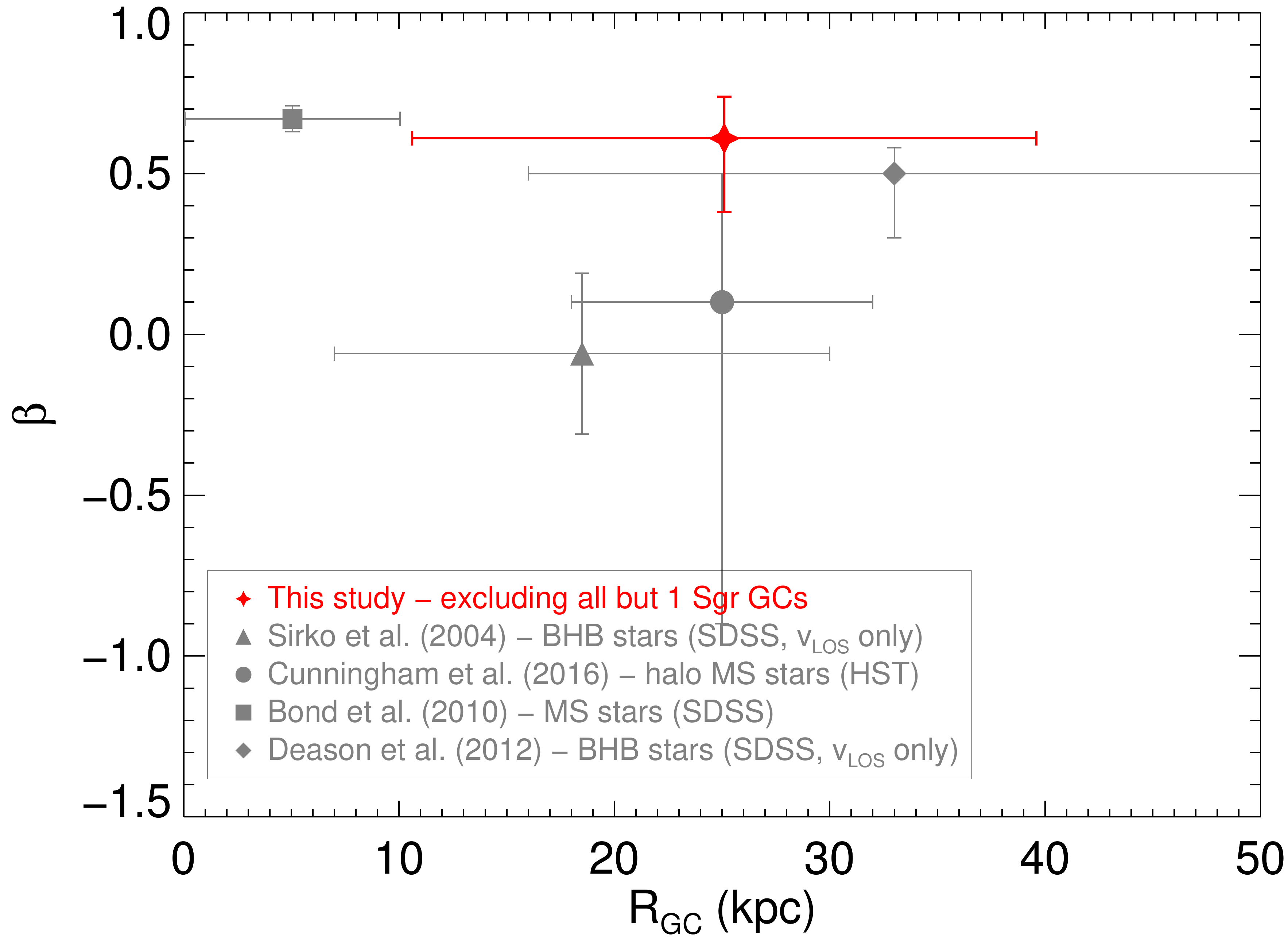}
\caption{Velocity anisotropy profile of the MW halo.
         \label{f:beta}
        }
\end{figure}
%

Figure~\ref{f:beta} shows our anisotropy parameter for GCs compared to 
those calculated from other studies using halo stars in the range 
$0 < \rgc < 50$~kpc. Our $\beta$ implies more radially-biased orbital 
motions than the results of \citet{sir04} and \citet{cun16} in a 
similar $\rgc$ range. \citet{sir04} used a large sample of SDSS halo 
stars spread across a significant portion of the sky, but only used 
$\vlos$ to constrain $\beta$. However, as shown by \citet{hat17}, 
measurements of $\beta$ from $\vlos$ alone can be unreliable beyond 
$\rgc \sim 15$ kpc. On the other hand, while \cite{cun16} used PMs 
of halo stars measured via \hst, their measurement is confined to a 
single line-of-sight, and is likely the result of including stars in 
a large shell-type structure such as the ``TriAnd'' overdensity, and 
therefore may not represent the relaxed halo system. If our $\beta$ 
from GCs represents the underlying halo system in the distance range 
$\rgc = 10-40$, the ``dip'' observed by \citet{cun16} can be 
interpreted as being a localized effect rather than a global 
decrease in $\beta$.

Our $\beta$ comfortably lies between those measured using stars in 
solar neighborhood \citep{bon10}, and distant halo stars \citep{dea12} 
(although the latter also potentially suffers from using only $\vlos$). 
Furthermore, the $\beta$ we measure for the halo GCs is in good 
agreement with simulations of stellar haloes, which generally predict 
radially anisotropic orbits at large distances \citep{die07,ras13,loe18}.

\subsection{Density and Potential}
\label{ss:gamma}

For the density slope, we use $\gamma = 3.53 \pm 0.01$, which we 
find by fitting to the halo clusters in the H10 catalog, and is in 
good agreement with previous values quoted in the literature 
\citep[e.g.][]{har01}. This we henceforth consider to be fixed.
For the mass estimates, we will use Sample2, that is the sample with 
NGC\,2419, Terzan\,7, Terzan\,8 and Pal\,12 excluded. These clusters 
span $10.6 \le r \le 39.5$~kpc, so we need to estimate the slope of 
underlying potential over this radial range. To do this, we consider 
a set of MW models for which the nucleus, bulge, and disk components 
are fixed, but the properties of the halo are allowed to vary -- 
we assume the halo is NFW \citep{nav96} in shape but vary both the 
mass and scale radius. The halos are further chosen to be consistent 
with both existing literature on theoretical MW potential profiles 
and the observed solar velocity. The resulting distribution of $\alpha$  
spans approximately $0.21 \lesssim \alpha \lesssim 0.52$. Both the 
density- and potential-fitting methods are described in detail 
in \citet{wat18} and are similar to the methods used in \citet{ann18}.

\subsection{Monte Carlo Simulations}
\label{ss:mcsims}

We also need to consider the shot noise in our calculations, that is 
how well the mass estimated using our sample of 16 clusters describes 
the underlying mass distribution from which they were drawn. 
In \citet{wat10}, we describe a suite of Monte Carlo simulations that 
were created for this purpose. We run 1000 simulations of 16 clusters, 
and calculate the mass in each case. Reassuringly, we find that the the 
fraction $f = M_\mathrm{est}/M_\mathrm{true}$ of the estimated mass to 
the true mass is $f = 1.00 \pm 0.22$, so our estimators do indeed recover 
the true mass on average. Further, we can use these simulations to 
provide accurate error bars on our mass estimates that correctly 
account for shot noise.

\subsection{Milky Way Mass}
\label{ss:mass}

Now that we have the ingredients in place, we proceed with estimating 
the mass of the MW within $\rmax = 39.5$~kpc. For each of the $\alpha$ 
values obtained from our model halos, we draw an anisotropy $\beta$ at 
random from the posterior distribution for the fit in Section~\ref{ss:beta} 
and calculate the mass $\Mrmax_{\rm est}$. Then we can infer the true mass 
$\Mrmax$ from this mass estimate by drawing a value of $f$ at random from 
the Monte Carlo simulation sample. We adopt the median of these masses as 
the best mass estimate, and use the 1-$\sigma$ (15.9 and 84.1) percentiles 
to estimate uncertainties. Thus, we estimate the mass of the MW to be
\begin{equation}
    \Mrmax = 0.61\substack{+0.18 \\ -0.12} \times 10^{12}\,\Msun .
\end{equation}
By drawing $f$ and $\beta$ values at random, shot noise and the uncertainties 
on the anisotropy are naturally folded into our result. From this, we can 
estimate the circular velocity at $\rmax$, for which we find 
\begin{equation}
\vcirc (\rmax) = 259^{+35}_{-26} \,\mathrm{km\,s}^{-1}.
\end{equation}

By comparing $\Mrmax$ and the virial mass $\Mvir$ for our model halos, 
we can also infer a virial mass for the MW from our estimate of the mass 
within $\rmax$. Thus we find the virial mass implied by our tracer mass 
estimates to be
\begin{equation}
    \Mvir = 2.05^{+0.97}_{-0.79} \times 10^{12}\,\Msun .
\end{equation}
Again, the uncertainties on the anisotropy have been propagated into these 
uncertainties. We note that $\Mvir$ has larger fractional uncertainty than 
$\Mrmax$ due to the uncertainty in the extrapolation of the dark halo 
mass profile to large radii.

\section{Conclusions}
\label{s:conclusions}

We present high-precision PM measurements of 20 GCs in the MW halo. 
The bulk motions of numerous GC stars were compared against distant 
background galaxies to achieve a median PM uncertainty of $0.06 \masyr$ 
per coordinate per cluster. Our PM results are not affected by possible 
rotation in the plane of the sky, and therefore represents the 
COM motions. For 10 out of 20 GCs, we compared our new PM results with 
previous measurements, which have much larger random (and systematic) 
uncertainties. The new and old results agree within the claimed 
1$\sigma$ uncertainties for four of these, while the rest 
are inconsistent with each other. We also find that the previous PM 
measurements systematically imply higher $\vtan$ and $\vtot$ 
(except for one GC we inspected).

We derive space motions of our target GCs in the Galactocentric frame 
by combining the newly measured PMs with existing $\vlos$ measurements.
We find a clear dichotomy in the Galactocentric $\vtan$ distribution 
such that there are five GCs with $\vtan > 290$~\kms\ 
(NGC\,6101, Arp\,2, Terzan\,7, Terzan\,8, and Pal\,12) while all 
other GCs have $\vtan < 200$~\kms. Four of these five GCs with high 
$\vtan$ are confirmed to be associated with the Sgr dSph based on 
comparing their observed properties to those of model particles of 
\citet{law10a}. We have tested the Sgr associations for three 
other GCs (NGC\,4147, NGC\,5024, and NGC\,5053) and found that 
they can be ruled out based on our PM measurements.
We also discuss the Sgr association of NGC\,2419, the most distant 
cluster in our sample. Our PM implies that NGC\,2419 seems 
to be associated with the Sgr stream debris recently found at distances 
similar to this cluster. However, we find an interesting offset 
in motion in the transverse direction with respect to the stream path.
We integrate orbits of our target GCs based on our PMs to explore 
their origins. We find that the young GCs can be separated into 
two groups, one with high orbital energy and angular momentum 
(which includes the Sgr GCs and Pyxis), and the other with low 
energy and momentum. This likely indicates two different origins 
among the young GCs, namely, accreted versus within-MW formation. 
Terzan\,8 which is older and more metal-poor than the other Sgr GCs, 
is also on a significantly less energetic orbit. 

We use a selected sample of GCs from our targets as dynamical tracers 
to provide a robust estimate of the MW mass. To represent the halo GC 
population in the range $\rgc = 10$--40 kpc, we remove NGC\,2419, and 
include only one out of the four confirmed Sgr cluster. We then calculate 
the anisotropy parameter for the 16 clusters as 
$\beta = 0.609\substack{+0.130 \\ -0.229}$. This implies 
significantly more radially-biased orbits than the $\beta$ measured 
using halo stars in similar Galactocentric distance ranges, which 
may be due to the fact that the halo star samples are biased by 
dynamically cold substructures in the halo, and that previous 
measurements relied on $\vlos$ alone to compute $\beta$, 
and this procedure can be unreliable at large Galactocentric 
distances ($> 15$ kpc).
We provide estimates of the power indices for the number density profile, 
as well as the underlying potential and calculate the MW mass using a 
tracer mass estimator that considers the full 6D phase-space information 
as inputs. Our best estimate for the MW mass within $\rgc = 39.5$~kpc is 
$M (<39.5~\mathrm{kpc}) = 0.61\substack{+0.18 \\ -0.12} \times 10^{12}\,\Msun$. 
We extrapolate our mass estimate to calculate the virial mass to be 
$M_\mathrm{vir} = 2.05\substack{+0.97 \\ -0.79} \times 10^{12}\,\Msun$.

PM measurements using multi-epoch \hst\ data have provided a breakthrough 
in understanding dynamics of tracer objects in the MW halo. Through 
our HSTPROMO collaboration \citep{vdm14}, we are continuing to make 
progress in improving the quality and quantity of PM measurements. 
While the release of {\it Gaia} DR2 (and subsequent data releases) 
will undoubtedly provide PMs for a wealth of halo objects, PMs with 
\hst\ will continue to be powerful for distant stellar systems. 
\hst\ results such as those presented in this paper should also prove 
to be very useful as an independent check for the upcoming {\it Gaia} 
data releases.
\smallskip

\acknowledgments
We thank the anonymous referee for constructive feedback that helped 
improve the quality of this paper.  
Support for this work was provided by NASA through grants for programs 
GO-14235, GO-12564, and AR-15017 from the Space Telescope Science Institute 
(STScI), which is operated by the Association of Universities for Research in 
Astronomy (AURA), Inc., under NASA contract NAS5-26555. A.D. is supported by a 
Royal Society University Research Fellowship and STFC grant ST/P000541/1.

\facility{HST (ACS/WFC and WFC3/UVIS)}.

\end{document}